\documentclass[twocolumn]{aastex7}
\usepackage{chemformula}
\usepackage{CJKutf8}
\usepackage{booktabs}
\usepackage{longtable}
\usepackage{wasysym}
\usepackage{hyperref}
\usepackage{lipsum}

\let\longtable*\relax

\received{Apr 28, 2025}
\revised{Jul 30, 2025}
\accepted{Aug 20, 2025}
\submitjournal{APJ}

\begin{document}
\begin{CJK*}{UTF8}{gbsn}

\title{The Cosmic Shoreline Revisited: A Metric for Atmospheric Retention Informed by Hydrodynamic Escape}

\author[0000-0002-1592-7832]{Xuan Ji(纪璇)}
\affiliation{Department of the Geophysical Sciences, University of Chicago, Chicago, IL 60637 USA}
\email{xuanji@uchicago.edu}

\author[0009-0008-8739-0932]{Richard D Chatterjee}
\affiliation{Department of Atmospheric, Oceanic and Planetary Physics, University of Oxford}
\email{richard.chatterjee@physics.ox.ac.uk}

\author[0000-0002-0508-857X]{Brandon Park Coy}
\affiliation{Department of the Geophysical Sciences, University of Chicago, Chicago, IL 60637 USA}
\email{bpcoy@uchicago.edu}

\author[0000-0002-1426-1186]{Edwin Kite}
\affiliation{Department of the Geophysical Sciences, University of Chicago, Chicago, IL 60637 USA}
\email{kite@uchicago.edu}


\begin{abstract}

The “cosmic shoreline,” a semi-empirical relation that separates airless worlds from worlds with atmospheres as proposed by K. J. Zahnle \& D. C. Catling, is now guiding large-scale JWST surveys aimed at detecting rocky exoplanet atmospheres. We expand upon this framework by revisiting the shoreline using existing hydrodynamic escape models applied to Earth-like, Venus-like, and steam atmospheres for rocky exoplanets, and we estimate energy-limited escape rates for \ch{CH4} atmospheres. We determine the critical instellation required for atmospheric retention by calculating time-integrated atmospheric mass loss. Our analysis introduces a new metric for target selection in the Rocky Worlds Director’s Discretionary Time and refines expectations for rocky planet atmosphere searches. Exploring initial volatile inventory ranging from 0.01\% to 1\% of planetary mass, we find that its variation prevents the definition of a unique clear-cut shoreline, though nonlinear escape physics can reduce this sensitivity to initial conditions. Additionally, uncertain distributions of high-energy stellar evolution and planet age further blur the critical instellations for atmospheric retention, yielding broad shorelines. Hydrodynamic escape models find atmospheric retention is markedly more favorable for higher-mass planets orbiting higher-mass stars, with carbon-rich atmospheres remaining plausible for 55 Cancri e despite its extreme instellation. We caution that our estimates are sensitive to processes with poorly understood dynamics, such as atomic line cooling. Finally, we illustrate how density measurements can be used to statistically test the existence of the cosmic shorelines, emphasizing the need for more precise mass and radius measurements.

\end{abstract}

\section{Introduction} \label{sec:intro}

Using 500 hours of James Webb Space Telescope (JWST) time and 240 orbits of the Hubble Space Telescope (HST), the Rocky Worlds\footnote{\url{https://rockyworlds.stsci.edu/}} Director's Discretionary Time (DDT) program \citep{redfield_report_2024} aims to find atmospheres on rocky exoplanets. An essential first step is identifying the most promising targets.

\subsection{The Cosmic Shoreline and Atmospheric Escape}

The Rocky Worlds DDT Targets Under Consideration (TUCs)\footnote{\url{https://outerspace.stsci.edu/pages/viewpage.action?pageId=257035126}} are currently organized using a priority metric based on the ``cosmic shoreline" proposed by \cite{zahnle_cosmic_2017}, a semi-empirical boundary that separates solar system bodies with atmospheres from those without. The shoreline is given by the relative cumulative XUV irradiation experienced by the body being proportional to its escape velocity to the fourth power. The cumulative XUV irradiation expected to be received by planets was reconstructed using a scaling law with bolometric flux and stellar mass, following Eq. 27 in \cite{zahnle_cosmic_2017}. 

Stellar X-ray ($<100$ \r{A}) and EUV (100–1000 \r{A}) photons (XUV) heat the upper atmospheres of planets, and when sufficiently intense, can directly drive atmospheric expansion out of the gravitational well to space in a hydrodynamic wind \citep[e.g.,][]{watson_dynamics_1981}. Young stars on the pre-main-sequence emit higher XUV flux, and M dwarfs, due to their slow evolution, sustain high XUV emission over much longer timescales compared to solar-mass stars \citep{preibisch_evolution_2005}. Consequently, planets orbiting lower‑mass stars tend to experience somewhat enhanced atmospheric escape at a given bolometric instellation. \citep{tian_history_2015, tian_water_2015}. Recently, \citet{pass_receding_2025} revised the estimated cumulative XUV flux for mid-to-late M-dwarfs and included corrections for both the pre-main-sequence phase and energetic flaring events, which roughly triples the cumulative XUV flux compared to the canonical estimate from \citet{zahnle_cosmic_2017}, placing the cosmic shoreline farther from the host star.

Apart from the physics of escape being much richer than the energy limit suggests and the variation of XUV evolution with stellar mass, the traditional ``cosmic shoreline" does not account for variations that are expected in volatile content or composition. With a larger initial volatile content, an atmosphere can be retained longer at the same escape rate \citep{owen_atmospheric_2019}. It is not known what controls initial volatile abundance on rocky planets. The volatiles of solar System terrestrial bodies are thought to derive from chondritic material \citep[e.g.,][]{marty_volatiles_1999, halliday_origins_2013, fischer-godde_ruthenium_2017} through planetesimal accretion \citep[e.g.,][]{bond_making_2010, rubie_accretion_2015} or pebble accretion \citep[e.g.,][]{johansen_anatomy_2023}, modified by metal-silicate differentiation and by impacts at all scales \citep[e.g.,][]{lichtenberg_water_2019, lichtenberg_bifurcation_2021, hirschmann_early_2021, liu_natural_2022, malamud_uranus_2024}. During accretion, atmospheres are temporarily puffy and may boil off \citep{owen_atmospheres_2016}. The volatiles available to build planetary atmospheres are further influenced by magma ocean degassing, atmospheric thermal loss, among other processes \citep[e.g.,][]{sossi_atmospheres_2021, sakuraba_numerous_2021, chen_impact_2022, gu_composition_2024, adibekyan_linking_2024, hasegawa_bulk_2024}.

These uncertainties suggest significant spread in initial volatile content for exoplanets. Theory suggests that rocky planets can form with substantial water inventories \citep{raymond_making_2004, kite_water_2021, wordsworth_atmospheres_2022}. The discovery of less-dense, highly irradiated, and potentially volatile-rich small planets (e.g., \citealt{brinkman2023,hu_secondary_2024}) has motivated theoretical studies of low-density lava worlds \citep{piette_rocky_2023} and carbon-rich puffy exo-Venuses \citep{peng_puffy_2024}. Theory suggests worlds forming between the soot line and water ice line could be C-rich (up to 1 wt\% carbon) \citep{bergin_tracing_2015, bergin_exoplanet_2023}, and worlds that are water-rich tend to be C-rich as well \citep{reynard_carbon-rich_2023}. 


Realistic estimates of escape rates are crucial for locating the cosmic shoreline, and constraining it requires sophisticated modeling efforts. Stellar photons and stellar wind drive atmospheric escape processes \citep[reviewed by][]{catling_atmospheric_2017, tian_atmospheric_2015, gronoff_atmospheric_2020}. A commonly used approximation is the energy-limited escape formulation \citep[e.g.][]{watson_dynamics_1981, lopez_how_2012}: the absorbed high-energy flux balances the work done to escape the gravitational potential of the planet, with an efficiency parameter (\(\epsilon\)) accounting for uncertainties of energy-gain and loss processes. \(\epsilon\) is not well constrained and varies significantly across different atmospheric compositions. For \ch{H2}-rich atmospheres, \cite{lopez_how_2012} assumes \(\epsilon = 0.1^{+0.1}_{-0.05} \), while \cite{erkaev_xuv-exposed_2013} uses \(\epsilon = 0.15 - 0.6\) in their model. Observations of Helium outflows from Mini-Neptunes suggest typical mass loss efficiencies around  \(\epsilon \sim 0.2\) \citep{caldiroli_irradiation-driven_2022, zhang_escaping_2022, zhang_more_2022, zhang_outflowing_2023}. For Na/K-rich atmospheres, \cite{ito_hydrodynamic_2021} show that \( \epsilon \sim 10^{-4} - 10^{-3} \) due to strong atomic line cooling. Constraining \(\epsilon\) requires comprehensive hydrodynamic modeling including heating efficiency, ionization, radiative cooling, and a wealth of other processes.

However, only a few studies have modeled secondary atmospheres loss at a high XUV flux for rocky planets. Models are generally of two forms. The first allows adiabatic expansion through a subsonic radial velocity profile but is limited to an adjusted Jeans escape formula as the upper boundary condition at the exobase \citep[e.g.][]{tian_hydrodynamic_2008-1,tian_thermal_2009-1,tian_thermal_2009, nakayama_survival_2022}. Hydrodynamic thermospheres are thought to correctly model the transition to a transonic wind \citep{tian_hydrodynamic_2008,tian_thermal_2009} though this has not been well tested. Transonic outflow of secondary atmospheres has been modeled with 1D first-principles time-stepping solutions of the hydrodynamic equations \citep{ johnstone_extreme_2019,johnstone_hydrodynamic_2020,  munoz_heavy_2021, garcia_munoz_heating_2023} and with analytic approximations \citep{chatterjee_novel_2024}. 

Over the years, thermosphere models for secondary atmosphere have incorporated more detailed heating and cooling processes. For \ch{CO2}, the model used in \cite{tian_thermal_2009} and \cite{tian_thermal_2009-1} includes both neutral and ion species, accounts for heating from collisions, chemical reactions, photoelectrons, and radiative line heating. For cooling, the model considers adiabatic cooling, thermal conduction, and radiative line cooling, including molecular and atomic-oxygen $63~\mu m$ cooling. The \ch{H2O}-atmosphere model used in \cite{johnstone_hydrodynamic_2020} adds $\sim$ 500 chemical reactions, including 56 photoreactions. However, both models consider atomic line cooling only from Lyman-$\alpha$ and a specific O ($63~\mu$m) line. \cite{nakayama_survival_2022} expands upon these by including chemical reactions involving internal excitation and ionization, along with a comprehensive list of atomic line cooling processes. Their results suggest that, for an Earth-like \ch{N2} atmosphere exposed to the young Sun’s radiation, atomic line cooling can suppress escape rates by around $10^4\times$ compared to models that omit these processes \citep{johnstone_extreme_2019}. 
It highlights the need for more realistic reassessments of \ch{CO2} and \ch{H2O} and \ch{CH4} atmospheres, for example, whether the models accurately track gas energy exchange and how to explicitly resolve the roles of molecular and atomic species \citep[e.g.][]{garcia_munoz_heating_2024, garcia_munoz_nlte_2024}

\cite{chatterjee_novel_2024} developed an analytical framework to describe atmospheric escape, capturing the transition from hydrostatic (Jeans) to hydrodynamic escape while considering atomic line cooling and ion-electron interactions. Building on simulations from \citet{Johnson2013}, \citet{chatterjee_novel_2024} derived a threshold XUV flux, dependent on planetary escape velocity, such that the upper atmosphere is collisional and hot enough for the onset of hydrodynamic escape. Above this threshold, escape rates scale linearly with XUV flux only when the atmosphere is weakly ionized and atomic line cooling is relatively minor. When the upper atmosphere is strongly ionized, their model predicts that a collisional-radiative thermostat, driven by strong atomic line cooling \citep{nakayama_survival_2022}, dominates the dynamics. This thermostat, along with ion-electron recombination and other processes, may limit the hydrodynamic escape rate to a relatively constant value, independent of further increases in XUV flux, and thus enhance atmosphere retention during the saturated X-ray phase.


In summary, studies on atmospheric hydrodynamic escape of high-mean-molecular-weight atmospheres lack consensus because key physical processes—such as atomic line cooling—are often omitted, as noted in later studies. Despite these uncertainties, identifying the broad trends in atmospheric retention as a function of fundamental planetary properties—i.e., the cosmic shoreline—is increasingly urgent to help prioritize promising targets for atmosphere searches, and maximize the scientific return of observations. 

Stellar wind-driven escape from rocky planets \citep{dong_atmospheric_2018} and impact erosion \citep{zahnle_cosmic_2017, sinclair_evolution_2020, krasnopolsky_detection_2004, schlichting_atmosphere_2018, denman_atmosphere_2020, wyatt_susceptibility_2020} also play a role in shaping cosmic shoreline. Volcanic outgassing or volatile replenishment from impacts, both time-dependent, can compete with atmospheric escape \citep{kite_geodynamics_2009, kral_cometary_2018, kite_exoplanet_2020, krissansen-totton_predictions_2022, krissansen-totton_implications_2023}. Though recent studies of present‐day escape rates for Earth, Mars, and Venus \citep{Strangeway2010, Gunell2018, Gronoff2020} have found that intrinsic magnetism is not necessarily protective against escape, we note that the strongly ionized hydrodynamic outflow proposed to shape super-Earths \citep{chatterjee_novel_2024}, may be particularly susceptible to magnetic suppression  (see \citealt{Owen2014} for the mechanism but in the hot Jupiter context).  Here, however, we restrict our scope to the unmagnetized scenario, for which existing hydrodynamic models apply.  In any case, planetary dynamos can take hundreds of millions of years to form and have variable lifetimes \citep[e.g.,][]{Labrosse2007, Zhang2022, Luo2024}.

Few studies have considered the time-integrated atmospheric loss to determine the cosmic shoreline, and they have either focused on specific targets \citep{ribas_habitability_2016}, assumed energy-limited loss \citep{zahnle_cosmic_2017}, or considered only hydrostatic Jeans escape \citep{looveren_habitable_2025}. Here, we calculate the atmospheric mass lost over a rocky planet's lifespan via thermal hydrodynamic escape driven by evolving stellar XUV. By assuming an initial volatile mass available for escape and modifying this initial parameter, we determine the parameter space where a planet can retain its atmosphere and then predict the cosmic shoreline.

\subsection{Statistical Test}

We suggest that, in addition to resource-intensive JWST searches, the existence of the cosmic shoreline(s) can be tested through population-level statistical analysis of planet density. The shorelines define the distribution of bare rocky planets versus those with secondary atmospheres across parameter space. Density measurements might be used to distinguish these two populations and then evaluate whether their distribution follows the cosmic shoreline predictions.

Similar to sub-Neptunes where the \ch{H2/He}-rich atmosphere can inflates the planet’s observable radius \citep{lopez_how_2012,lopez_understanding_2014, luque_density_2022}, recent studies on Puffy Venus scenarios suggest that a sufficiently thick, high-mean-molecular-weight atmosphere on hot rocky exoplanets can inflate planetary radii, lowering density \citep{peng_puffy_2024}. By normalizing observed planetary densities against Earth-like composition models (ratio'd density), planets with extended atmospheres can potentially be distinguished from bare rocky planets. 

Atmospheric height increases with temperature and decreases with gravity, making radius boosting more pronounced for smaller, highly irradiated planets \citep{peng_puffy_2024}. Without volatile loss, ratio'd density is expected to decrease with increasing instellation and increase with planetary mass. However, stronger atmospheric escape on highly irradiated, low-mass planets leads to preferential atmospheric survival on larger, less-irradiated planets. Thus, analyzing whether ratio'd density increases or decreases with instellation and gravity provides a potential test for the existence of the cosmic shorelines.

The correlation between bulk density and instellation could also result from the solid body formation process, for example, Mercury’s high metal-to-silicate ratio compared to outer terrestrial planets. Several models predict metal-silicate separation in the innermost protoplanetary disk disk, leading to iron-enriched inner planets \citep[see the review by][]{ebel_elusive_2018}. These scenarios predict higher densities compared to Earth-like compositions for close-in planets. In contrast, our hypothesis assumes a random distribution centered on Earth-like compositions for bare rocks. Thus, the absolute bulk densities of innermost planets can help distinguish between iron-enriched gradient and atmospheric loss trends and test the shoreline-based interpretation. The steepness of the slope of the density–instellation trend may also serve as a diagnostic and warrants further study.

\subsection{Structure}

In this study, we combine a stellar evolution model, X-ray parameterization, EUV extrapolation, and atmospheric loss rate prescriptions (Sec. \ref{sec:method}). In Sec. \ref{sec:stats}, we present our method for estimating the cosmic shorelines. Sec. \ref{sec:results} presents our revised cosmic shorelines and the re-ranking of target exoplanets. We introduce the density trend test for the cosmic shorelines in Sec. \ref{sec:density}.  Finally, we discuss limitations of this work in Sec. \ref{sec:discussion}.

\section{Model} \label{sec:method}

\subsection{Assumptions and Initial Conditions}

We adopt the mass-radius relation for Earth-like composition from Fig.~1 of \cite{zeng_growth_2019} (Fig.~\ref{fig:MR}) to determine the planetary radius (\(R_p\)). Here, \(R_p\) refers to the radius at the solid surface, excluding the atmosphere. We consider atmospheric compositions that are \ch{CO2}-dominated, \ch{CH4}-dominated, \ch{H2O}-dominated, or \ch{N2}-dominated. 

The volatile inventory available for escape is defined as the total volatile content in the molten silicate and atmosphere after the planet forms. Our reasoning is as follows: accretional energy melts the silicate, distributing volatiles between the magma ocean and the atmosphere. The sequestration of volatiles into the solid mantle during the solidification of the primary magma ocean is small relative to their exsolution to the atmosphere \citep{elkins-tanton_linked_2008} (and Appendix \ref{sec:app}). For hot rocky lava worlds, as atmosphere escapes, the greenhouse effect weakens, leading to surface cooling and magma ocean freezing. This process sequesters even fewer volatiles into the mantle than the result derived in Appendix \ref{sec:app}, since volatile concentration in the magma decreases as the atmosphere depletes. Therefore, our assumption of small sequestration and small volcanic revival is reasonable for the scope of this study, though more detailed investigations are warranted. Thus almost all initial volatiles are available to escape. We assume no subduction nor carbon cycle, focusing on planets too hot for surface liquid water.


For \ch{CO2}- and \ch{CH4}-dominated atmospheres, we assume an initial carbon mass and calculate the carbon loss rate. For \ch{N2}-dominated and \ch{H2O}-dominated atmospheres, we assume an initial nitrogen mass and calculate the nitrogen loss rate. \ch{H2O} is very soluble in magma and has non-negligible solubility in solid silicate \citep{kite_water_2021}. As a result, the actual initial \ch{H2O} concentration in the bulk silicate portion of the planet is likely higher than the values presented here.

To quantify volatile loss, we prescribe the atmospheric escape loss rate using hydrodynamic models. (\ch{CO2}: \cite{tian_thermal_2009-1, tian_thermal_2009}; \ch{N2}-dominated: \cite{nakayama_survival_2022, chatterjee_novel_2024}; and \ch{H2O}: \cite{johnstone_hydrodynamic_2020}). We consider stellar masses (\(M_*\)) ranging from 0.1 to 1 $M_\odot$. Starting at a certain stellar age, the total volatile losses integrated over the lifetime of the planet are compared to the initial volatile content to assess if the atmosphere is retained.

Solar wind-driven escape from rocky planets and impact erosion are not included in this initial investigation. Volatile replenishment from impacts is assumed to be small relative to the initial volatile inventory. We also assume a fixed semi-major axis throughout the planet's lifespan and do not consider the planetary migration.

\subsection{Stellar XUV evolution}

All atmospheric escape mechanisms are influenced by upper atmosphere temperature. The primary source of heating in the thermospheres of planets is XUV flux. To calculate the total atmospheric loss a planet may experience, it is essential to account for the evolution of XUV flux over time. To do this, we adopt bolometric luminosity evolution models and scale the X-ray and EUV components accordingly.

\subsubsection{Stellar evolution}\label{sec:star}

We use the bolometric luminosity and stellar radius evolution from \citep{baraffe_new_2015} (Fig.~\ref{fig:baraffel}). We neglect this uncertainty as it is significantly smaller than the uncertainties introduced by the X-ray models and EUV extrapolation. We adopt a stellar age distribution from \cite{berger_gaiakepler_2020}, and apply an upper age cutoff that scales with stellar mass according to the stellar lifetime relation: \(\tau_{max}(M_*)=10^{10}/(M_*/M_\odot)^{2.5} \text{yr}\). We also tested the influence stellar age distributions by combining the distribution from \citet{swastik_age_2023} for FGK stars and \citet{Gaidos2023} for M dwarfs. Although \citet{Gaidos2023} provides a distribution skewed toward younger stars, the resulting cosmic shorelines closely overlap with those from our default case based on \citet{berger_gaiakepler_2020}.


\subsubsection{X-ray}\label{sec:method-xray}

Stellar X-ray luminosities decrease over time as magnetic braking slows stellar rotation. For young stars, stellar X-ray emission is ``saturated"  with a constant ratio X-ray luminosity to bolometric luminosity, \( L_X/L_{bol} \) \citep[e.g.,][]{vilhu_nature_1984}. After this phase, \( L_X/L_{bol} \) declines exponentially with time, following a power-law decay characterized by a decline index \(L_X/L_{bol}  \propto t^{-\alpha}\). To determine \( L_X/L_{bol} (t) \), we use two approaches. The first method (S07) follows the results of Figure ~5 from \citet{selsis_habitable_2007}, which is based on ROSAT observations, assuming a constant \( L_X/L_{bol} = 10^{-3.2} \) during saturation, followed by a power-law decline with fixed index \( \alpha \). The duration of the saturation phase varies as a function of stellar mass for \( 0.1 < M_\odot < 1 \), with lower-mass stars staying saturated for longer. This method provides a simplified yet widely used prescription for estimating cumulative XUV exposure across different stellar masses. The second (J12+G16) combines parameterizations from \citep{jackson_coronal_2012} for stellar masses above 0.5 \( M_\odot \), based on ROSAT, Chandra, and XMM–Newton data, and from \citep{guinan_living_2016} for stellar masses below 0.5 \( M_\odot \), derived from HST and Chandra observations. Notably, in J12+G16, the saturated X-ray-to-bolometric flux ratio and decline index (\(\alpha\)) vary with stellar mass, while they remain relatively constant in S07 (Fig. \ref{fig:X-ray}). In our Monte Carlo simulations (Sec.~\ref{sec:stats}), we randomly select between the two models with equal probability.

\subsubsection{EUV}


High-energy solar corona observations suggest that solar EUV emission remains relatively strong even as X-ray surface flux decreases on a timescale of hours, which suggests that EUV may decline more slowly over time compared to X-rays, though the timescale here is in Gyr \citep{chadney_xuv-driven_2015, king_xuv_2018, johnstone_active_2021}. However, EUV fluxes for most stars are difficult to measure due to strong interstellar absorption in this wave band. 

To estimate EUV flux based on X-ray flux, we use the scaling relation from \cite{king_euv_2020} (their Eq. 3 \& Fig.~2):

\begin{equation}
\frac{L_{EUV}}{L_{bol}} = \beta \left( \frac{L_{bol}}{A} \right)^\gamma \left( \frac{L_X}{L_{bol}} \right)^{\gamma+1}
\label{eq: euv}
\end{equation}
where $A$ is the stellar surface area, calculated using the stellar radius from \cite{baraffe_new_2015}. The parameters vary for different EUV bands. \cite{king_xuv_2018} provide parameters for different EUV bands, with lower bounds ranging from 51 to 124 \r{A} and the upper bound fixed at 912 \r{A}. \cite{king_euv_2020} refines the parameters, determining that for wavelengths 100–360 \r{A}, $\gamma_{\text{hard}} = -0.35^{+0.07}_{-0.15}$ and $\beta_{\text{hard}} = 116$ (erg s$^{-1}$ cm$^{-2})^{\gamma_{\text{hard}}}$, while for wavelengths 360–920 \r{A}, $\gamma_{\text{soft}} = -0.76^{+0.16}_{-0.04}$ and $\beta_{\text{soft}} = 3040$ (erg s$^{-1}$ cm$^{-2})^{\gamma_{\text{soft}}}$, where the uncertainties cover the variations by different bands introduced in \cite{king_xuv_2018}. We adopt these values as input (Fig.~\ref{fig:EUV}).

Models uncertainties include: (1) different X-ray models and (2) uncertain EUV extrapolation parameters.

\subsection{Atmospheric Escape Rates} \label{sec:hydro}

Several studies modeled the transition to hydrodynamic escape for high-mean-molecular-weight atmospheres. In this study, we adopt atmospheric loss rates as a function of XUV flux for atmospheres with different compositions from previous studies (\ch{CO2}: \cite{tian_thermal_2009-1, tian_thermal_2009}; \ch{N2}-dominated: \cite{nakayama_survival_2022, chatterjee_novel_2024}; and \ch{H2O}: \cite{johnstone_hydrodynamic_2020}).  

\subsubsection{\textsc{\ch{CO2}} - with detailed simulations for high-mass super-Earths}

For a \ch{CO2}-dominated atmosphere, we obtain escape rates (kg/s) from Fig.~6 of \cite{tian_thermal_2009} by converting the escape flux per unit surface area (kg/s/m²) (at the solid surface with radius of \(R_p\)) into the total atmospheric loss rate. We also extract escape rates from Fig.~4 of \cite{tian_thermal_2009-1}, converting the time axis into XUV flux using their assumption that \( F_{XUV} = 29.7 \cdot t^{-1.23} \) erg cm\(^{-2}\)s\(^{-1}\), where \( t \) is the stellar age in billions of years. We then interpolate the data to obtain a continuous function for the escape rate as a function of XUV flux. Beyond the XUV flux limits (\(F_{XUV}  \gtrapprox 1000 \times\) present-day Earth value), we extrapolate the loss rates linearly on a logarithmic scale of $F_{XUV}$. \cite{tian_thermal_2009-1} provides output for a Mars-sized planet, while \cite{tian_thermal_2009} presents results for planets with 5.9$\times$, 7.5$\times$, and 10$\times$ Earth masses, and there was no experiment conducted for Earth-sized planets. Consequently, the atmospheric escape rate for an Earth-sized planet is highly sensitive to the interpolation approach used. We illustrate various approaches in Fig.~\ref{fig:CO2-interpolation} and adopt the maximum and minimum values for a given XUV flux as the uncertainty range.



\subsubsection{\textsc{\ch{H2O}} - featuring transonic hydrodynamics}

For steam (\ch{H2O}) atmospheres, we use the results from \cite{johnstone_hydrodynamic_2020}'s Kompot Code. We interpolate the escape rate as a function of \( F_X + F_{EUV} \), using data from their Table 1.  Notably, their \( F_{EUV}/F_X \) ratio is $10\times$ higher than ours (Fig.~\ref{fig:EUV}). If we instead interpolate the escape rate using X-ray flux alone, the resulting escape flux would increase by a factor of 3. However, we also acknowledge that a subsequent study by the same author, \cite{johnstone_active_2021}, reports a \( F_{EUV}/F_X \) ratio more consistent with our adopted scaling law (Fig.~\ref{fig:EUV}). Their results suggest that hundreds of bars of \ch{O2} can accumulate due to the different loss efficiencies of H and O, though not for the highest XUV fluxes, but we do not account for this in our model.

\subsubsection{\textsc{\ch{N2}} - accounting for atomic line cooling}\label{sec:N2O2-method}

For an \ch{N2}-dominated atmosphere, we consider two studies with differing conclusions. We firstly consider the results from \cite{nakayama_survival_2022} (hereafter N2022, 79.58\% \ch{N2}, 20.40\% \ch{O2}, 0.02\% \ch{CO2}, and 6$\times 10^{-6}$ \ch{H2O} as lower boundary condition). We convert the erosion timescale from their Fig.~8 into flux by dividing the mass of a 1-bar atmosphere by the timescale. \cite{nakayama_survival_2022} found that an \ch{N2}-dominated atmosphere is protected by atomic line cooling up to $>1000\times$ present-day Earth's $F_{XUV}$. Their model includes the process in \cite{johnstone_upper_2018}, and a more comprehensive treatment of atomic line cooling using nitrogen (N) and oxygen (O) (ionized and neutral). They only explore the Jeans escape as their results show the atomic line cooling is strong enough to prohibit the onset of hydrodynamic escape. 

We also use results from \cite{chatterjee_novel_2024} (hereafter CP2024) to determine the escape rate for an \ch{N2}-dominated atmosphere across various planet masses, considering atomic line cooling from nitrogen. In their simulations, oxygen cooling is considered for Mars- to Earth-sized planets but not for super-Earths. CP2024 considers a threshold in XUV flux to drive transonic outflow based on an onset in collisionality for Mars-to-Earth sized planets (when advective cooling dominates) and an onset for Super-Earths based on the collisional radiative thermostat (when line cooling dominates).



In CP2024, the escape rate as a function of XUV flux is divided into three regimes: 1. weak XUV flux: escape is negligible; 2. intermediate XUV flux: the escape rate per unit surface area is energy-limited and increases proportionally with XUV flux (energy-limited regime, hereafter); 3. high XUV flux: due to the``thermostat" effect of strong atomic line cooling and the restricted penetration of XUV radiation, the escape rate are assumed to be approximated as reaching a maximum value independent of further XUV increases. We refer to this regime as the ``high-XUV plateau" hereafter. In summary, our implementation of CP2024 will serve to illustrate the result of energy-limited behavior restricted to a range of XUV fluxes. The inclusion of a threshold and plateau for hydrodynamic escape also depends on planet mass - the XUV range of energy-limited regime and the effective efficiency decreases for more massive planets with larger escape velocities (Fig.~\ref{fig:N2O2-loss-rate}).

The bottom two rows of Table \ref{tab:CP24} show the range of XUV flux in which one can apply the energy-limited approximation. In this regime, the loss rate follows the scaling relation \(\dot{C}/\dot{C}_{\star} = F_{\text{XUV}}/F_{\text{XUV},\star}\). For XUV fluxes exceeding this range, the loss rate remains constant at its highest value within this regime. For fluxes below the energy-limited regime, we arbitrarily set the loss rate per surface area to a very low value of \(10^{-17}\) kg/s/m\(^2\). To generalize the loss rate for any planetary mass, we perform a 2D interpolation for the energy-limited escape rate per surface area. Additionally, we use lower and upper limits to estimate the escape rate, as listed in the bottom two rows of Table~\ref{tab:CP24}, and vary the rate randomly within this range in our simulations (see Fig.~\ref{fig:N2O2-loss-rate-frac}). The lower limit accounts for the possibility of subsonic outflow as discussed in \cite{chatterjee_novel_2024}.

\begin{table*}[htbp]
    \centering
    \resizebox{\textwidth}{!}{
    \begin{tabular}{lcccc}
        $M_{p,\star}~(M_\oplus)$ & 0.11 & 1.00 & 1.76 & 5.90 \\
        \hline
        $F_{xuv,\star}~(F_{xuv,\oplus})$ & 50 & 400 & 1000 & 2000 \\
        Loss Rate, $\dot{C}_{\star}$ (kg/s) & $3\times 10^5 $ & $10^6$ & $2.2\times10^5$ & $3.4\times10^5$ \\
        Loss Rate per Surface Area (kg/m$^2$/s) & $2.3\times10^{-9}$ & $2.0\times10^{-9}$ & $3.1\times10^{-10}$ & $2.6\times10^{-10}$ \\
        Energy Limited Range of $F_{xuv}~(F_{xuv,\oplus})$ & $5 - 500$ & $126.5 - 1264.9$ & $501.2 - 1995.2$ & $1954.5 - 2046.6$ \\
         & $5 - 5000$ & $126.5 - 12649.1$ & $501.2 - 19952.6$ & $1954.5 - 20465.8$ \\
        \bottomrule
    \end{tabular}
    }
    \caption{Adjusted escape rates from  \cite{chatterjee_novel_2024} for constraining an approximately energy limited regime of an \ch{N2}-dominated atmosphere. Atmospheric escape rate as a function of XUV flux follows three regimes: negligible at low flux, energy-limited and proportional at intermediate flux (energy limited regime), and plateau at high flux due to atomic cooling. \( F_{\text{xuv},\star} \) (in units of Earth’s present-day XUV flux, \( F_{\text{xuv},\oplus} \)) and \( \dot{C}_{\star} \) represent a reference point within the energy-limited regime. For $F_{xuv}$ within the energy limited regime, the escape rate scales linearly: $\dot{C} = \dot{C}_{\star}\cdot \left(F_{xuv}/F_{xuv,\star}\right)$. The last two rows indicate the upper and lower limits of where the energy-limited regime transitions to the saturation regime, which corresponds to an uncertainty of a factor of 10. }
    \label{tab:CP24}
\end{table*}

\subsubsection{\textsc{\ch{CH4}} - in the energy limit}

We also consider pure \ch{CH4} atmospheres. In the absence of detailed hydrodynamic escape modeling specifically for \ch{CH4}, we use an energy-limited loss rate estimation \citep[e.g.][]{watson_dynamics_1981,lopez_how_2012}

\begin{equation}
    \dot{C}_{CH4} = \epsilon\frac{\pi F_{XUV} R_{p}^3}{GM_{p}} \label{eq:EL1}
\end{equation}
where \( G \) is the gravitational constant, and \( M_p \) and \( R_p \) are the planetary mass and radius. The efficiency \( \epsilon \), which measures the cooling effect and other uncertain factors, is not well constrained as discussed in Sec. \ref{sec:intro}. Here, we adopt a uniform distribution for the logarithm of efficiency, \( \epsilon \sim 5\times 10^{\mathcal{U}(-2,-1)} \) based on the line cooling effects explored in \citep{yoshida_suppression_2024}. For very low XUV flux, H will escape leaving C behind (e.g. Titan), but we do not model this. 

The escape of \ch{CH4} is also constrained by photodissociation driven by Lyman-alpha radiation (\ch{CH4} + h$\nu$ (1216 \r{A}) $\rightarrow$ products) \citep{krasnopolsky_detection_2004, wordsworth_transient_2017}. Using Earth’s mean Lyman-alpha photon flux ($I_{\text{Ly}\alpha} = 3.7 \times 10^{11} \ \text{photons cm}^{-2} \text{ s}^{-1}$) \citep{woods_validation_1996}, the maximum carbon mass loss rate can be estimated as $I_{\text{Ly}\alpha} \cdot m_C \cdot \pi R_\oplus^2 \sim 10^4$ kg s$^{-1}$. This estimate is one order of magnitude higher than the energy-limited escape rate with an efficiency $\epsilon = 0.1$ and present-day Earth’s XUV flux of 0.00464 $\text{W}/\text{m}^2$. Thus, as long as Lyman-alpha scales with the total XUV flux, methane photolysis does not further limit the escape rate, which remains energy-limited.

Eq.~\ref{eq:EL1} assumes the XUV absorption radius \( R_{XUV} = R_p \), at which the atmosphere becomes optically thick to XUV photons, following \cite{ito_hydrodynamic_2021, stafne_predicting_2024}, though $R_{XUV}$ can be computed by integrating the XUV absorption cross-section \citep{murray-clay_atmospheric_2009,lopez_how_2012, owen_mapping_2024, yoshida_less_2022}. The tidal enhancement factor \( K_{\text{tide}} \), which accounts for the reduced energy required for atmospheric escape within the planet's Roche lobe  \citep{erkaev_euv-driven_2016}, is also not explicitly computed. For \( R_{XUV}\) smaller than twice Hill radius, \( K_{\text{tide}} < 2\). In this study, the effects of varying \( R_{XUV} \) and \(K_{\text{tide}}\) are incorporated into the uncertainty range of \(\epsilon\).

\subsubsection{Summary}

Hydrodynamic simulations for heavy-mean-molecular-weight atmospheres remain limited, preventing meaningful inter-model comparisons. This paper serves as an initial analysis to investigate how the cosmic shorelines can be determined using more realistic atmospheric escape rates derived from hydrodynamic simulations. We emphasize the need for more modeling work.

\begin{figure*}[h!]
    \centering
    \includegraphics[width=0.8\linewidth]{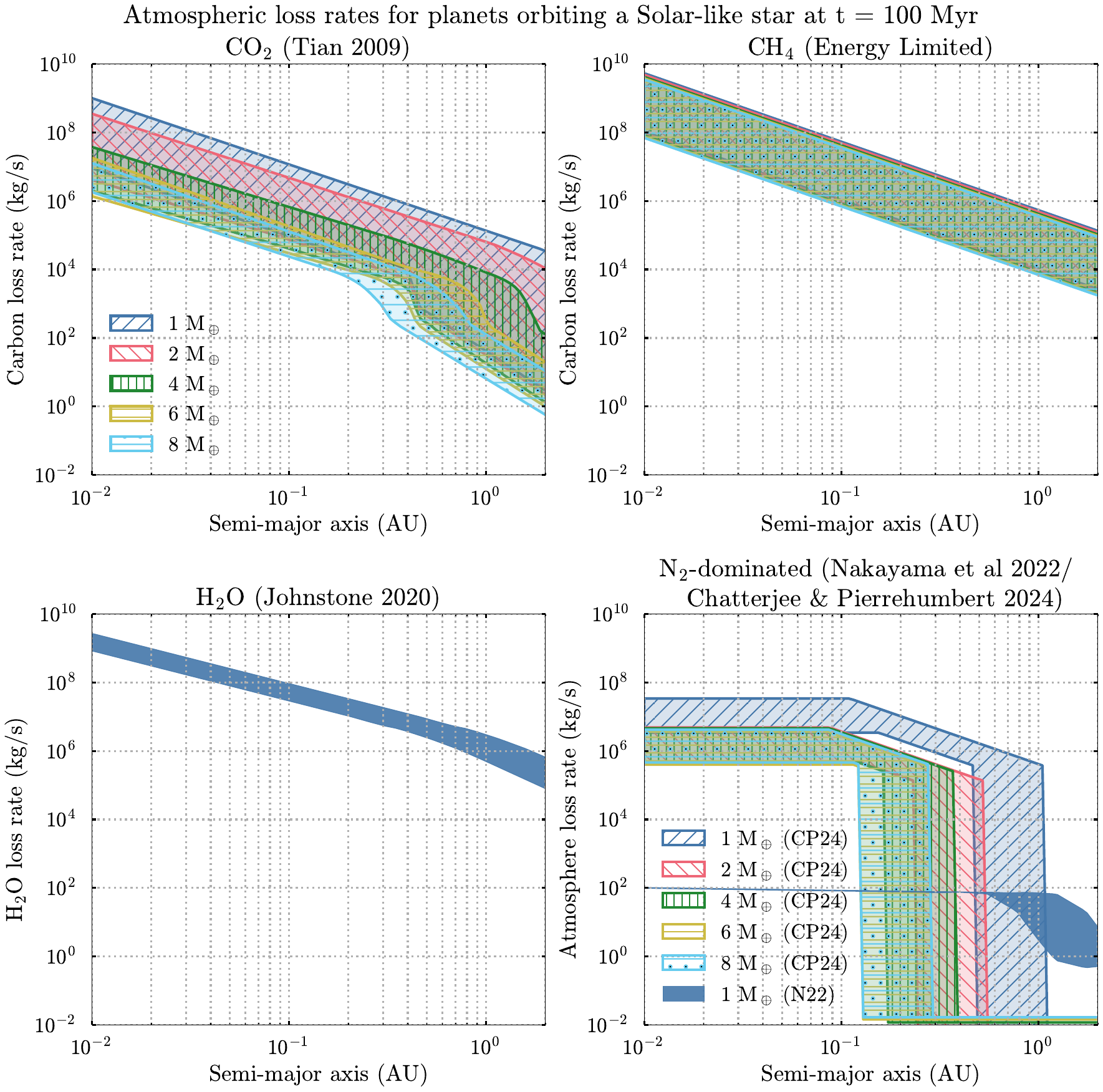}
    \caption{\emph{Atmospheric loss rates for planet of different masses orbiting a Sun-like star at \( t = 100 \) Myr for a range of escape models/compositions}. For \ch{CO2}-dominated, and \ch{CH4}-dominated, the escape rates for varying $M_p$ are color-coded, as indicated in the legend. For \ch{N2}-dominated atmospheres, we use results from two different studies. CP2024 refers to the analytical framework developed by \citet{chatterjee_novel_2024} which provides rates for varying $M_p$, while N2022 is an 1D numerical model from \citet{nakayama_survival_2022}, which only applies to Earth-mass planets. The shaded regions represent the range of atmospheric loss rates, including uncertainties in XUV flux. Additional uncertainties are included for specific atmospheres, such as \ch{CO2} interpolation methods, the \( F_X/F_{EUV} \) ratio (\ch{H2O} atmosphere), and efficiency (\(\epsilon\)) (\ch{CH4} atmosphere) (Table \ref{tab:monte_carlo}). For \ch{CH4} atmospheres, the loss rate weakly depends on $M_p$, whereas \ch{CO2} loss is strongly mass-dependent. For \ch{N2}-dominated atmospheres, the only two studies that consider atomic line cooling, show the escape rates stay constant at higher XUV flux, but the values differ significantly: \citet{nakayama_survival_2022} suggests that \ch{N2}-dominated atmospheres escape very slowly, whereas \citet{chatterjee_novel_2024} predicts high escape rates at high XUV flux.}
    \label{fig:atm_loss}
\end{figure*}

By combining the escape rate prescription with the stellar model described above, the atmospheric escape rate as a function of time for a planet orbiting a specific type of star at a given semi-major axis can be determined, though orbits can change over time. Instead of providing a single value, we show a range of atmospheric escape rates for a planet with a given mass and orbital distance, incorporating all previously discussed uncertainties (Fig.~\ref{fig:atm_loss}). The contributions of each uncertainty component are illustrated in Fig.~\ref{fig:CO2-unc}, using \ch{CO2} as an example. The key assumptions and the sources of uncertainties in the escape models are listed in Table~ \ref{tab:model_assumptions}.

For \ch{CO2} and \ch{CH4} atmospheres, the results apply to planets with masses ranging from 0.5 to 10 \( M_\oplus \), while for \ch{H2O}atmospheres, hydrodynamic simulations are only available for planet with  \( 1 M_\oplus \). For \ch{N2}-dominated atmosphere, we show  \( 1 M_\oplus \) results from both N2022 and CP2024, and apply to other masses using the CP2024 prescription.

\section{Statistical Method}\label{sec:stats}

\begin{table*}
\caption{Model input parameters used for the Monte Carlo simulations.}
\label{tab:monte_carlo}
\centering
\begin{tabular}{lll}
\hline
\textbf{Parameter}        & \textbf{Distribution / Choices}            & \textbf{Description} \\
\hline
$S/S_0$        & $ 10^{\mathcal{U}(-2,5)}$    & Bolometric instellation scaled to Earth's value. \\
$\tau_*$     & Histogram from Fig.1 in \cite{berger_gaiakepler_2020} & Stellar age distribution. \\
$\tau_0$  & $10^{\mathcal{U}(6,8)}$ yr 
                                                                   & Planet formation time \footnote{\cite{righter_terrestrial_2011}}. \\
\texttt{xray\_model}      & \{\texttt{"Selsis"}, \texttt{"Jackson"}\} 
                                                                   & Choice of X-ray flux evolution model (Sec.~\ref{sec:method-xray}). \\
$\gamma_{\text{hard}}$      & $\mathcal{U}(-0.5, -0.28)$ & Power-law index for EUV extrapolation (Eq. \ref{eq: euv}). \\
$\gamma_{\text{soft}}$       & $\mathcal{U}(-0.8, -0.6)$  & Power-law index for EUV extrapolation (Eq. \ref{eq: euv}). \\
\texttt{CO2\_method}         & \{\texttt{"linear"}, \texttt{"log"}, \texttt{"GP"}\} 
                                                                   & \ch{CO2} escape rate interpolation method (Sec. \ref{sec:hydro}). \\
$\epsilon$  & $5\times 10^{\mathcal{U}(-2, -1)}$ & Escape efficiency for energy-limited \ch{CH4} escape. \\
\hline
\end{tabular}
\end{table*}

Given the many uncertainties of our model assumptions (e.g., stellar XUV flux, interpolation methods, and evolutionary timescales), we use a Monte Carlo approach to statistically evaluate the cosmic shorelines. Here, the cosmic shoreline is defined as the critical bolometric instellation (\(S^*_{bol}\)) below which a planet has more than a 90\% probability of retaining its atmosphere, given an initial volatile inventory \(M_{0,\text{v}}\). We explore how the cosmic shoreline depends on stellar mass and planetary mass, among other parameters that are harder to measure precisely (e.g., stellar age and formation time). Each uncertain parameter is drawn from an assumed probability distribution (Table~\ref{tab:monte_carlo}), and for each draw, we calculate the total atmospheric loss (\(M_{\text{loss}}\)) over the planet’s lifetime. For Monte Carlo simulations at fixed \(M_p\) and \(M_*\), we conduct \(10^4\) trials, labeling each outcome \(i\) according to whether the planet retains its atmosphere:
\begin{equation}
Y_i = 
\begin{cases}
1, & \text{if } M_{\text{loss}} \geq M_{0,v}, \\
0, & \text{if } M_{\text{loss}} < M_{0,v}.
\end{cases}
\end{equation}
Thus, \(Y_i = 1\) indicates that the initial atmosphere is totally lost, whereas \(Y_i = 0\) indicates it is retained.

We then employ logistic regression to estimate the relationship between the logarithm of bolometric instellation \(\log S_i\) and the probability of losing the atmosphere. Concretely, we model
\begin{equation}
    P(Y_i = 1 \mid S_i) \;=\; 
\frac{p_{max}}{1 + \exp\!\bigl(-(\alpha + \beta\, \log S_i)\bigr)}
\end{equation}
where \(\alpha\) and \(\beta\) are parameters fit to our Monte Carlo dataset \(\{(S_i, Y_i)\}\). The $p_{max}$ is a free parameter for CP2024 \ch{N2}-dominated simulations but is forced to be 1 for other cases (see Fig.~\ref{fig:logistic}). When averaging \(\hat{\alpha}\) and \(\hat{\beta}\) are obtained, we solve for the critical instellation \(S^*_{bol}\) that yields a chosen probability \(p_0\) of complete atmospheric loss:
\begin{equation}
p_0 
\;=\; 
P(Y_i=1 \mid S^*_{bol}) 
\;=\; 
\frac{p_{max}}{1 + \exp\!\bigl(-(\hat{\alpha} + \hat{\beta}\,\log S^*_{bol})\bigr)},
\end{equation}
which implies
\begin{equation}
    \log  S^*_{bol} \;=\; \frac{\ln\!\bigl(\tfrac{p_0}{p_0-p_{max}}\bigr) - \hat{\alpha}}{\hat{\beta}}.
\end{equation}

We then obtain the critical instellation (\(S^*_{bol}\)) with a given loss probability \(p_0\). We adopt \(p_0 = 10\%\) as the central definition of the cosmic shoreline, meaning that below \(S^*_{bol}\), there is a 90\% probability that the atmosphere is retained. We also show the range corresponding to \(p_0 = 50\%\) and \(p_0 = 1\%\), representing 50\% and 99\% retention probabilities, respectively. 

From our 10,000-point Monte Carlo sample for a given planetary mass and stellar mass, we then employ a bootstrap approach to obtain a robust estimate of the \( S^*_{bol} \) for a chosen probability level \(p_0\). Specifically, we randomly draw \(5{,}000\) data points from the original set and we refit the logistic model to this bootstrap sample to obtain parameters and compute the corresponding \( S^*_{bol} \).  We repeat it for 50 times and then average the 50 resulting \( S^*_{bol} \) values to get the final cosmic shoreline. Fig.~\ref{fig:logistic} illustrates example logistic fits. By repeating this procedure for various stellar and planetary masses while incorporating the uncertainties in formation time, age, and other factors, we trace out the cosmic shoreline in parameter space.

\section{Cosmic Shorelines} \label{sec:results}

The cosmic shoreline in this study is defined as the critical bolometric instellation (\( S^*_{bol} \)) below which a planet is highly likely (\(>90\%\)) to retain its atmosphere, given the mass of volatiles available for loss. We focus on thermal escape driven by stellar XUV flux. \( S^*_{bol} \) depends on $M_*$, as the cumulative XUV flux varies with stellar type, and on $M_p$, which affects gravitational binding energy and escape physics. 

Additionally, \( S^*_{bol} \) depends on the initial volatile mass. For \ch{CO2}- and \ch{CH4}-dominated atmospheres, we assume an initial C mass ranging from \( 10^{-4} \times\), which is roughly carbon mass ratios in the atmosphere and silicate of Earth and Venus, to \( 10^{-2} \times \) planetary mass. For \ch{N2}-dominated atmospheres, we assume a fixed available \ch{N} mass ranging from \( 10^{-4} \times\) to \( 10^{-2} \times \) planetary mass. which may not fully capture reality, as such atmospheres may not originate from the initial volatile partitioning during planet formation, but could be replenished later through a water loss redox pump mechanism which is time-dependent and outgassing flux-limited \citep{wordsworth_atmospheric_2016, schaefer_predictions_2016, wordsworth_redox_2018, kite_water_2021}. To reflect this uncertainty, we use a dashed line to represent the cosmic shoreline for \ch{N2}-dominated atmospheres with a fixed initial nitrogen mass fraction. For \ch{H2O} atmospheres, we consider an initial water mass between \(10^{-4}\) and \(10^{-2}\) of the planetary mass. If these volatiles remain in the atmosphere, this corresponds to surface pressures of approximately \(10^2\) and \(10^4\) bar.

Note that differences in the cosmic shorelines across atmospheric compositions may partially reflect real physical processes, but they also depend strongly on model assumptions. For example, if future models for \ch{CO2} atmospheres incorporate atomic line cooling, the corresponding shoreline would likely shift closer to the star.

\subsection{Instellation vs Stellar Mass}

\begin{figure*}
    \centering
    \includegraphics[width=0.9\linewidth]{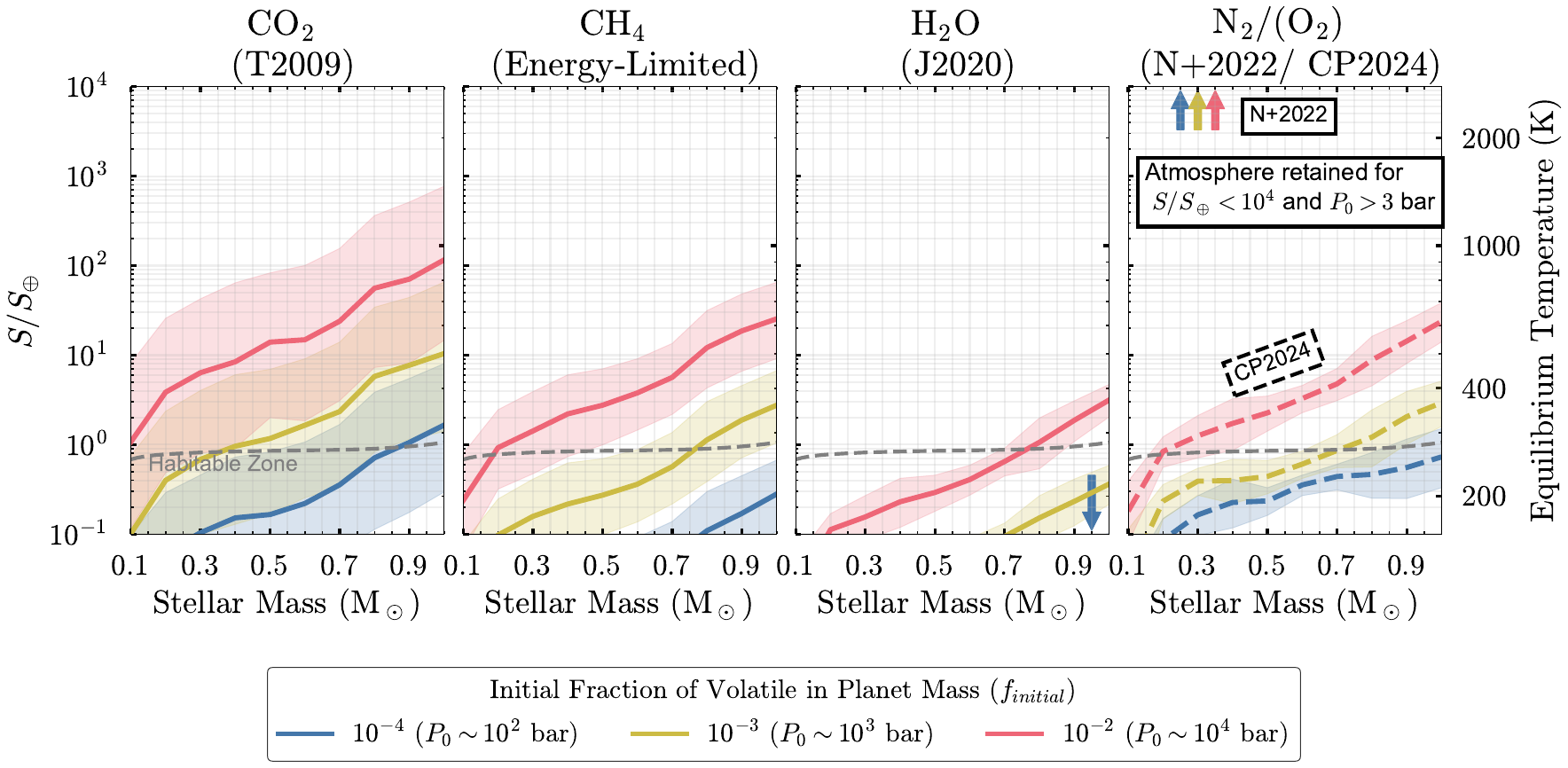}
    \caption{\emph{Critical instellations for atmospheric retention vs stellar mass for Earth-mass planets across different escape models/compositions and probing a range of volatile abundances.} The critical bolometric instellation (\(S^*_{bol}\)) is normalized to Earth's insolation and stellar mass relative to the Sun. The right y-axis also shows the corresponding equilibrium temperature ($T_{eq}$), assuming no albedo and perfect redistribution. Blue, yellow, and red lines correspond to initial volatile fractions of \(10^{-4}\), \(10^{-3}\), and \(10^{-2}\) in planetary mass, respectively. The shaded regions represent a 50–99\% probability range of atmospheric retention output by our model, and the solid curves represent 90\% probability of atmospheric retention. For \ch{H2O} atmospheres, the blue line lies below \(1\, S_{\oplus}\). For an \ch{N2}-dominated atmosphere, predictions from \cite{nakayama_survival_2022} (N2022) suggest that all three cosmic shorelines exceed \(10^4\, S_{\oplus}\), indicating extreme resilience to atmospheric escape. In contrast, \cite{chatterjee_novel_2024} (CP2024) predicts a less permissive cosmic shorelines. The horizontal dashed line marks the runaway greenhouse limit from \citet{kopparapu_habitable_2013}.}
    \label{fig:cs_stmass}
\end{figure*}

Fig.~\ref{fig:cs_stmass} shows various cosmic shorelines for Earth-mass planets as a function of host-star mass. The blue lines represent an initial volatile fraction of \(\sim 10^{-4}\, M_p\) in planetary mass. Exoplanets located below blue lines for \ch{CO2} or \ch{CH4} atmosphere could retain an corresponding atmosphere if they possessed Earth-like carbon inventories or more. The red line corresponds to an initial volatile fraction of \(10^{-2}\), a possible upper limit for volatile mass. Exoplanets lying above this red shaded region have low \(<50\%\) probability of retaining an atmosphere, even under highly volatile-rich conditions. 

The critical instellation is lower for lower mass stars because of their extended pre-main-sequence phase of saturated XUV emission and proportionally higher XUV fluxes for the same instellation. For the escape of \ch{CH4} atmospheres with the energy-limited estimate, this shift reflects the greater cumulative XUV received by planets orbiting lower-mass stars, while for escaping \ch{CO2} and \ch{N2}-dominated atmospheres, the nonlinear dependence of escape rates on XUV flux means that not only the cumulative flux, but also the time evolution of the stellar XUV luminosity, influences cumulative atmospheric loss. 

The broader shaded regions for \ch{CO2}-dominated atmospheres compared to other cases are due mostly to the uncertain \ch{CO2} interpolation method for \ch{CO2}-dominated atmospheres, emphasizing the need for future modeling for Earth-sized planets. 


For \ch{H2O} atmospheres, the escape rate is around one order of magnitude higher than \ch{CH4} at relatively low XUV flux. Only for \( M_*>0.7M_{\odot}\) can a planet at around Earth's instellation retain its atmosphere, and only if the initial \ch{H2O} mass is 1\% of the planetary mass. Note that \ch{H2O} has a non-negligible solubility in solid silicate compared to carbon, implying that even if the atmosphere is completely stripped, a substantial amount of water may still be retained in the mantle. The difference between the \ch{CH4} and \ch{H2O} can be explained by the hydrodynamic simulations of \cite{johnstone_hydrodynamic_2020} finding a much higher effective efficiency. However, if the line cooling effect explored in \citet{yoshida_less_2022} was included, the shorelines may get closer to the star.

A higher \ch{H2O} content, for example, 30\% of planetary mass, \citep{raymond_making_2004, luque_density_2022, moore_water_2024} can push the cosmic shorelines to higher instellations. However, such water-rich atmospheres may be unstable due to the boil-off effect described by \cite{owen_atmospheres_2016} and a more realistic shoreline should account for this instability. Our results indicate that Earth would have lost an equivalent of 200 bars of \ch{H2O} if fully evaporated, which contradicts Earth's history. This discrepancy could be attributed to the assumption of a pure \ch{H2O} atmosphere in \citet{johnstone_hydrodynamic_2020}; in reality, the presence of a background atmosphere could enable cold trap from condensation, which could limit the escape rate and protect surface liquid water on habitable planets \citep{krissansen-totton_implications_2023}. 

Additionally, as water escapes from water-rich worlds, \ch{O2} may accumulate in the atmosphere due to fractionation in the escaping outflow \citep{luger_extreme_2015, schaefer_predictions_2016, johnstone_upper_2018, kite_water_2021, cherubim_oxidation_2025}. Our current shorelines do not account for the hydrodynamic escape of any residual \ch{O2} atmospheres. Including this effect would lead to a higher atmospheric retention probability, though in the high XUV-flux regime this fractionation may not occur \citep{johnstone_hydrodynamic_2020}.

For \ch{N2}-dominated atmospheres, predictions vary significantly between reference studies. The estimates based on \cite{nakayama_survival_2022} suggest that a planet with an initial atmospheric mass of \(10^{-4} M_p\) can retain its atmosphere even at instellations of 10,000 times Earth's insolation. In contrast, predictions derived from \cite{chatterjee_novel_2024} indicate that atmospheric stripping could occur at much lower instellations (dashed curves in the lower right corner of Fig.~\ref{fig:cs_stmass}). The discrepancy can be be explained by assumptions for the hydrostatic equilibrium in \cite{nakayama_survival_2022} neglecting the ion-exobase and ambipolar electrostatic fields, though oxygen line cooling may also play a role. Resolving this discrepancy requires further work on full-set atomic line cooling and ionization effects to provide a more reliable constraint on \ch{N2}-dominated atmospheric escape. 


The CP2024 shorelines are pinched relatively close together, especially for the late M-dwarfs, due to the nonlinear behavior of the threshold XUV flux for rapid escape. For late M-dwarfs, the XUV saturation phase is long enough and the XUV decline thereafter gradual enough that only a doubling of instellation is required to evaporate 1000 bars rather than 100 bars, or 10,000 bars rather than 1000 bars. So, for the two orders of magnitude range in volatile mass \(10^{-4} M_p\) and \(10^{-2} M_p\), the critical instellations vary by less than an order of magnitude. 




\subsection{Instellation vs Escape Velocity}

\begin{figure*}
    \centering
    \includegraphics[width=1.0\linewidth]{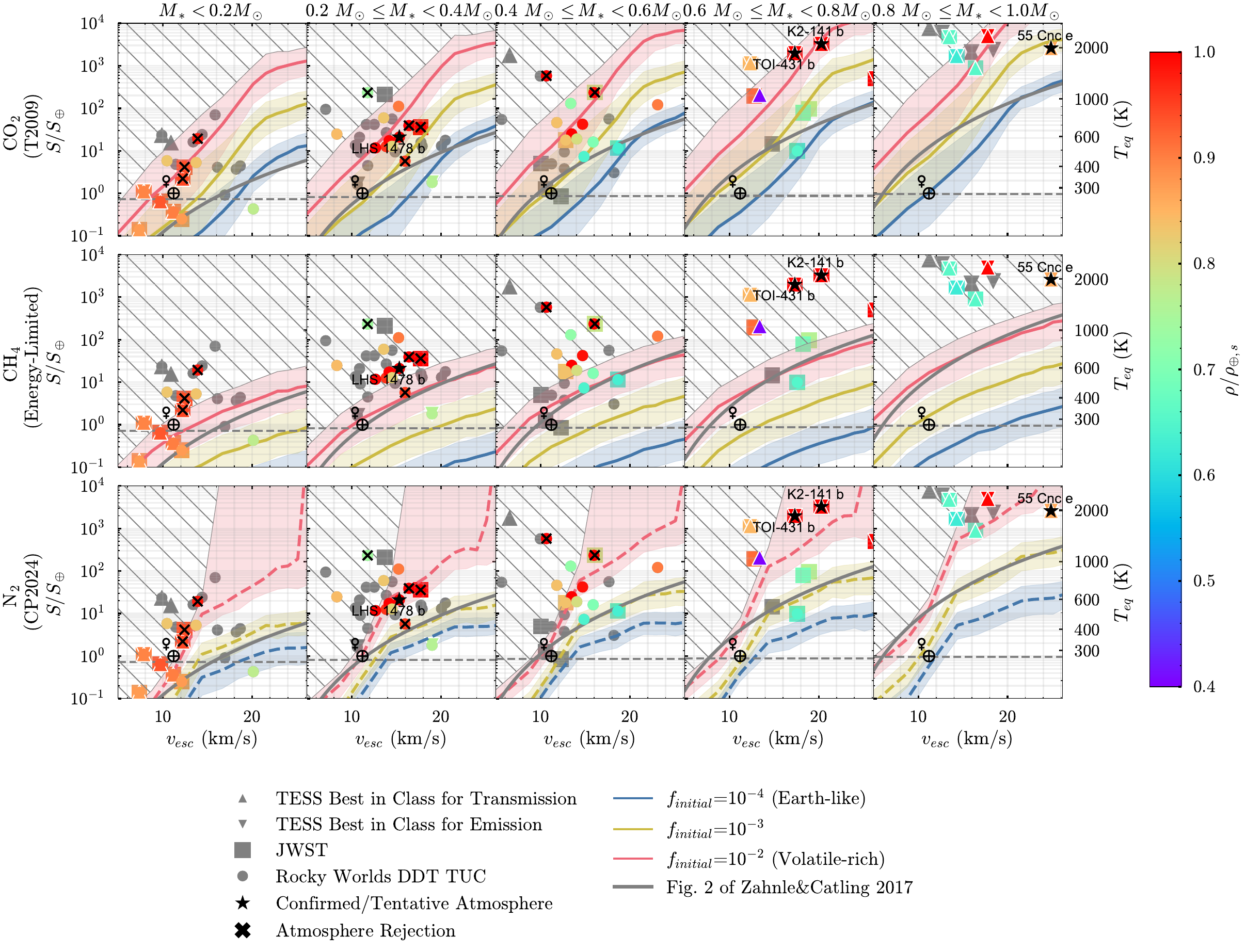}
    \caption{\emph{Cosmic Shoreline Revisited: critical instellations for atmospheric retention vs. planetary escape velocity across different escape models/compositions binned by host-star mass and probing a range of volatile abundances.} The right y-axis also shows the corresponding equilibrium temperature ($T_{eq}$), assuming zero albedo and full heat redistribution. The top row shows results for \ch{CO2}-dominated atmospheres, the middle row shows \ch{CH4}-dominated atmospheres, and the bottom row shows \ch{N2}-dominated atmospheres. Each column corresponds to a specific range of host stellar masses. Blue, yellow and red lines indicate a 90\% probability of atmosphere retention with initial volatile fractions of \(10^{-4}\), \(10^{-3}\) and \(10^{-2}\) of the planetary mass, respectively. The shaded regions in blue, yellow and red represent a 50–99\% probability of atmosphere retention. The hatched region above the 50\% red line represents conditions where planets are unlikely to retain an atmospheres even if volatile-rich. For comparison, the thick gray line reproduces the XUV-driven cosmic shoreline from Fig.~2 of \cite{zahnle_cosmic_2017}, with cumulative XUV flux converted to bolometric flux to match our y-axis, using the relation \( S = F_{\text{XUV}} (L_*/L_\odot)^{0.6} \) (their Eq. 27). The horizontal dashed gray lines mark the runaway greenhouse limit from \citet{kopparapu_habitable_2013}. Venus (\(\venus\)), Earth(\(\oplus\)) are plotted for reference. The symbols mark exoplanet targets from four different samples (Sec.~\ref{sec:data}). For planets with both radius and mass measurements, the density—scaled to that of a planet with Earth-like composition (\(\rho_{\oplus,s}(R_p)\))—is color-coded. Lower densities may suggest thicker atmospheres or higher volatile content. Planets with confirmed (55 Cnc e) and tentative (TOI-431 b, LHS 1478 b) atmosphere detections are labeled. Crosses denote planets with thick atmosphere ruled out: TRAPPIST-1 c, b, GJ 1132 b, GJ 1252 b, LTT 1445 A b, GJ 486 b, TOI-1468 b, GJ 367 b and TOI-1685 b (left to right). See Table~\ref{tab:targets} and Sec.~\ref{sec:metric} for planet-by-planet details.}
    \label{fig:CS_pl_vesp}
\end{figure*}

Fig.~\ref{fig:CS_pl_vesp} shows the cosmic shorelines for \ch{CO2}, \ch{CH4}, and \ch{N2} atmospheres (using CP2024) as a function of planetary escape velocity ($v_{esc}$) for a variety of stellar mass ranges. We exclude \ch{H2O} atmospheres as hydrodynamic simulations for steam atmospheres do not provide results for varying $M_p$. The loss rates from hydrodynamic simulations are functions of $M_p$, and we calculate $v_{esc}$ for each $M_p$ assuming an Earth-like composition \citep{zeng_growth_2019}.

The thick gray curve represents the traditional cosmic shoreline of \citet{zahnle_cosmic_2017} for reference, adopting their cumulative XUV scaling \( S = F_{\text{XUV}} (L_*/L_\odot)^{0.6} \) (their Eq. 27). The red line corresponds to an initial volatile mass of \(10^{-2} \times\) planetary mass as an upper limit, while the blue line represents \(10^{-4}\) planetary mass as a lower limit.

Our results, combined with recent atmospheric detection observations, suggest that the cosmic shoreline is not a sharp dividing line but rather a transition zone—where planets shift from likely retaining atmospheres to losing them. The width of this zone is influenced by the initial volatile fraction, atmospheric composition, and potentially variations in stellar activity, which warrant further investigation in future studies \citep{pass_receding_2025}.

The overall pattern of ratio'd density reveals that blue and green dots, representing less-dense planets, tend to cluster at the atmosphere-retaining side of cosmic shorelines compared to red and orange dots, suggesting that planets with lower scaled densities are less susceptible to atmospheric loss according to our shorelines. Further implications of this trend will be explored in detail in Sec. \ref{sec:density}.

The cosmic shorelines move to lower instellation for low $M_*$, and most Rocky Worlds DDT TUCs orbit M-dwarfs. Many of these planets fall within the red shaded region, indicating a potential to detect atmospheres around them if their initial volatile content exceeds 1 wt\% of their planetary mass. In contrast, planets in the hashed zone have a probability of less than 50\% of retaining an atmosphere, even with an initial volatile content of 1 wt\%. These planets are therefore unlikely to be suitable candidates for atmospheric retention.

The cosmic shorelines for both \ch{CO2} and \ch{CH4} atmospheres shift inward with increasing planetary escape velocity, indicating that more massive planets are better at retaining their atmospheres against hydrodynamic escape. For \ch{CH4}, the energy-limited escape rate is given by \(\dot{C}\,(\text{kg}\cdot\text{s}^{-1}) \sim F_{\text{XUV}} \cdot R_p^3 / M_p\). The mass–radius relationship is approximately \(M_p \sim R_p^{3.7}\) \citep{zeng_growth_2019}. It leads to \(\dot{C}\,(\text{kg}\cdot\text{s}^{-1}) \sim F_{\text{XUV}} \cdot R_p^{-0.7}\), which varies by at most a factor of two across the rocky planet in the range \(0.5R_\oplus  \lesssim  R_p \lesssim 1.6 R_\oplus\). Therefore, the increase in the critical instellation for more massive planets is primarily due to their larger volatile reservoirs, as we assume a fixed initial volatile mass fraction: (\(C_{total} = f_{initial}\cdot M_p \sim R_p^{3.7}\)). In contrast, for \ch{CO2}, the critical instellation rises a lot for more massive planets. This stronger gravity requires higher upper-atmosphere temperatures for efficient thermal escape to occur, which continues until adiabatic cooling effects become dominant and suppress further escape. 

The slope of the cosmic shorelines determined using energy-limited estimates for \ch{CH4} closely resembles the traditional cosmic shoreline of \citet{zahnle_cosmic_2017}. However, the shoreline derived from hydrodynamic simulations for \ch{CO2} atmospheres shifts closer to the star for more massive planets, becoming generally more favorable for atmosphere retention at all stellar masses. 

\begin{figure}
    \centering
    \includegraphics[width=1.0\linewidth]{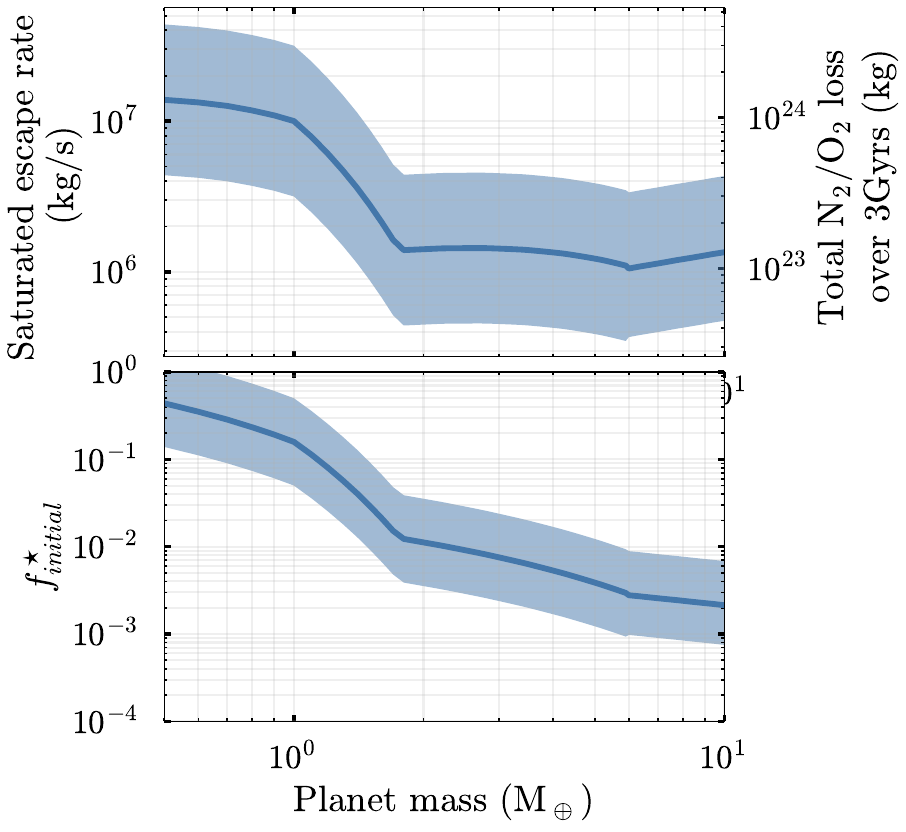}
    \caption{Upper panel: Escape rate at high-XUV plateau of \ch{N2}-dominated atmospheres as a function of planetary mass, based on \cite{chatterjee_novel_2024}. The right y-axis represents the total atmospheric loss over 3 Gyr, which also defines the critical initial volatile content required for atmospheric retention over this timescale—planets with an initial volatile mass above this threshold can retain their atmosphere regardless of past XUV flux intensity. Lower panel: the critical initial volatile fraction (relative to planetary mass) ($f^*_{initial}$) sufficient to sustain an atmosphere over 3 Gyr, independent of historical XUV flux intensity, shown as a function of planetary mass. The shaded region represents uncertainty in the critical flux at which the escape rate plateaus (see Table~ \ref{tab:CP24}).}
    \label{fig:N2O2-max-loss}
\end{figure}

The shape of the cosmic shorelines for \ch{N2}-dominated atmospheres varies wildly with initial volatile content. When the initial nitrogen and oxygen mass is \(\leq 10^{-3} M_p\), the shape of the cosmic shorelines largely follows the traditional energy-limited estimate; however, the slope differs between super-Earths and Earth-sized planets due to the varying role of line cooling. For the volatile-rich case with \( f_{\text{initial}} = 10^{-2} \), the 50\%-probability upper boundary initially follows the energy-limited curvature but rises steeply beyond a certain planetary mass. This behavior reflects the plateau regime, where the escape rate becomes independent of XUV flux (see Sec. \ref{sec:N2O2-method} and Fig. \ref{fig:N2O2-max-loss}). Since our sample spans a broad range of stellar ages, planets at the same orbital distance may either retain or lose their atmospheres depending on how long atmospheric loss has been occurring (see Fig.~\ref{fig:logistic}). For more massive planets with larger initial volatile inventories, the probability of retention exceeds 50\% regardless of XUV flux, i.e. the instellation or orbital distance. For low-mass stars with prolonged X-ray saturation phases, atmospheric loss is more dependent on stellar age. In contrast, for more massive stars with shorter saturation phases, the total loss is less strongly age-dependent, reducing uncertainties.

The high-XUV plateau explored in this study, motivated by the collisional-radiative thermostat from \cite{chatterjee_novel_2024}, can limit the total atmospheric loss, regardless of the historical XUV exposure. The maximum atmospheric loss can be estimated by multiplying the plateaued escape rate by the planet's age. In Fig.~\ref{fig:N2O2-max-loss}, we show the atmospheric escape rate at plateau and the maximum atmospheric loss over 3 Gyr (upper panel) along with the corresponding critical initial volatile fraction (\( f_{\text{initial}} \)) (lower panel), which represents the minimum volatile inventory required to make atmospheric retention over this timescale independent of historical XUV flux. As shown in Fig.~\ref{fig:N2O2-max-loss}, for planets with \( M_p > 1.5 M_\oplus \), \( f_{\text{initial},\star} < 10^{-2} \), confirming that the upper-limit case can always retain its atmosphere, even at extremely close orbital distances.

\subsection{Best Planets for Atmospheric Retention}

\subsubsection{Rocky Planets Sample}\label{sec:data}

We select rocky planets from the confirmed exoplanet catalog on the NASA Exoplanet Archive \citep{nasa_exoplanet_archive_confirmed_2019}, using the following criteria: (1) for planets orbiting FGK stars (\(M_* \geq 0.6 M_\odot\)), we select those with radii below the radius gap, defined by \(\log_{10}(R_p/R_\oplus) = 0.11 \log_{10}(P/\text{days}) + 0.37\) (Eq. 4 of \cite{ho_deep_2023}); (2) for M dwarfs, we include planets with scaled densities greater than 0.6 (measured density divided by the theoretical density assuming an Earth-like composition) following \citet{luque_density_2022}. For those without mass measurements, we include planets with \(R_p < 1.6 R_\oplus\) a possible dividing line between `rocky' and volatile-rich worlds \citep{rogers_most_2015, cloutier2020evolution, parviainen_span_2023}.


Planetary parameters were obtained via the NASA Exoplanet Archive and processed using the NASA ExoArchive Aggregator\footnote{\url{https://github.com/lkreidberg/TSM}}. We checked for and removed mass values for planets that only have maximum or minimum reported masses. For planets without mass data and \(R_p < 1.6 R_\oplus\), properties were estimated assuming an Earth-like composition using the mass-radius relation from \cite{zeng_growth_2019}. For planets with both radius and mass measurements, densities were calculated and scaled relative to an Earth-like composition at the same radius, defined as ratio'd density ($\rho_{\oplus,s}(R_{p})$). A low density may indicate a thick atmosphere, higher volatile content, or measurement uncertainty.

In Figure. \ref{fig:CS_pl_vesp}, we compare our newly derived cosmic shorelines to potentially rocky planets from three sources: a subset of TESS Best in Class for transmission and emission spectroscopy with JWST, as updated by Tom Evans-Soma (private communication) \citep{kempton_framework_2018,hord_identification_2024}, selected JWST targets\footnote{\url{https://www.stsci.edu/~nnikolov/TrExoLiSTS/JWST/trexolists.html}}, and JWST Rocky Worlds DDT Targets Under Consideration \footnote{\url{https://outerspace.stsci.edu/pages/viewpage.action?pageId=257035126}}.

\subsubsection{Updated Priority Metric and Targets}\label{sec:metric}

The initial priority metric for Rocky Worlds DDT TUCs was calculated from the tangential distance between the star and the traditional cosmic shoreline. Given the changing curvature of our cosmic shorelines, we propose refining the priority metric to measure the vertical distance, specifically, the difference in the base-10 logarithm of instellation between the planet and the $90\%$ cosmic shorelines for a 1 wt\% volatile mass fraction at a given planetary mass. Although Fig.~\ref{fig:CS_pl_vesp} shows only five panels corresponding to different stellar mass bins, we computed shorelines for stellar masses ranging from 0.1 \(M_\odot\) to 1 \(M_\odot\) in increments of 0.01 \(M_\odot\), which is roughly the highest level of precision for reported stellar masses (see Fig.~\ref{fig:CS_0d01st_mass}). A priority score of zero implies that the planet is $90\%$ likely to retain a \ch{CO2}-dominated atmosphere in our model framework, assuming an initial volatile content of $1\%\,M_{p}$. While this is likely an over-optimistic assumption for most rocky planets, changing the initial volatile content leads to relatively similar ordering in priority score. The updated priority scores are presented in Table~\ref{tab:targets}, where the first three columns list scores separately for \ch{CO2}, \ch{CH4}, and \ch{N2} atmospheres, sorted by \ch{CO2} priority score. Here, we focus on planets with \ch{CO2} priority scores greater than zero whereas those less than zero are included in Appendix Table \ref{tab:targets_appendix}.

For ease of use, we also provide a polynomial fit to the 90\% \ch{CO2} cosmic shoreline assuming 1 wt\% initial volatile fraction, the one used in our priority metric, as a function of planetary escape velocity and stellar mass:

\begin{widetext}
\begin{equation}
    \begin{split}
        \log_{10}(S/S_\oplus) =& -1.613 \\
        &-5.447 \times 10^{-2} \cdot v_{esc} + 4.568 \cdot M_* \\
        & + 2.270\times 10^{-2} \cdot v_{esc}^2 - 6.991 \times 10^{-3} \cdot v_{esc}M_* - 5.769 \cdot M_*^2\\
        &-5.517 \times 10^{-4} \cdot v_{esc}^3 +2.702 \times 10^{-4}  \cdot v_{esc}^2M_* +2.289 \times 10^{-3}  v_{esc}M_*^2 + 3.516 \cdot M_*^3
    \end{split}
\end{equation}
\end{widetext}
where $S/S_\oplus$ is the bolometric instellation relative to Earth, $v_{\rm esc}$ is the planetary escape velocity in $\text{km/s}$, and $M_*$ is the stellar mass in $M_\odot$.

Several of these planets have low ratio'd densities that suggest highly extended atmospheres or volatile-rich layers and thus may not be considered `rocky' in the traditional solar system sense and comprise much of the sample most likely to retain atmospheres (e.g., LHS~1140~b, TOI~1452~b, TOI-776~b and TOI-260~b). Therefore, we bold planets with high observability metrics that are likely ``rocky''--those with a radius below $1.6\,R_{\oplus}$ and a ratio'd density above $0.8\,\rho_{\oplus,s}$--and that have a \ch{CO2} priority metric greater than zero. We consider planets with an emission spectroscopy metric (ESM, \citealt{kempton_framework_2018}) above that of TRAPPIST-1 c (ESM=1.7, amenable to atmospheric reconnaissance with MIRI F1500W) or a transmission spectroscopy metric (TSM) greater than 10 (the threshold for atmospheric characterization suggested by \citealt{kempton_framework_2018}). This list includes, in order of decreasing \ch{CO2} priority score: TOI-4559~b, TOI-711~b, LTT~1445~A~b, TOI-1693~b, TRAPPIST-1~g, TOI-1468~b, TRAPPIST-1~f, GJ~486~b, LHS~1140~c, LHS~1478~b, Gliese~12~b, GJ~3929~b, K2-141~b, TRAPPIST-1~h, HD~260655~b, LHS~1815~b, TRAPPIST-1~e, LTT~1445~A~c, and TRAPPIST-1~c. Several of these planets have been observed in emission, suggesting a lack of thick atmospheres [LTT~1445~A~b \citep{wachiraphan_thermal_2024}, TOI-1468~b \citep{valdes2025hot}, GJ~486~b \citep{mansfield_no_2024}, and TRAPPIST-1~c \citep{zieba_no_2023}, cf. \citep{lincowski_potential_2023}]. Thus, if rocky planets generally form with similar volatile inventories, which is uncertain, we can consider LTT~1445~A~b as a loose lower bound of where rocky planet atmospheres might be retained. Indeed, LTT~1445~A~b's \ch{CO2} priority score drops to 0.0 given a more conservative initial volatile fraction of $0.1\textrm{wt}\%$. TOI-4559~b and TOI-771~b stand out as potentially rocky planets with high observability metrics above this threshold. In particular, TOI-4559~b orbits an earlier-type M dwarf ($\sim$M2V) and may not be subject to the extended pre-main sequence mass loss for planets around mid-to-late M dwarfs proposed by \citet{pass_receding_2025}. However, both planets currently lack published mass values and thus it is uncertain whether they are truly `rocky', highlighting the need for radial velocity characterization.

While several additional potentially `rocky' planets exist above this threshold, including TOI-198~b, TOI-1680~b, TOI-237~b, LP~890-9~c, TOI-700~d, and Kepler-167~d, their low observability metrics means that atmospheric detection while likely take a very large time investment with JWST. For example, LP~980-9~c will be observed over 18 transits in Cycle 4 (PID: 7073, PIs: Lustig-Yaeger and Stevenson). In addition, some of these planets' temperate equilibrium temperatures mean that thin ($\lesssim1\,$bar) atmospheres may be subject to atmospheric collapse if they are tidally locked \citep{wordsworth2015}. Whether these planets are, in fact, tidally locked is unknown, but planets around low-mass stars experience much higher tidal forces that promote efficient tidal locking (e.g., \citealt{barnes2017}).

Given that the vast majority of our planet sample falls below LTT~1445~A~b in priority score, it is likely that many rocky planets that will be observed in JWST Cycles 1-4 will be bona fide `bare rocks' lacking thick atmospheres. However, bare rock observations remain useful in constraining the escape environments of different star types while providing population-level upper bounds on volatile inventories for rocky planets. In addition, eclipse observations are particularly powerful for providing useful geological information for planets without atmospheres (e.g., \citealt{coy2025,paragas25}).




\begin{longtable}{llllllllllllllll}
\toprule
\caption{Atmospheric Retention Targets Ranked by Priority Metric for \ch{CO2}-Dominated Atmospheres}\\
\hline
Planet & \multicolumn{3}{c}{Priority Metric} &  $R_p$ & $M_p$ & $v_{esc}$ & $\rho$ \footnote{the density scaled to that of a planet with Earth-like composition} & $M_*$ & $T_{eq}$ & \textbf{K}-mag & $(R_{p}/R_{s})^{2}$ & ESM & TSM\\
& [\ch{CO2}] & [\ch{CH4}] & [\ch{N2}]&  $(R_\oplus)$ & $(M_\oplus)$ & (km/s) & ($\rho_{\oplus,s}$) & $(M_\odot)$ & (K)  & & (ppm) &  & \\
\midrule
LHS 1140 b $^\#$ & 3.13 & 1.14 & 2.30 & 1.73 & 5.60 & 20.1 & 0.77 & 0.18 & 225 & 8.8 & 5397 & 0.1 & 67 \\
TOI-1452 b & 2.38 & 0.59 & 1.63 & 1.67 & 4.82 & 19.0 & 0.77 & 0.25 & 322 & 9.7 & 3107 & 0.6 & 39 \\
TOI-198 b & 2.33 & 0.65 & 1.74 & 1.44 & 3.82$^{\dagger}$ & 18.2$^{\dagger}$ & 1$^{\dagger}$ & 0.47 & 368 & 7.9 & 896 & 0.6 & 5$^{\dagger}$ \\
TOI-776 b & 1.89 & 0.15 & 1.31 & 1.80 & 5.00 & 18.7 & 0.63 & 0.54 & 512 & 7.6 & 908 & 2.7 & 50 \\
TOI-260 b & 1.79 & 0.26 & 1.35 & 1.71 & 4.23 & 17.6 & 0.65 & 0.62 & 493 & 6.6 & 667 & 2.6 & 67 \\
TOI-1680 b & 1.73 & 0.00 & 1.00 & 1.47 & 4.08$^{\dagger}$ & 18.7$^{\dagger}$ & 1$^{\dagger}$ & 0.18 & 403 & 10.8 & 4073 & 1.3 & 6$^{\dagger}$ \\
TOI-237 b & 1.69 & 0.04 & 1.01 & 1.44 & 3.82$^{\dagger}$ & 18.2$^{\dagger}$ & 1$^{\dagger}$ & 0.18 & 386 & 10.9 & 3915 & 1.0 & 6$^{\dagger}$ \\
LP 890-9 c & 1.67 & 0.27 & 1.22 & 1.37 & 3.13$^{\dagger}$ & 16.9$^{\dagger}$ & 1$^{\dagger}$ & 0.12 & 271 & 11.3 & 6487 & 0.2 & 6$^{\dagger}$ \\
TOI-1634 b & 1.59 & -0.64 & 0.90 & 1.77 & 7.57 & 23.1 & 0.90 & 0.45 & 922 & 8.6 & 1305 & 14.0 & 54 \\
\textbf{TOI-4559 b} & 1.36 & -0.24 & 0.81 & 1.42 & 3.56$^{\dagger}$ & 17.7$^{\dagger}$ & 1$^{\dagger}$ & 0.39 & 554 & 8.6 & 1203 & \textbf{3.1} & 8$^{\dagger}$ \\
TOI-1075 b & 1.34 & -0.89 & 1.15 & 1.79 & 9.95 & 26.4 & 1.06 & 0.60 & 1321 & 9.1 & 799 & 12.0 & 29 \\
TOI-700 d $^\#$ & 1.33 & 0.57 & 0.81 & 1.07 & 1.29$^{\dagger}$ & 12.3$^{\dagger}$ & 1$^{\dagger}$ & 0.41 & 267 & 8.6 & 546 & $<$0.1  & 3$^{\dagger}$ \\
Kepler-167 d & 1.28 & 0.22 & 1.46 & 1.24 & 2.16$^{\dagger}$ & 14.8$^{\dagger}$ & 1$^{\dagger}$ & 0.78 & 542 & 11.8 & 230 & 0.1 & 1$^{\dagger}$ \\
55 Cnc e & 1.25 & -0.97 & 0.96 & 1.86 & 9.38 & 25.1 & 0.92 & 0.91 & 1958 & 4.0 & 326 & 67.7 & 211 \\
TOI-178 c & 1.20 & -0.62 & 0.61 & 1.67 & 4.77 & 18.9 & 0.76 & 0.65 & 873 & 8.7 & 552 & 3.9 & 34 \\
\textbf{TOI-771 b} & 1.18 & -0.42 & 0.58 & 1.42 & 3.63$^{\dagger}$ & 17.9$^{\dagger}$ & 1$^{\dagger}$ & 0.22 & 526 & 9.7 & 2902 & \textbf{4.5} & \textbf{11$^{\dagger}$} \\
TOI-836 b & 1.15 & -0.52 & 0.62 & 1.70 & 4.53 & 18.2 & 0.69 & 0.68 & 828 & 6.8 & 552 & 7.5 & 79 \\
WASP-47 e & 1.13 & -1.11 & 0.66 & 1.83 & 9.00 & 24.8 & 0.93 & 1.06 & 2325 & 10.2 & 211 & 3.1 & 11 \\ \hline\hline
\textbf{LTT 1445 A b}$^{*}$ & 1.08 & -0.15 & 0.77 & 1.34 & 2.73 & 16.0 & 0.95 & 0.26 & 430 & 6.5 & 2055 & \textbf{6.0} & \textbf{34} \\
TOI-244 b & 0.99 & -0.08 & 0.85 & 1.52 & 2.68 & 14.9 & 0.64 & 0.43 & 457 & 8.0 & 1070 & 2.0 & 71 \\
\textbf{TOI-1693 b} & 0.94 & -0.62 & 0.48 & 1.41 & 3.51$^{\dagger}$ & 17.7$^{\dagger}$ & 1$^{\dagger}$ & 0.49 & 764 & 8.3 & 790 & \textbf{6.1} & 8$^{\dagger}$ \\
\textbf{TRAPPIST-1 g} $^\#$& 0.93 & 0.14 & 0.35 & 1.13 & 1.32 & 12.1 & 0.88 & 0.09 & 197 & 10.3 & 7540 & $<$0.1 & \textbf{15} \\
HD 260655 c & 0.91 & -0.31 & 0.67 & 1.53 & 3.09 & 15.9 & 0.70 & 0.44 & 557 & 5.9 & 1025 & 8.9 & 196 \\
\textbf{TOI-1468 b}$^{*}$ & 0.89 & -0.66 & 0.36 & 1.28 & 3.21 & 17.7 & 1.24 & 0.34 & 681 & 8.5 & 1164 & \textbf{6.1} & \textbf{10} \\
TOI-700 e & 0.82 & 0.18 & -0.03 & 0.95 & 0.85$^{\dagger}$ & 10.6$^{\dagger}$ & 1$^{\dagger}$ & 0.41 & 295 & 8.6 & 431 & 0.1 & 4$^{\dagger}$ \\
LP 890-9 b & 0.78 & -0.47 & 0.48 & 1.32 & 2.74$^{\dagger}$ & 16.1$^{\dagger}$ & 1$^{\dagger}$ & 0.12 & 395 & 11.3 & 6049 & 1.6 & 9$^{\dagger}$ \\
K2-239 d & 0.73 & -0.08 & 0.30 & 1.10 & 1.41$^{\dagger}$ & 12.6$^{\dagger}$ & 1$^{\dagger}$ & 0.40 & 399 & 10.0 & 785 & 0.3 & 4$^{\dagger}$ \\
\textbf{TRAPPIST-1 f} $^\#$ & 0.57 & -0.15 & -0.25 & 1.04 & 1.04 & 11.2 & 0.90 & 0.09 & 217 & 10.3 & 6460 & 0.1 & \textbf{17} \\
TOI-2096 b & 0.54 & -0.52 & 0.37 & 1.24 & 2.19$^{\dagger}$ & 14.8$^{\dagger}$ & 1$^{\dagger}$ & 0.23 & 487 & 11.0 & 2352 & 1.4 & 6$^{\dagger}$ \\
\textbf{GJ 486 b}$^{*}$ & 0.46 & -0.84 & 0.07 & 1.29 & 2.77 & 16.4 & 1.09 & 0.31 & 696 & 6.4 & 1328 & \textbf{21.1} & \textbf{35} \\
L 98-59 c & 0.45 & -0.54 & 0.37 & 1.34 & 2.25 & 14.5 & 0.82 & 0.31 & 526 & 7.1 & 1550 & 7.3 & 28 \\
K2-129 b & 0.38 & -0.28 & -0.21 & 1.04 & 1.15$^{\dagger}$ & 11.8$^{\dagger}$ & 1$^{\dagger}$ & 0.36 & 405 & 8.9 & 701 & 0.5 & 7$^{\dagger}$ \\
GJ 357 b & 0.38 & -0.56 & 0.19 & 1.20 & 1.84 & 13.9 & 0.96 & 0.34 & 524 & 6.5 & 1066 & 6.3 & 28 \\
\textbf{LHS 1140 c} & 0.35 & -0.51 & 0.19 & 1.27 & 1.91 & 13.7 & 0.83 & 0.18 & 422 & 8.8 & 2917 & \textbf{3.0} & \textbf{22} \\
K2-239 b & 0.33 & -0.48 & -0.10 & 1.10 & 1.41$^{\dagger}$ & 12.6$^{\dagger}$ & 1$^{\dagger}$ & 0.40 & 502 & 10.0 & 785 & 0.8 & 5$^{\dagger}$ \\
\textbf{LHS 1478 b} & 0.32 & -0.79 & 0.11 & 1.24 & 2.33 & 15.3 & 1.05 & 0.24 & 595 & 8.8 & 2142 & \textbf{7.0} & \textbf{18} \\
TOI-406 c & 0.32 & -0.63 & 0.25 & 1.32 & 2.08 & 14.0 & 0.80 & 0.41 & 580 & 8.9 & 871 & 2.5 & 8 \\
K2-239 c & 0.29 & -0.34 & -0.45 & 1.00 & 1.01$^{\dagger}$ & 11.2$^{\dagger}$ & 1$^{\dagger}$ & 0.40 & 427 & 10.0 & 649 & 0.4 & 4$^{\dagger}$ \\
\textbf{Gliese 12 b} & 0.27 & -0.32 & -0.60 & 0.90 & 0.69$^{\dagger}$ & 9.8$^{\dagger}$ & 1$^{\dagger}$ & 0.24 & 313 & 7.8 & 941 & 0.3 & \textbf{16$^{\dagger}$} \\
\textbf{GJ 3929 b} & 0.25 & -0.71 & 0.15 & 1.09 & 1.75 & 14.2 & 1.23 & 0.31 & 567 & 7.9 & 975 & \textbf{4.1} & \textbf{14} \\
Kepler-1107 b & 0.23 & -1.50 & -0.22 & 1.45 & 3.92$^{\dagger}$ & 18.4$^{\dagger}$ & 1$^{\dagger}$ & 0.86 & 1943 & 12.2 & 291 & 1.4 & 1$^{\dagger}$ \\
\textbf{K2-141 b} & 0.21 & -1.88 & -0.53 & 1.51 & 4.97 & 20.3 & 1.06 & 0.71 & 2101 & 8.4 & 413 & \textbf{14.9} & \textbf{60} \\
\textbf{TRAPPIST-1 h} $^\#$ & 0.20 & -0.35 & -0.82 & 0.76 & 0.33 & 7.4 & 0.88 & 0.09 & 171 & 10.3 & 3372 & $<$0.1  & \textbf{16} \\
\textbf{HD 260655 b} & 0.18 & -0.82 & 0.11 & 1.24 & 2.14 & 14.7 & 0.99 & 0.44 & 709 & 5.9 & 671 & \textbf{11.8} & \textbf{29} \\
\textbf{LHS 1815 b} & 0.17 & -0.70 & -0.01 & 1.09 & 1.58 & 13.5 & 1.14 & 0.50 & 617 & 8.0 & 396 & \textbf{1.9} & 7 \\
TOI-700 b & 0.13 & -0.49 & -0.77 & 0.91 & 0.73$^{\dagger}$ & 10.0$^{\dagger}$ & 1$^{\dagger}$ & 0.41 & 415 & 8.6 & 396 & 0.4 & 6$^{\dagger}$ \\
TOI-270 b & 0.09 & -0.74 & -0.31 & 1.21 & 1.58 & 12.8 & 0.83 & 0.40 & 581 & 8.3 & 847 & 3.2 & 13 \\
\textbf{TRAPPIST-1 e} $^\#$& 0.07 & -0.58 & -0.92 & 0.92 & 0.69 & 9.7 & 0.93 & 0.09 & 249 & 10.3 & 5007 & 0.2 & \textbf{20} \\
\textbf{LTT 1445 A c} & 0.02 & -0.79 & -0.43 & 1.07 & 1.37 & 12.7 & 1.06 & 0.26 & 514 & 6.5 & 1310 & \textbf{7.6} & \textbf{41} \\
\textbf{TRAPPIST-1 c}$^{*}$ & 0.01 & -0.79 & -0.54 & 1.10 & 1.31 & 12.2 & 0.95 & 0.09 & 339 & 10.3 & 7119 & \textbf{1.7} & \textbf{25} \\ \hline
\bottomrule
\caption{The scores for each atmospheric composition represent the difference in the base-10 logarithm of instellation between the planet and the cosmic shorelines for a 1 wt\% volatile mass fraction at a given planetary mass. More \textbf{positive} values are more atmosphere-favorable. We bold planets with high observability metrics that are likely ``rocky''. $^{\dagger}$These planets lack a measured mass value and we assume an Earth-like composition for these calculations. $^{*}$These planets have been suggested to have no thick atmosphere based on thermal emission observations. $\#$ These planets lie exterior to the habitable‐zone inner edge \citep{kopparapu_habitable_2013}, and \ch{H2O} cold trap may inhibit escape and boost retention probabilities \citep{krissansen-totton_implications_2023}. We do not take the individual constrained stellar age into account, and the score is obtained by assuming an stellar age distribution from \cite{berger_gaiakepler_2020}.}
\label{tab:targets}
\end{longtable}

\subsubsection{Comparison to Planets with Atmospheric Detection/Rejection}

Emission observations have began to provide tentative detections for atmospheres around rocky planets, including LHS~1478~b \citep{august_hot_2025}, TOI-431~b \citep{monaghan2025}, 55 Cnc e \citep{demory_map_2016,hu_secondary_2024}, and K2-141~b \citep{zieba2022}. 3 of 4 of these planets lie beyond the temperature threshold for forming thin silicate vapor atmospheres from vapor pressure equilibrium with a dayside magma pool (e.g., \citealt{kite2016atmosphere}) and thus may not fall under the traditional `cosmic shoreline' framework of initial volatile atmosphere retention. However, the tentative detection of a carbon-rich atmosphere on 55 Cnc e \citep{hu_secondary_2024} may imply that these atmospheres are not only sustained by vapor pressure equilibrium, but rather formed with a significant fraction of volatiles. Detection or nondetection of atmospheres could also be used to refine interpretations of the cosmic shoreline framework \citep{bertathompson2025}. We discuss these planets and several `bare rock' detections in context of our shorelines below.

55 Cnc e, which likely has an atmosphere despite its very high insolation \citep{hu_secondary_2024}, lies well below the cosmic shorelines for \( f_{\text{initial}} = 10^{-2} \) for \ch{CO2} and \ch{N2}-dominated atmospheres (thick pink line of Fig. \ref{fig:CS_pl_vesp}), while it is beyond the energy-limited estimate in the \ch{CH4} case. TOI-431 b, another hot rocky super-Earth with low dayside emission—implying a \textit{tentative} atmospheric detection \citep{monaghan2025}—falls within the upper boundary of the shaded region for \(f_{\text{initial}} = 10^{-2}\) (the upper bound implies 50\% chance of atmosphere retention). Notably, both planets are massive super-Earths orbiting more massive stars, and our shorelines favor atmospheric retention in such cases. 

LHS 1478 b, which has a \textit{tentatively} detected atmosphere \citep{august_hot_2025}---suggested by a possible shallow eclipse depth---falls near the 90\% cosmic shorelines for \( f_{\text{initial}} = 10^{-2} \) for both \ch{CO2} and \ch{N2}-dominated atmospheres. GJ 486 b, located near LHS 1478 b in parameter space, likely lacks a thick atmosphere \citep{mansfield_no_2024}. This atmosphere rejection can help constrain GJ~486~b's initial volatile inventory: if there is a carbon-dominated oxidized environment, the total carbon mass is likely to be less than \(0.01\, M_p\); if it initially possessed a \ch{N2}-dominated atmosphere, its mass is likely to be below \(0.01\, M_p\), within our model framework. GJ 1132 b, another likely bare rock \citep{xue_jwst_2024}, lies above all our shorelines but very close to the \ch{CO2} red line, suggesting that its initial carbon fraction was less likely to be more than \(0.01~M_p\) if carbon-dominated oxidized environment.

For the TRAPPIST-1 system, both TRAPPIST-1 b and c lie on or beyond the \(f_{\text{initial}} = 10^{-2}\) boundary for all three atmospheric compositions. This means that even if these planets formed super volatile-rich that they are probably unable to retain atmospheres. Notably, their host star has an extremely low mass (\(M_* = 0.089 M_\odot\)), at the lower edge of the stellar mass range considered in the panel, where shorelines are derived by randomly sampling stellar masses within that range. As a result, the actual shoreline for TRAPPIST planets is even farther from the star than shown in Fig. \ref{fig:CS_pl_vesp}. Conversely, TOI-431 b orbits a star with mass \(0.78\,M_\odot\) at the high end of the panel, meaning its more accurate shoreline would lie closer to the star than shown. To address this, we reduced the stellar mass bin size to \(0.01\,M_\odot\) when generating Table~\ref{tab:targets}. We also do not account for the age dependence of the cosmic shorelines. While escape is largely dominated by the early saturated phase, TRAPPIST-1 is relatively old and lies at relative high-end tail of our stellar age distribution, meaning its planets are even more likely to have lost their atmospheres than suggested by our average predictions. Note that this single‐species framework does not address the habitable‐zone TRAPPIST‑1 planets well, where cold‐trap effects in mixed atmospheres may be significant \citep{krissansen-totton_implications_2023}.

Earth and Venus straddle the 90\%-probability \(f_{\text{initial}} = 10^{-4}\) shoreline, yet there is no evidence that thermal escape has removed such a large amount of volatiles \citep[e.g.][]{johnson_earthn_2018, zahnle_elemental_2023}. This implies that our shorelines may underestimate atmospheric retention, and that rocky planets could retain atmospheres more readily than predicted. We use this discrepancy as an example to discuss our model’s assumptions and limitations; see Sec. \ref{sec:discussion} for details.

\section{Trend of ratio'd density}\label{sec:density}

Since planetary mass (\(M_p\)) can be measured through radial velocity or transit-timing variation (TTV) observations, planet density can be calculated using the measured transit radius: \(\rho = M_p/\left(4/3 \pi R_{\text{transit}}^3\right)\). The transit radius \( R_{\text{transit}} \) corresponds to the altitude where the atmosphere becomes sufficiently opaque in a slant viewing geometry, for $\sim (e-1)/e$ starlight will be blocked. If opaque clouds or hazes are present, transit radius then corresponds to the altitude of the cloud tops \citep{lopez_how_2012,gao_aerosol_2020, gao_universal_2021}. Because the transit radius includes the atmospheric contribution, the measured density (\(\rho\)) can be lower than the density of the solid portion of the planet alone. By comparing the observed density with theoretical density estimates for rocky planets, we can distinguish planets with substantial atmospheres from bare rocky worlds \citep{rogers_most_2015,jontof-hutter_compositional_2019,luque_density_2022}.

As the cosmic shorelines separate bare rocky planets from those with significant atmospheres in planetary-mass–instellation space, we expect statistical trends in the observed densities of exoplanets. Specifically, density should increase with instellation and decrease with planetary mass because planets that experience weaker XUV flux and have higher gravity are more likely to retain their atmospheres and have lower densities. This trend provides a new observational test for the cosmic shorelines hypothesis. 

We build a model, with several simplifying assumptions, intended to demonstrate the concept rather than provide precise predictions. For this study, we focus on \ch{CO2} atmospheres as a representative case, given that MIRI on JWST is most sensitive to \ch{CO2} detection at 15 $\mu$m, and given the importance of \ch{CO2} for solar system rocky planets. 

\subsection{Vertical Thermal Structure}

To see how atmosphere boosts planetary radius, we construct a one-dimensional vertical structure model that considers the thermal structure of the atmosphere and the underlying magma ocean, if present. Using this, we estimate the transit radius of a planet with a given volatile content by accounting for radiative and convective layers of atmosphere, volatile partitioning between the atmosphere and magma, and the effect of hydrostatic equilibrium on atmospheric extent. We then derive the density ratio'd to an Earth-like composition and perform statistical analysis to investigate how the cosmic shorelines influence the trend of the ratio'd density.

\begin{figure}
    \centering
    \includegraphics[width=1.0\linewidth]{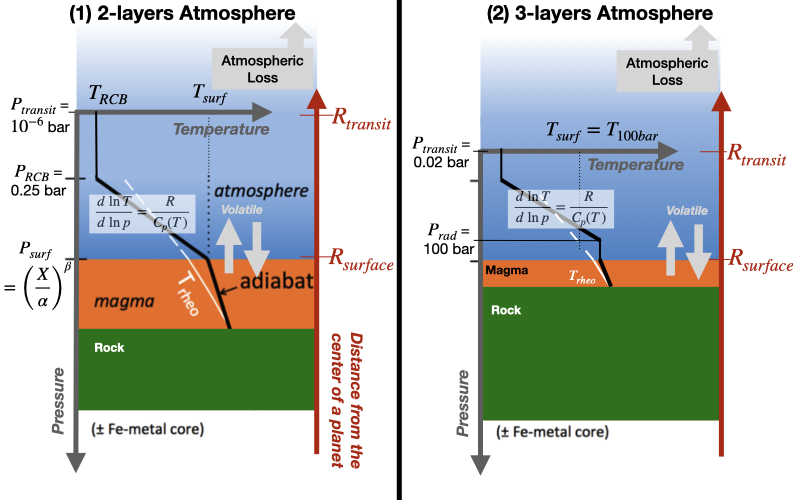}
    \caption{Planetary thermal structure and two end-member estimates for transit radius. The thin black line denotes an isothermal atmosphere in the radiative layer, while the thick black line represents an adiabatic profile in the convective layer. The white line corresponds to the rheological transition ($\sim$40\% melt fraction) P-T relation for rock. If \(T_{\text{surf}} > T_{\text{rheo}}(P = P_{\text{surf}})\), the surface is a low-viscosity magma ocean. The intersection of the thick black line with the white line marks the depth at which magma crystallizes. Volatiles are assumed to be partitioned between the magma and atmosphere, with surface pressure governed by the solubility law. The pressure level probed during a transit (\(P_{\text{transit}}\)) is set to a constant value. (1) Upper-limit estimate of atmospheric height: Assumes the entire atmosphere below the RCB is fully convective, following a dry adiabatic profile. \(P_{\text{transit}} = 10^{-6} \text{ bar}\).  (2) Lower-limit estimate of atmospheric height: Assumes a radiative layer extends from the surface up to \(P = 100\) bar. \(P_{\text{transit}} = 0.02\) bar.}
    \label{fig:1D-profile}
\end{figure}

We first consider a two-layer atmospheric structure, as illustrated in panel (1) of Fig.~\ref{fig:1D-profile}: 1. a radiative upper atmosphere, where the temperature is isothermal above the radiative-convective boundary (RCB), and 2. a convective lower atmosphere, where the temperature profile follows an adiabatic gradient. RCB temperature is the skin temperature determined by radiative equilibrium \citep{catling_atmospheric_2017}, assuming an albedo of 0.3:

\begin{equation}
    T_{\text{RCB}} = \left(\frac{(1-0.3)S}{8\sigma}\right)^{1/4}
\end{equation}

where \( S \) is the instellation, and \( \sigma \) is the Stefan-Boltzmann constant. This formulation assumes that heat is efficiently redistributed across the planet by a thick atmosphere, leading to a uniform temperature distribution. The RCB pressure is fixed as \(p_{\text{RCB}} = 0.25 \text{bar}\), which can correctly predict the surface temperature of Venus. Below the RCB, the surface temperature (\(T_{\text{surf}}\)) is obtained by integrating a dry adiabatic profile down to the surface. In the convective region, pressure and temperature obey the adiabatic relation $d\ln{T}/d\ln{P} = R/C_p$, where R is the ideal gas constant and $C_p$ is the specific heat at constant pressure. We employ Shomate polynomials to model the temperature dependence of $C_p$. \citep{chase_nist-janaf_1998}

For thick \ch{CO2}-dominated atmospheres, however, the convective zone may not extend all the way to the surface. Instead, a radiative, isothermal layer might develop near the surface due to strong absorption of stellar radiation and internal heating \citep{selsis_cool_2023, peng_puffy_2024}. Since a full radiative transfer calculation is beyond the scope of this model, we adopt a simplified assumption: in the lower-limit estimate of transit radius (panel (2) of Fig.~\ref{fig:1D-profile}), we assume the convective-radiative transition in the deep atmosphere starts from 100 bar. This assumption allows us to explore the potential range of atmospheric extent while acknowledging uncertainties in radiative transfer.

The extent of a possible magma ocean is constrained by the rheological transition temperature (\(T_{\text{rheo}}\)) at a given pressure, corresponding to $\sim$40\% melt fraction for silicate rock (following the method of \cite{kite_exoplanet_2020}, which is based on \cite{andrault_solidus_2011}). If \(T_{\text{surf}} > T_{\text{rheo}}(P = P_{\text{surf}})\), the surface remains molten, and volatiles can continue to exchange between the atmosphere and magma ocean. The partitioning of volatiles is governed by a solubility law that relates surface pressure to the volatile concentration in the magma (\(X_{\text{magma}}\)(kg/kg)):  

\begin{equation}
    P_{\text{surf}} = \left( X_{\text{magma}} / \alpha \right)^{\beta}
\end{equation}

where \(\alpha = 1.94\times 10^{-3} \text{Pa}^{1/\beta}\) and \(\beta=0.714\) are empirical parameters for \ch{CO2} obtained from \cite{lichtenberg_vertically_2021}. We acknowledge that solubility depends on both temperature and pressure; our model considers only pressure.

To link this atmospheric model to transit observations, we assume \(P_{\text{transit}}\) as a constant value obtained from previous studies. The altitude at this level is derived by assuming hydrostatic balance. Two end-member scenarios are considered for estimating the atmospheric height (Fig. \ref{fig:TP-profile}):

\begin{enumerate}
    \item \textit{Upper-limit estimate:} The atmosphere below the RCB is assumed to be fully convective, following a dry adiabatic profile. The transit pressure is set to \(P_{\text{transit}} = 10^{-6} \text{ bar}\), which is set by cloud top \citep{gao_aerosol_2020}.
    \item \textit{Lower-limit estimate:} A radiative layer is assumed to develop near the surface, limiting atmospheric expansion. The transit pressure is set to \(P_{\text{transit}} = 0.02\) bar \citep{lopez_how_2012}, obtained by integrating molecular gas opacity without clouds.
\end{enumerate}

\begin{figure}
    \centering
    \includegraphics[width=1.0\linewidth]{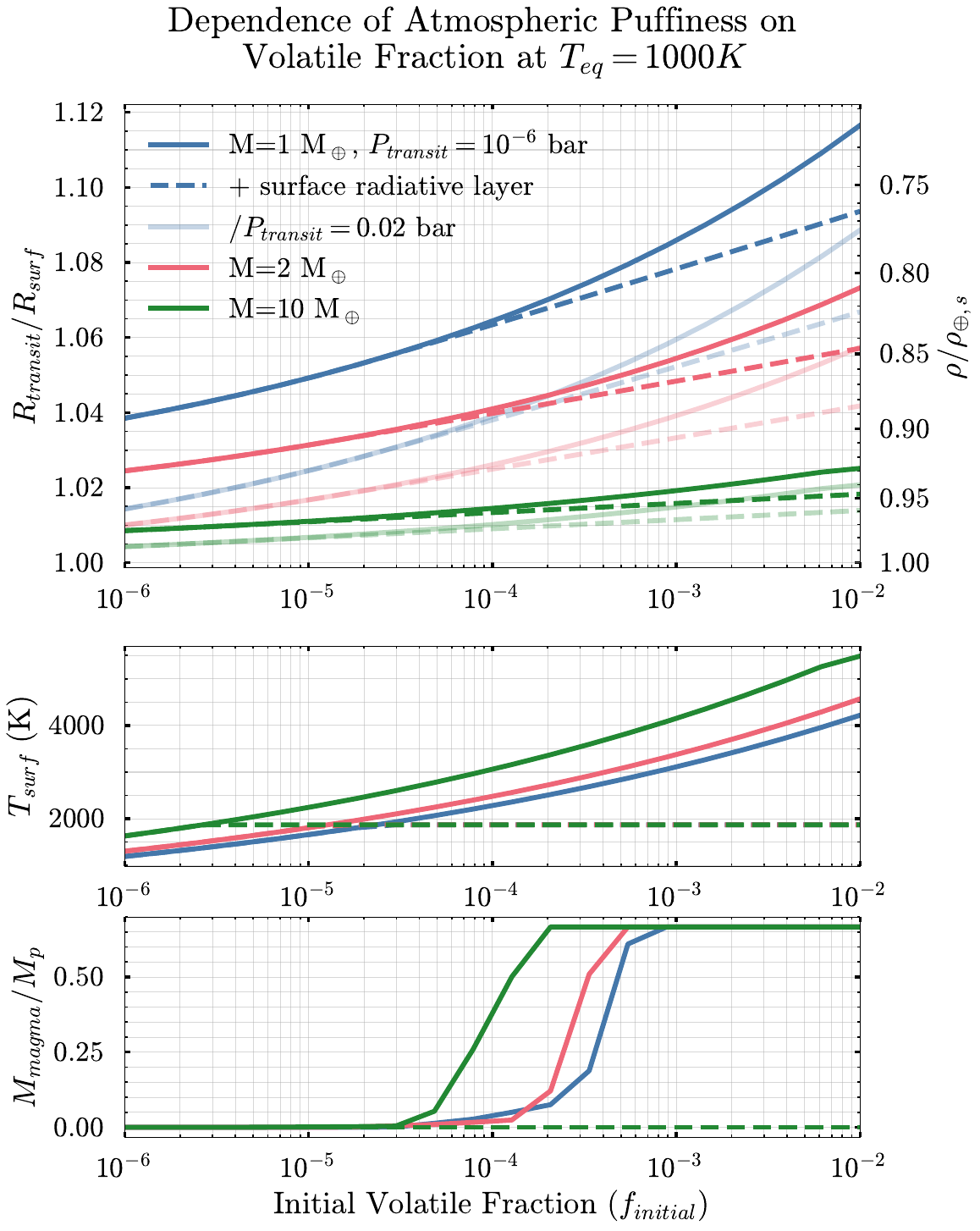}
    \caption{Atmospheric Radius Boosting effect. Top: The ratio of the transit radius to the radius of the solid surface (including the magma layer if present) as a function of the initial volatile fraction (\( f_{\text{initial}} \)). The right y-axis shows the corresponding ratio'd density. Solid lines correspond to the two-layer model (Scenario (1) in Fig.~\ref{fig:1D-profile}) with \( P_{\text{transit}} = 10^{-6} \) bar, where the atmosphere consists of a convective lower layer and an isothermal upper layer beyond the radiative-convective boundary (RCB). Dashed lines represent the three-layer model (Scenario (2) in Fig.~\ref{fig:1D-profile} but with adjusted \(P_{\text{transit}}\)), which includes an additional radiative layer near the surface. The lighter-colored lines indicate the same atmospheric conditions, but the transit radius is measured at \( P_{\text{transit}} = 0.02 \) bar. Middle: Surface temperature for the different atmospheric scenarios described above. At \(T  \gtrapprox 10^4\)K, the silicate and \ch{Fe} will expand \citep{lock_structure_2017}, which is not considered in our model. Bottom: Magma ocean mass fraction for different atmospheric structures. The layered structure of the atmosphere influences the transit radius only when \( f_{\text{initial}} \) is sufficiently high for the surface pressure to exceed 100 bar. The choice of \( P_{\text{transit}} \) has a much greater impact on the derived transit radius than the atmospheric structure alone.}
    \label{fig:transit-radius}
\end{figure}

By applying this model, we can investigate how instellation and volatile content influence the transit radius \( R_{\text{transit}} \), which accounts for both the solid body (including magma layer) and the atmospheric thickness at $P_{transit}$ (Fig.~\ref{fig:transit-radius}). We do not account for the potential puffiness of the magma layer relative to the solid mantle.

Since planetary mass (\(M_p\)) can often be constrained through radial velocity measurements or TTVs, we can get planet density using the transit radius: \(\rho = M_p/\left(\frac{4}{3} \pi R_{\text{transit}}^3 \right)\). Because the transit radius includes the atmospheric contribution, the measured density (\(\rho\)) will always be lower than the density of the solid portion of the planet alone for worlds with thick atmospheres. Comparing the observed density to theoretical models of rocky planets can help distinguish planets with substantial atmospheres from bare rocky worlds.

\subsection{Monte Carlo Simulations}

We use a Monte Carlo approach similar to the one described in Sec. \ref{sec:method}, but with additional variables, incorporating the final atmospheric height after loss and the corresponding planetary density. The newly introduced variables and their distributions are listed in Table \ref{tab:monte_carlo_2}.

To quantify the variation in the bulk density of the solid portion of the planet \citep{xu_exogeology_2021}, we generate three samples under different assumptions about the distribution of planetary radii. Given a planetary mass (\(M_p\)), the radius is assigned according to the following distributions::

\begin{itemize}
    \item (a) No compositional Variance.: Fixed Earth-like Composition: The radius is set to \(R_{\oplus,s}(M_p)\), assuming a composition of 32.5\% \ch{Fe} + 67.5\% \ch{MgSiO3}.  
    \item (b) Narrow Gaussian distribution: The radius is drawn from a Gaussian distribution centered at \(R_{\oplus,s}(M_p)\), with a standard deviation equal to one-third of the difference between \(R_{\oplus,s}(M_p)\) and \(R_{\text{silicate},s}(M_p)\), where \(R_{\text{silicate},s}(M_p)\) corresponds to a pure \ch{MgSiO3} composition.  
    \item (c) Wide Gaussian distribution: The radius is drawn from a Gaussian distribution centered at \(R_{\oplus,s}(M_p)\), with a standard deviation of the difference between \(R_{\oplus,s}(M_p)\) and \(R_{\text{silicate,s}}(M_p)\).
\end{itemize}. 

All calculations are based on the mass-radius relations provided by \cite{zeng_growth_2019}. Since the maximum volatile fraction in our simulations is limited to 1\% of the planetary mass, we neglect any contribution of atmospheric mass to \(M_p\). The mass-radius diagram for the solid body of the planets, showing these compositional variations, is presented in Fig.~\ref{fig:MR_solid}. The symmetric sampling of radii introduces a bias, leading to a fat tail of low-density planets. Among the tested scenarios, scenario (b) is the most realistic. 

For each assumption regarding solid-body density variation, we generate \(10^4\) data points and compute their time-integrated atmospheric loss. Using the remaining atmosphere, we then determine the corresponding transit radius using both upper- and lower-limit estimates and density. The density is then normalized by a theoretical model of an Earth-like composition (\(\rho/\rho_{\oplus,s}\)) \citep{zeng_growth_2019}.

\begin{table*}
\caption{Model input parameters used for Monte Carlo simulations of density changes driven by atmospheric loss.}
\label{tab:monte_carlo_2}
\centering
\begin{tabular}{lll}
\hline
\textbf{Parameter}        & \textbf{Distribution / Range}            & \textbf{Description} \\
\hline
$M_*$    &      $\mathcal{U}(0.1,1)$    & Stellar mass. \\
$M_p$    &  $\mathcal{U}(0.5,1)$         & Planetary mass. \\
$S/S_0$        & $ 10^{\mathcal{U}(1,4)}$ \footnote{To account for tidal disruption, we impose an upper‐instellation cutoff at $(S/S_0)_{max} = 2\times 10^4 \cdot M_*^{2.8}$.}    & Bolometric instellation scaled to Earth value. \\
$f_{\text{initial}}$ & $ 10^{\mathcal{U}(-4,-2)}$ & Initial volatile fraction.\\
& (a) $R_{\oplus,s} (M_p)$ & \\ 
$R_p$ & (b) $\mathcal{N}(R_{\oplus,s}(M_p),\sigma_b);\sigma_b = (R_{\text{silicate},s}(M_p)-R_{\oplus,s}(M_p))/3$ & Planetary radius\\
& (c)  $\mathcal{N}(R_{\oplus,s}(M_p),\sigma_c);\sigma_c = (R_{\text{silicate},s}(M_p)-R_{\oplus,s}(M_p))$ &\\
\hline
\end{tabular}
\end{table*}

\subsection{Predictions with Cosmic Shoreline}

\begin{figure*}
    \centering
    \includegraphics[width=0.9\linewidth]{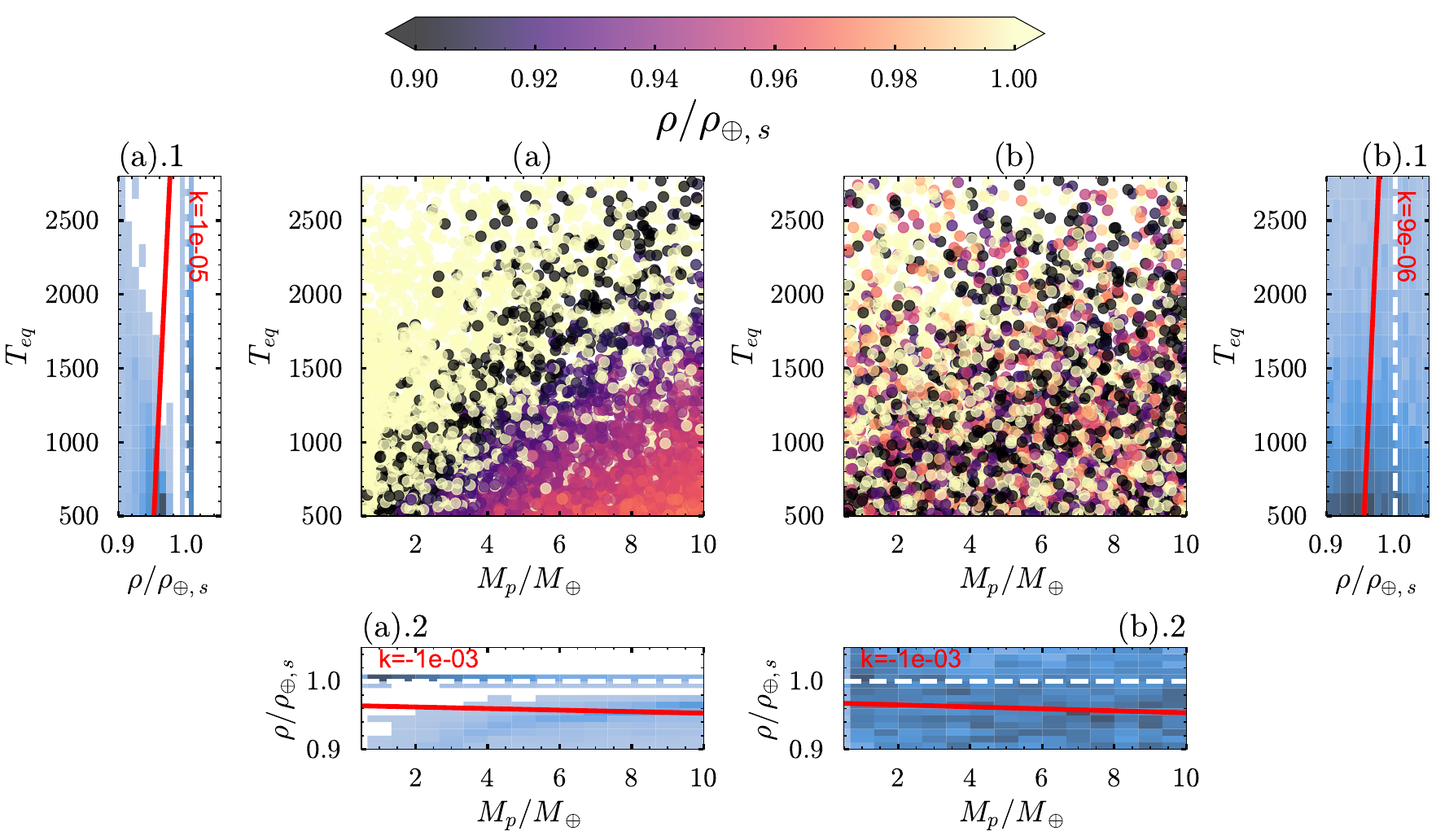}
    \caption{How the cosmic shoreline shapes the trend of ratio'd density (\(\rho/\rho_{\oplus,s}\)) for different assumptions of solid-body density distribution (Scenario (a) and (b) as seen in Table \ref{tab:monte_carlo_2}). Panels (a) and (b) present scatter plots of a subset of $10^3$ planets, plotted as planetary mass ($M_p$) against equilibrium temperature($T_{eq}$)—an alternative representation of the traditional cosmic shoreline format, since $v_{esc}$ scales with $M_p$, and $T_{eq}$ is derived from instellation ($S/S_0$). The dots are color-coded by the ratio'd density with the upper-limit estimate, where atmosphere has two layers and \(P_{transit}=10^{-6}\) bar. In (a), the cosmic shoreline shapes the boundary between yellow dots (bare rocks without atmosphere) and black dots (planets where the radius-boosting effect of an atmosphere is most distinguishable). In (b), where solid-body density varies, the transition is less distinct, but a visual trend remains, with more yellow points clustered in the upper-left region (high-temperature, low-mass planets), indicating that atmospheric loss is more severe in this regime. For better visualization, the color scale is limited between 0.9 and 1.0, although actual density values can extend beyond this range. Panels (a).1 and (b).1: A bivariate histogram is computed and visualized as a heatmap, where darker colors indicate higher number density. Linear regression of $\rho/\rho_{\oplus,s}$ as a function of equilibrium temperature (\(T_{eq}\)) for the full 10,000-planet sample. A decreasing trend is observed, as planets at high \(T_{eq}\) are more likely to be bare rocks with higher densities. Panels (a).2 and (b).2: Linear regression of $\rho/\rho_{\oplus,s}$ as a function of planetary mass (\(M_p\)), also showing a decreasing trend. This is because lower-mass planets are more likely to be stripped of their atmospheres, leading to a larger fraction of high-density bare rocky planets at lower masses. } 
    \label{fig:rho_ratio_pattern}
\end{figure*}

We present the results of Monte Carlo simulations under assumptions (a) and (b) in Fig.~\ref{fig:rho_ratio_pattern}. The two middle panels show planetary mass against equilibrium temperature, an alternative representation of the traditional cosmic shoreline format, as escape velocity scales with planetary mass, and equilibrium temperature is derived from instellation.

In panel (a)., where all planets are assumed to have an Earth-like composition, the ratio'd density \(\rho/\rho_{\oplus,s}\) does not exceed 1. The light yellow dots, representing planets with a ratio'd density of 1, correspond to bare rocky planets that have completely lost their atmospheres. These planets are adjacent to the black dots, which exhibit the lowest ratio'd density due to a substantial atmospheric contribution and a more pronounced radius-boosting effect. The transition zone from yellow to black dots follows the shape of the cosmic shoreline, as shown in Fig.~\ref{fig:CS_pl_vesp}. This boundary is not a strict cutoff but rather a transition zone influenced by factors such as initial volatile content and other parameters listed in Table \ref{tab:monte_carlo}.

Starting from the transition zone, for planets with \(\rho/\rho_{\oplus,s} < 1\), the ratio'd density increases toward the lower right (higher mass, lower irradiation planets). Two key factors contribute to this effect: 1. Atmospheric survival constraint: Retaining an atmosphere requires that the initial volatile content exceeds the cumulative atmospheric mass loss over time. For planets that retain their atmospheres and lie near the cosmic shoreline corresponding to a given \( f_{\text{initial}} \), their volatile inventories must surpass this threshold. The mass of the remaining atmosphere is therefore expected to be of the same order, or greater. Otherwise, the planet would have become a bare rock. From the upper left to the lower right, among the planets with atmosphere retained, planets with thinner atmospheres increasingly dominate over those with thick atmospheres. 2. Atmospheric radius-boosting effect: At higher \(T_{\text{eq}}\), atmospheric scale heights are larger due to increased thermal expansion. This amplifies the difference in transit radius between atmosphere-bearing and stripped planets, making the contrast in ratio'd density more prominent even if the absolute atmospheric mass is similar. For smaller planets, the radius-boosting effect leads to a lower ratio'd density compared to more massive planets with the same volatile fraction (Fig.~\ref{fig:transit-radius}).


We show bivariate histograms and perform linear regression for ratio'd density as a function of \(T_{\text{eq}}\) in panel (a.1) and as a function of planetary mass (\(M_p\)) in panel (a.2). The horizontal line at \(\rho/\rho_{\oplus,s} = 1\) consists of planets that have completely lost their atmospheres. Along this line, the color transitions from dark to light as \(M_p\) increases and \(T_{\text{eq}}\) decreases, indicating that fewer planets experience complete atmospheric loss, while more planets retain sufficient volatiles, resulting in \(\rho/\rho_{\oplus,s} < 1\). Consequently, the linear regression exhibits a decreasing trend with increasing \(M_p\) and decreasing \(T_{\text{eq}}\).


For assumption (b) of the narrow Gaussian distribution, the variation in the solid portion’s density introduces scatter,  making the distinct trends observed in (a) less visually apparent. The transition zone between atmosphere-bearing and stripped planets becomes more diffuse. Despite this, the linear regression still reveals a consistent decreasing trend in ratio'd density with decreasing \(T_{\text{eq}}\) and increasing \(M_p\), similar to (a). This suggests that despite the uncertainty in solid-body density, the underlying effect of atmospheric retention and loss, as shaped by the cosmic shoreline, remains statistically significant. We do not present results for assumption (c), as it follows a similar pattern with even greater scatter while preserving the same overall statistical behavior.
 
The decreasing trend in ratio'd density is statistically significant for a sample of 10,000 planets, but it may not be reliably detected with smaller sample sizes. To estimate the minimum number of observed exoplanets required to confirm this trend, we conduct a 500-time bootstrap analysis with varying sample sizes.

\begin{figure*}
    \centering
    \includegraphics[width=0.9\linewidth]{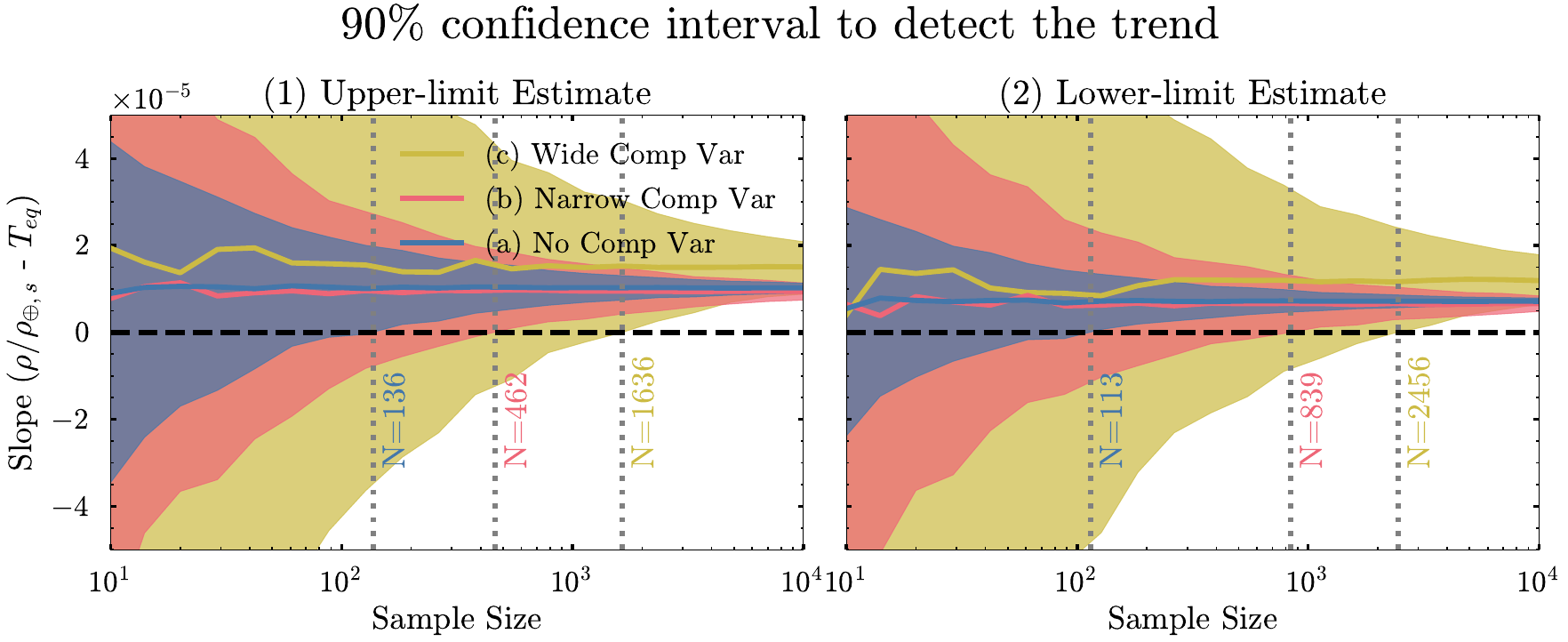}
    \caption{Constraining the slope of ratio'd density trends with varying sample sizes. The y-values correspond to the slope of the scaled density (\(\rho/\rho_{\oplus,s}\)) as a function of planetary mass (\(M_p\)), which is negative when considering the full sample of \(10^4\) simulated planets. Panel (1) corresponds to the upper-limit estimate of the transit radius, where atmosphere has two layers and \(P_{transit}=10^{-6}\)bar, while panel (2) represents the lower-limit estimate, where atmosphere has three layers and \(P_{transit}=0.02\)bar. The solid lines indicate the median slope values derived from bootstrap resampling, while the shaded regions represent the 90\% confidence intervals. Different colors represent assumptions for solid-body density: (a, blue) fixed Earth-like composition; (b, red) Gaussian scatter with \(\sigma\) = 1/3 the difference between pure-silicate and earth-like; (c, orange) wider Gaussian with \(\sigma\) = full silicate–Earth difference. The results illustrate how increasing the sample size improves the precision of the slope estimation. The wiggles are artifacts of the Monte Carlo procedure.}
    \label{fig:slope-size}
\end{figure*}

We draw sub-samples from our 10,000-planet sample without replacement. For each subsample of a given size, we perform a linear regression to obtain the slope of the ratio'd density trend. Repeating this process 1000 times, we construct a distribution of slope values for each sub-sample size. The mean slope from the 500 bootstrap trials is shown as the solid line in Fig.~\ref{fig:slope-size}, which remains approximately constant across sample sizes. The shaded region indicates the 90\% confidence interval of the slope distribution at each sample size. The location where the whole shaded region crosses the line slope=0 corresponds to an estimate of the minimum number of observed exoplanets needed to confirm the expected density trend influenced by atmospheric loss and the cosmic shoreline, with a 90\% confidence.

In Fig.~\ref{fig:slope-size}, we show our bootstrap results for both the lower- and upper-limit estimates of the transit radius. For assumption (a), where all planets have an Earth-like composition, there is no significant difference between the two estimates. However, for assumption (b), where the solid density follows a Gaussian distribution, the required sample size for detecting the trend in the lower-limit estimate is $\sim 2 \times$ for the upper-limit estimate.

For assumptions (a) and (b), the decreasing trend of \(\rho/\rho_{\oplus,s}\) with \(T_{\text{eq}}\) could be constrained with a feasible sample size. However, constraining the trend with planetary mass (\(M_p\)) requires a larger, less feasible sample size. If the variation in solid composition is as large as in assumption (c), the required sample size exceeds 1000, making it difficult to detect the trend with near-future exoplanet observations.

Since the instellation dependence of a planet’s iron core size \citep{ebel_elusive_2018} may also contribute to the correlation between bulk density and instellation, accurately constraining the bulk densities of the innermost planets is required for distinguishing between the two hypotheses. If these planets are significantly iron-rich, the trend could be explained by a solid-body compositional gradient; otherwise, the cosmic shoreline interpretation is favored, though the degeneracy still warrants further investigation.

In summary, we propose a novel approach to examine the cosmic shoreline, but it remains challenging to achieve with current exoplanet data. Transit and emission spectroscopy remain the most effective methods for constraining the shoreline.

\subsubsection{Impact of Observational Uncertainties}

To assess the impact of observational uncertainties on the detectability of the density trend, we incorporated simulated measurement errors into our Monte Carlo analysis. For each planet, we randomly perturbed its mass and radius using Gaussian noise (e.g. \( M_{p,obs} = \mathcal{N}(M_p,\delta_{M_p})\)). We vary relative uncertainties for mass and radius (\(\delta_{M_p}/M_p\) and \(\delta_{R_p}/R_p\)) to represent different levels of measurement precision. This approach allows us to quantify how increasing or decreasing measurement precision affects the minimum sample size needed to statistically detect the trend in ratio'd density driven by atmospheric loss. 

\begin{figure}
    \centering
    \includegraphics[width=1.0\linewidth]{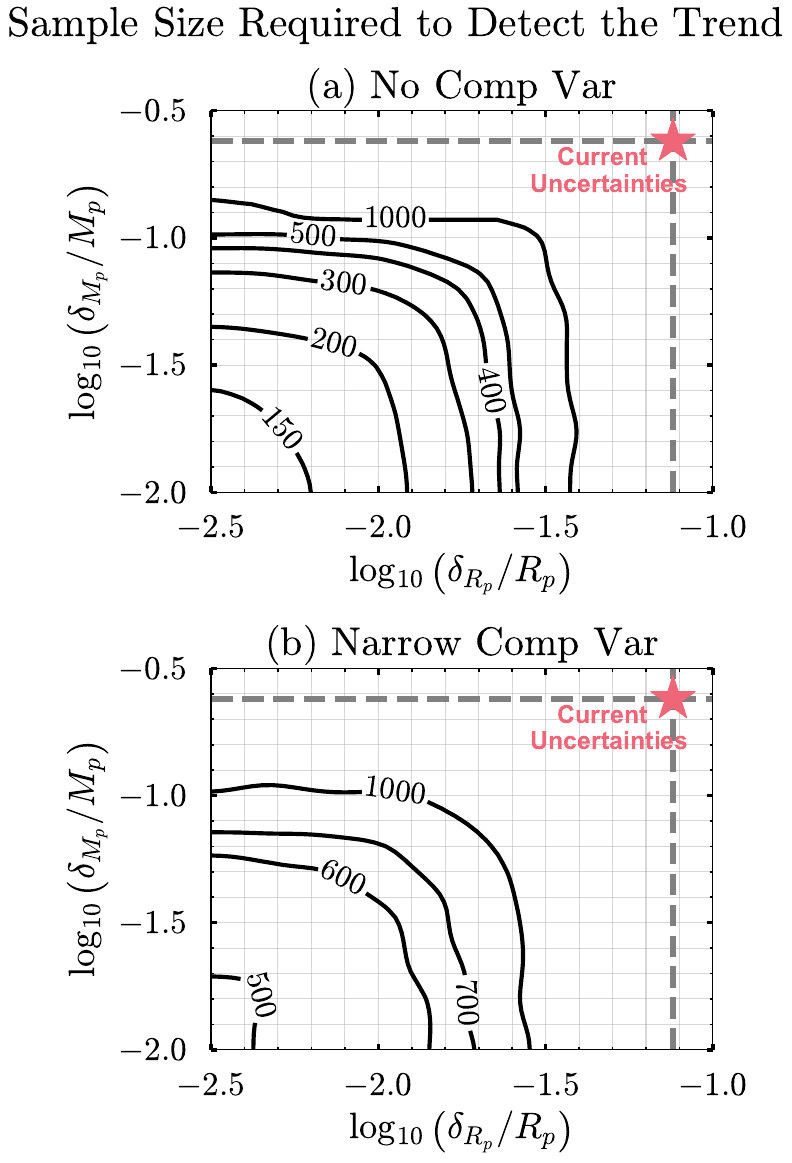}
    \caption{Sample size required to detect the trend ---measured bulk densities increase with increasing $T_{eq}$--- as a function of relative measurement uncertainties in radius and mass, for two different assumptions of the variance of the solid-body's density (Comp Var): (a) fixed Earth-like composition; (b) Gaussian scatter with \(\sigma\) = 1/3 the difference between pure-silicate and earth-like. The vertical and horizontal dashed lines indicate the average relative uncertainties in mass and radius for current-day observations  (see Sec. \ref{sec:data}) for sample details. The contour lines are smoothed by a Gaussian filter, and the remaining wiggles are artifacts of the Monte Carlo and smoothing.}
    \label{fig:size-unc}
\end{figure}

The results, shown in Fig.~\ref{fig:size-unc}, indicate that when there is no variation in solid density (panel a), or when the variance is reasonably small (panel b), improving the precision of both radius and mass measurements can reduce the required sample size to several hundreds to robustly detect the expected density trend. However, if the standard variance exceeds the difference between pure-silicate and Earth-like compositions (assumption (c)), improving measurement precision has almost no effect. In this case, the measurement uncertainty is smaller than the intrinsic variation introduced by the solid component, rendering the atmospheric contribution undetectable.



\section{Discussion}\label{sec:discussion}

\subsection{Cosmic Shorelines}

The hydrodynamic simulations discussed in Sec. \ref{sec:hydro} have many uncertainties. These simulations provide atmospheric loss rates as a function of XUV flux; however, the XUV wavelength range varies across models. For example, \cite{tian_thermal_2009} and \cite{johnstone_hydrodynamic_2020} consider XUV wavelengths up to 1050 \r{A} and 1000 \r{A}, respectively, while \cite{nakayama_survival_2022} restricts the range to 910 \r{A} to exclude Lyman-\(\alpha\) emission. Although the resulting impact is not explicitly included in our model, it is expected to fall within the uncertainties of our EUV extrapolation, as \cite{ribas_evolution_2005} shows that the flux ratio between 920–1180 \r{A} and 1–920 \r{A} is 0.03 for a 0.1 Gyr star and 0.2 for the Sun. Additionally, Table 1 in \cite{johnstone_hydrodynamic_2020} provides an X-ray-to-EUV relation that deviates from our EUV extrapolation, suggesting differences in spectral shapes between our model and their input, as well as potential differences from other hydrodynamic simulations. We do not know the cause of the difference. However, \cite{johnstone_active_2021} show a relation consistent with our model (Fig.~\ref{fig:EUV}).

The EUV extrapolation in our model assumes that the solar spectrum is representative of other stars, which may overlook the diversity in stellar properties and activity levels. For X-ray flux, we adopt standard evolutionary models, though stellar behavior can vary among stars of the same mass \citep{penz_influence_2008}, for example, differences in initial rotation rates can result in varying durations of the saturated phase \citep{tu_extreme_2015, johnstone_active_2021}. Additionally, our model does not account for contributions from energetic flares \citep{do_amaral_contribution_2022, pass_receding_2025}. We sample EUV fluxes over a range spanning more than a factor of 20 (Fig.~\ref{fig:EUV}; \citealt{king_euv_2020}), which might encompass these uncertainties. 

\cite{nakayama_survival_2022} show that atomic line cooling of C, N, and O can cool the upper atmosphere sufficiently to prevent hydrodynamic escape. The high efficiency of those metal emissions was earlier emphasized in \cite{Liu_Tian_2018}. Lyman-\(\alpha\) cooling (\citealp[e.g.,][]{murray-clay_atmospheric_2009, zhang_escaping_2022} is prominent in \ch{H2}-rich atmospheres, along with molecular cooling \citep{yoshida_less_2022,garcia_munoz_heating_2023,yoshida_suppression_2024,garcia_munoz_heating_2024,garcia_munoz_nlte_2024} and a range of cooling lines in rock vapour atmospheres \citep{ito_hydrodynamic_2021}. However, \cite{chatterjee_novel_2024} suggest that in the scenarios modeled by \cite{nakayama_survival_2022}, considering an ionized exobase would push the atmosphere into a hydrodynamic regime, enhancing escape. Furthermore, including ambipolar diffusion could increase the escape rate. 


The hydrodynamic escape simulations used here assume a single‐component atmosphere without a cold trap. In a more realistic mixed atmosphere, however, photo‑dissociation of \ch{H2O}, and the subsequent H escape and hydrodynamic drag of heavier species could complicate escape. The \ch{H2O} cold trap can create a dichotomy between habitable‑zone and hotter planets \citep{krissansen-totton_implications_2023}. 


In this study, we argue that the solid mantle volatile content is much smaller than the volatiles in magma and atmosphere. However, if the initial volatile inventory is high, even a small fraction trapped in the mantle could later form a detectable atmosphere via volcanism (Fig. \ref{fig:CS_delay}). Also, many M‑dwarf planets likely accreted thick \ch{H2}-rich atmospheres, and the \ch{H2} can also react with the magma ocean to make \ch{H2O}. At the same time, the hydrodynamic loss of \ch{H} will drag away heavy species. This primary‐atmosphere evolution may define a shoreline distinct from those we model here \citep{kite_exoplanet_2020, krissansen-totton_erosion_2024, cherubim_oxidation_2025}. Additionally, for \ch{N2}-dominated atmospheres, revival through a water loss redox pump is more likely, which is time-dependent and supply-limited \citep{wordsworth_atmospheric_2016, schaefer_predictions_2016, kite_water_2021}. Thus models that incorporate hydrodynamic escape-driven magma ocean solidification, chemical equilibrium, mantle outgassing, and atmospheric evolution, along with stellar evolution. might be considered for future work.

We tested a nonthermal ion escape rate for \ch{CO2}-dominated atmospheres using the results from  \cite{chin_role_2024, dong_is_2017, wood_new_2021}, and found that it is two orders of magnitude lower than thermal escape rates and therefore should be excluded. The planetary magnetic fields could influence atmospheric escape, particularly on hot planets experiencing global ion outflows \citep[e.g.,][]{cohen_magnetospheric_2014, ramstad_intrinsic_2021}). Moreover, the timing of dynamo onset may play a critical role in atmospheric retention, in the context of stellar evolution. Additionally, impact erosion can significantly strip atmospheres, particularly during the early stages of planetary formation \citep[e.g.,][]{schlichting_atmosphere_2018, kegerreis_atmospheric_2020,wyatt_susceptibility_2020}. The influence of the role of occasional but intense stellar activity cycles in modulating escape rates also warrant further investigation \citep{wagner_observations_1988,chadney_xuv-driven_2015,levine_exoplanet_2024, pass_receding_2025}.

Given the uncertainties in our hydrodynamic modeling and statistical assumptions, the cosmic shorelines in Figs. \ref{fig:cs_stmass} and \ref{fig:CS_pl_vesp} are best used to compare relative atmospheric‐retention probabilities across planets and should be applied to any single planet with caution. For example, our model predict Earth and Venus are near the 90\% \(f_{\text{initial}} = 10^{-4}\) shoreline, implying hundreds of bars of potential loss. However, there is no evidence of such extensive thermal escape. Several factors could reconcile this discrepancy: (1) Probabilistic interpretation: The 90\% retention shoreline indicates that, at that orbit, 90\% of modeled cumulative lose less than $10^{-4}M_p$, while 10\% lose more. The distribution of cumulative loss spans orders of magnitude (Fig.~\ref{fig:logistic}). (2). Stellar Rotation: We adopt an average X‑ray prescription, but slow‑rotating stars can exhibit saturated X‑ray fluxes an order of magnitude below our prescription (\citealt{tu_extreme_2015}). Our Sun may itself have been a slow rotator (\citealt{2008A&A...489L..53B, 2019ApJ...876L..16S, 2020Sci...368..518R, 2021MNRAS.500.1158J}), reducing historical loss rates. (3). Atomic line cooling, omitted in the \ch{CO2} model applied here \citep{tian_thermal_2009,tian_thermal_2009-1}, could efficiently suppress hydrodynamic escape. In mixed atmospheres, this cooling might be stronger than in the \ch{N2} model used here\citep{chatterjee_novel_2024}. (4). Regime transitions: our \ch{CO2} shoreline ignores the shift from hydrodynamic to hydrostatic escape at low XUV fluxes. In contrast, the \ch{N2}–CP24 model explicitly incorporates this transition, yielding a negligible minimum integrated loss if the stellar XUV history never exceeds the hydrodynamic threshold. (5). Hundreds of bars of mass loss may have occurred yet remain unrecognized, for example, if it occurred at the embryo stage. 

\citet{zahnle_cosmic_2017} emphasized the unknown unknowns in their cosmic shoreline hypothesis. The calculations reported here have explored the effects of the known uncertainties and complexities on the cosmic shoreline. However, we have also discussed known processes that we have neglected due to both issues of scope and lack of available simulations. 


\subsection{Trend of ratio'd density}

This study provides a first-step analysis of how density trends can be used to test the cosmic shoreline (Sec. \ref{sec:density}), but it is based on simplified assumptions and considers only \ch{CO2}-dominated atmospheres. Future work might expand to include atmospheres with varied compositions and incorporate a more realistic atmospheric structure, especially for the possible thermal inversion under high-pressure and high-temperature conditions \citep{selsis_cool_2023}, and determine transit radii using detailed radiative transfer calculations.

We considered density variations in the solid body while centering the composition around an Earth-like composition. We also did a sensitivity test with an alternative assumption where planetary radii are uniformly distributed between \( R_{\text{silicate}}(M_p) \) and \( R_{\text{iron}}(M_p) \). Under this assumption, the scaled density \(\rho/\rho_{\oplus}^*\) retains the same decreasing trend with decreasing equilibrium temperature. However, it exhibits an increasing trend with planetary mass. This arises from the intrinsic mass-radius relation of the solid body alone (see Fig.~\ref{fig:solid-only}), where more massive planets tend to have higher solid-body densities. This introduces a potential complication in interpreting the trend, as any dependence of solid-body composition on \(M_p\) and/or \(T_{eq}\) could contribute to or alter the pattern. Further investigation is needed to disentangle the influence of atmospheric retention from intrinsic density variations of the planetary core \citep{valencia_detailed_2007, rogers_framework_2010, plotnykov_observation_2024, rice_uncertainties_2025}.

Even if the density variations are centered around an Earth-like composition, our results indicate that achieving statistically significant conclusions with near-future datasets will be challenging, but improved observational uncertainties in the future might reduce the required sample size. In addition, as noted by \citet{Valencia_2025}, the pure‑silicate density defines a rocky threshold radius. Planets with radii above the threshold must contain volatiles and thus either retain an atmosphere or have the potential to regenerate one. Applying this constraint in statistical analyses could further reduce the required sample size.

In summary, JWST transmission and emission spectroscopy remain the most effective methods for detecting atmospheres on rocky planets and testing the cosmic shoreline.

\section{Conclusions}

With the cosmic shoreline defined as the critical bolometric instellation (\( S^*_{bol} \)) above which a planet is highly unlikely to retain its atmosphere, given the mass of volatiles available for loss, we combine the results of previous hydrodynamic atmospheric models \citep{tian_thermal_2009,tian_thermal_2009-1,johnstone_hydrodynamic_2020,nakayama_survival_2022,chatterjee_novel_2024} and stellar evolution models \citep{baraffe_new_2015,jackson_coronal_2012,selsis_habitable_2007,king_euv_2020} and find:
\begin{enumerate}
    \item \emph{Rocky Worlds DDT:} Most targets under consideration fall on the unfavorable side of \cite{zahnle_cosmic_2017}'s original, XUV-driven semi-empirical cosmic shoreline. In contrast, our revised shoreline---based on existing hydrodynamic escape models---are more optimistic, indicating that tens of known exoplanets may be capable of retaining atmospheres if they began with sufficient volatile inventories. For example, 55 Cnc e, which has a detected atmosphere, lies beyond the traditional cosmic shoreline but remains within our revised \ch{CO2} and \ch{N2} boundaries, assuming it began with a volatile mass fraction of 1\%. Future models that account for atomic line cooling more fully would likely shift the shoreline even further toward higher retention.
    \item \emph{Cosmic shoreline as a transition zone:} If planets form with variable initial volatile contents and compositions, then we expect a gradual transition between airless worlds and planets retaining substantial atmospheres. Atmosphere nondetection can help constrain the maximum volatile content, and it may apply to other planets in the system if volatile variation within a single system is small.
    \item \emph{Dependence on stellar mass:} The cosmic shoreline shifts toward higher instellation with increasing stellar mass, due not only to the lower cumulative XUV output of higher-mass stars \citep{zahnle_cosmic_2017, pass_receding_2025}, but also to nonlinearities in the relationship between escape rate and XUV flux. We assessed how the cosmic shoreline’s dependence on stellar mass varies across hydrodynamic escape models. 
    \item \emph{Dependence on planetary mass:} With hydrodynamic escape-based models, the scaling of the shoreline with escape velocity is nonuniform. Due to the effects of atomic line cooling, volatile-rich super-Earths may be highly resilient to catastrophic photoevaporation even on close-in orbits around low mass stars. Studies of the collisional-radiative nonequilibrium of super-Earth ionospheres under intense XUV irradiation would help constrain this regime.
    \item \emph{New metric for atmosphere retention:} We introduce a priority metric that quantifies the difference between a planet's instellation and the critical instellation required for atmospheric retention at a given planetary mass. A ranked list of select exoplanets based on this metric is provided in Table \ref{tab:targets}. Our results highlight TOI-4559~b and TOI-771~b as potentially rocky planets with high observability metrics that are likely to retain atmospheres.
    \item \emph{Testing the cosmic shoreline with ratio'd density:} Due to the radius-boosting effect of atmospheres, the ratio'd density (\(\rho/\rho_{\oplus}^*\)) provides an alternative way to test the cosmic shoreline. Since atmospheric retention becomes less likely at higher instellation, an increasing trend in scaled density with increasing instellation could be tested. Improving radius and mass measurement precision could reduce the required sample size to hundreds of exoplanets.    
\end{enumerate}
Our detailed reanalysis of the cosmic shoreline provides a foundation for effective refinement of the hypothesis through both modeling studies and observational surveys. 

\begin{acknowledgments}
This work was supported by NASA award No.80NSSC21K1718, which is part of the Habitable Worlds program. R.D.C also acknowledges support from the UK Science and Technology Facilities Council (STFC) and the Alfred P. Sloan Foundation under grant G202114194 (AEThER). This research has made use of the NASA Exoplanet Archive, which is operated by the California Institute of Technology, under contract with the National Aeronautics and Space Administration under the Exoplanet Exploration Program. We thank the anonymous reviewer and Antonio Garc\'{\i}a Mu\~{n}oz for comments that greatly improved the quality of the manuscript. We thank Michael Zhang, Bowen Fan, David Catling, Ted Bergin, Dorian Abbot, RJ Graham, Robin Wordsworth, Jegug Ih, Madison Brady, Caroline Piaulet, Lily Zhao, Rafa Luque, Tim Lichtenberg, Ray Pierrehumbert and Oli Shorttle for insightful discussions. We thank the anonymous reviewer for their constructive feedback and insightful suggestions, which significantly improved the clarity and quality of this manuscript. All code associated with this paper is available via the Zenodo (\url{https://zenodo.org/records/17216167}) or upon request from the lead author.
\end{acknowledgments}

\software{Numpy \citep{harris2020array}, Matplotlib \citep{Hunter:2007}, statsmodels.api \citep{seabold2010statsmodels}, Scipy \citep{2020SciPy-NMeth}, pandas \citep{reback2020pandas}, SciencePlots \citep{SciencePlots}}

\bibliographystyle{aasjournal7}
\bibliography{Cosmic}

\begin{thebibliography}{}
\expandafter\ifx\csname natexlab\endcsname\relax\def\natexlab#1{#1}\fi
\providecommand{\url}[1]{\href{#1}{#1}}
\providecommand{\dodoi}[1]{doi:~\href{http://doi.org/#1}{\nolinkurl{#1}}}
\providecommand{\doeprint}[1]{\href{http://ascl.net/#1}{\nolinkurl{http://ascl.net/#1}}}
\providecommand{\doarXiv}[1]{\href{https://arxiv.org/abs/#1}{\nolinkurl{https://arxiv.org/abs/#1}}}

\bibitem[{V. Adibekyan {et~al.}(2024)Adibekyan, Deal, Dorn, Dittrich, Soares, Sousa, Santos, Bitsch, Mordasini, Barros, Bossini, Campante, Delgado~Mena, Demangeon, Figueira, Moedas, Martirosyan, Israelian, \& Hakobyan}]{adibekyan_linking_2024}
Adibekyan, V., Deal, M., Dorn, C., {et~al.} 2024, \bibinfo{title}{Linking the primordial composition of planet building disks to the present-day composition of rocky exoplanets,} Astronomy and Astrophysics, 692, A67, \dodoi{10.1051/0004-6361/202452193}

\bibitem[{D. Andrault {et~al.}(2011)Andrault, Bolfan-Casanova, Nigro, Bouhifd, Garbarino, \& Mezouar}]{andrault_solidus_2011}
Andrault, D., Bolfan-Casanova, N., Nigro, G.~L., {et~al.} 2011, \bibinfo{title}{Solidus and liquidus profiles of chondritic mantle: {Implication} for melting of the {Earth} across its history,} Earth and Planetary Science Letters, 304, 251, \dodoi{10.1016/j.epsl.2011.02.006}

\bibitem[{P.~C. August {et~al.}(2025)August, Buchhave, Diamond-Lowe, Mendonça, Gressier, Rathcke, Allen, Fortune, Jones, Valdés, Demory, Espinoza, Fisher, Gibson, Heng, Hoeijmakers, Hooton, Kitzmann, Prinoth, Eastman, \& Barnes}]{august_hot_2025}
August, P.~C., Buchhave, L.~A., Diamond-Lowe, H., {et~al.} 2025, \bibinfo{title}{Hot {Rocks} {Survey} {I}: {A} possible shallow eclipse for {LHS} 1478 b,} Astronomy \& Astrophysics, 695, A171, \dodoi{10.1051/0004-6361/202452611}

\bibitem[{I. Baraffe {et~al.}(2015)Baraffe, Homeier, Allard, \& Chabrier}]{baraffe_new_2015}
Baraffe, I., Homeier, D., Allard, F., \& Chabrier, G. 2015, \bibinfo{title}{New evolutionary models for pre-main sequence and main sequence low-mass stars down to the hydrogen-burning limit,} Astronomy \& Astrophysics, 577, A42, \dodoi{10.1051/0004-6361/201425481}

\bibitem[{R. Barnes(2017)Barnes}]{barnes2017}
Barnes, R. 2017, \bibinfo{title}{Tidal locking of habitable exoplanets,} Celestial mechanics and dynamical astronomy, 129, 509

\bibitem[{T.~A. Berger {et~al.}(2020)Berger, Huber, Gaidos, Van~Saders, \& Weiss}]{berger_gaiakepler_2020}
Berger, T.~A., Huber, D., Gaidos, E., Van~Saders, J.~L., \& Weiss, L.~M. 2020, \bibinfo{title}{The {Gaia}–{Kepler} {Stellar} {Properties} {Catalog}. {II}. {Planet} {Radius} {Demographics} as a {Function} of {Stellar} {Mass} and {Age},} The Astronomical Journal, 160, 108, \dodoi{10.3847/1538-3881/aba18a}

\bibitem[{E.~A. Bergin {et~al.}(2015)Bergin, Blake, Ciesla, Hirschmann, \& Li}]{bergin_tracing_2015}
Bergin, E.~A., Blake, G.~A., Ciesla, F., Hirschmann, M.~M., \& Li, J. 2015, \bibinfo{title}{Tracing the ingredients for a habitable earth from interstellar space through planet formation,} Proceedings of the National Academy of Sciences, 112, 8965, \dodoi{10.1073/pnas.1500954112}

\bibitem[{E.~A. Bergin {et~al.}(2023)Bergin, Kempton, Hirschmann, Bastelberger, Teal, Blake, Ciesla, \& Li}]{bergin_exoplanet_2023}
Bergin, E.~A., Kempton, E. M.-R., Hirschmann, M., {et~al.} 2023, \bibinfo{title}{Exoplanet {Volatile} {Carbon} {Content} as a {Natural} {Pathway} for {Haze} {Formation},} The Astrophysical Journal Letters, 949, L17, \dodoi{10.3847/2041-8213/acd377}

\bibitem[{Z.~K. Berta-Thompson {et~al.}(2025)Berta-Thompson, Wachiraphan, \& Murray}]{bertathompson2025}
Berta-Thompson, Z.~K., Wachiraphan, P., \& Murray, C. 2025, \bibinfo{title}{The 3D Cosmic Shoreline for Nurturing Planetary Atmospheres,} \doarXiv{2507.02136}

\bibitem[{I. Blanchard {et~al.}(2022)Blanchard, Rubie, Jennings, Franchi, Zhao, Petitgirard, Miyajima, Jacobson, \& Morbidelli}]{blanchard_metalsilicate_2022}
Blanchard, I., Rubie, D., Jennings, E., {et~al.} 2022, \bibinfo{title}{The metal–silicate partitioning of carbon during {Earth}'s accretion and its distribution in the early solar system,} Earth and Planetary Science Letters, 580, 117374, \dodoi{10.1016/j.epsl.2022.117374}

\bibitem[{J.~C. Bond {et~al.}(2010)Bond, Lauretta, \& O’Brien}]{bond_making_2010}
Bond, J.~C., Lauretta, D.~S., \& O’Brien, D.~P. 2010, \bibinfo{title}{Making the {Earth}: {Combining} dynamics and chemistry in the {Solar} {System},} Icarus, 205, 321, \dodoi{10.1016/j.icarus.2009.07.037}

\bibitem[{J. {Bouvier}(2008){Bouvier}}]{2008A&A...489L..53B}
{Bouvier}, J. 2008, \bibinfo{title}{{Lithium depletion and the rotational history of exoplanet host stars},} \aap, 489, L53, \dodoi{10.1051/0004-6361:200810574}

\bibitem[{C.~L. Brinkman {et~al.}(2023)Brinkman, Weiss, Dai, Huber, Kite, Valencia, Bean, Beard, Behmard, Blunt, {et~al.}}]{brinkman2023}
Brinkman, C.~L., Weiss, L.~M., Dai, F., {et~al.} 2023, \bibinfo{title}{TOI-561 b: A Low-density Ultra-short-period “Rocky” Planet around a Metal-poor Star,} The Astronomical Journal, 165, 88

\bibitem[{A. Caldiroli {et~al.}(2022)Caldiroli, Haardt, Gallo, Spinelli, Malsky, \& Rauscher}]{caldiroli_irradiation-driven_2022}
Caldiroli, A., Haardt, F., Gallo, E., {et~al.} 2022, \bibinfo{title}{Irradiation-driven escape of primordial planetary atmospheres: {II}. {Evaporation} efficiency of sub-{Neptunes} through hot {Jupiters},} Astronomy \& Astrophysics, 663, A122, \dodoi{10.1051/0004-6361/202142763}

\bibitem[{D.~C. Catling \& J.~F. Kasting(2017)Catling \& Kasting}]{catling_atmospheric_2017}
Catling, D.~C., \& Kasting, J.~F. 2017, Atmospheric {Evolution} on {Inhabited} and {Lifeless} {Worlds} (Cambridge University Press).
\newblock \url{https://ui.adsabs.harvard.edu/abs/2017aeil.book.....C}

\bibitem[{J.~M. Chadney {et~al.}(2015)Chadney, Galand, Unruh, Koskinen, \& Sanz-Forcada}]{chadney_xuv-driven_2015}
Chadney, J.~M., Galand, M., Unruh, Y.~C., Koskinen, T.~T., \& Sanz-Forcada, J. 2015, \bibinfo{title}{{XUV}-driven mass loss from extrasolar giant planets orbiting active stars,} Icarus, 250, 357, \dodoi{10.1016/j.icarus.2014.12.012}

\bibitem[{M. Chase(1998)Chase}]{chase_nist-janaf_1998}
Chase, M. 1998, {NIST}-{JANAF} {Thermochemical} {Tables}, 4th {Edition} (American Institute of Physics, -1)

\bibitem[{R.~D. Chatterjee \& R.~T. Pierrehumbert(2024)Chatterjee \& Pierrehumbert}]{chatterjee_novel_2024}
Chatterjee, R.~D., \& Pierrehumbert, R.~T. 2024, \bibinfo{title}{Novel {Physics} of {Escaping} {Secondary} {Atmospheres} {May} {Shape} the {Cosmic} {Shoreline},} arXiv, \dodoi{10.48550/arXiv.2412.05188}

\bibitem[{H. Chen \& S.~A. Jacobson(2022)Chen \& Jacobson}]{chen_impact_2022}
Chen, H., \& Jacobson, S.~A. 2022, \bibinfo{title}{Impact induced atmosphere-mantle exchange sets the volatile elemental ratios on primitive {Earths},} Earth and Planetary Science Letters, 594, 117741, \dodoi{10.1016/j.epsl.2022.117741}

\bibitem[{C. Cherubim {et~al.}(2025)Cherubim, Wordsworth, Bower, Sossi, Adams, \& Hu}]{cherubim_oxidation_2025}
Cherubim, C., Wordsworth, R., Bower, D.~J., {et~al.} 2025, \bibinfo{title}{An {Oxidation} {Gradient} {Straddling} the {Small} {Planet} {Radius} {Valley},} The Astrophysical Journal, 983, 97, \dodoi{10.3847/1538-4357/adbca9}

\bibitem[{L. Chin {et~al.}(2024)Chin, Dong, \& Lingam}]{chin_role_2024}
Chin, L., Dong, C., \& Lingam, M. 2024, \bibinfo{title}{Role of {Planetary} {Radius} on {Atmospheric} {Escape} of {Rocky} {Exoplanets},} The Astrophysical Journal Letters, 963, L20, \dodoi{10.3847/2041-8213/ad27d8}

\bibitem[{R. Cloutier \& K. Menou(2020)Cloutier \& Menou}]{cloutier2020evolution}
Cloutier, R., \& Menou, K. 2020, \bibinfo{title}{Evolution of the radius valley around low-mass stars from Kepler and K2,} The Astronomical Journal, 159, 211, \dodoi{10.3847/1538-3881/ab8237}

\bibitem[{O. Cohen {et~al.}(2014)Cohen, Drake, Glocer, Garraffo, Poppenhaeger, Bell, Ridley, \& Gombosi}]{cohen_magnetospheric_2014}
Cohen, O., Drake, J.~J., Glocer, A., {et~al.} 2014, \bibinfo{title}{Magnetospheric Structure and Atmospheric Joule Heating of Habitable Planets Orbiting M-dwarf Stars,} The Astrophysical Journal, 790, 57, \dodoi{10.1088/0004-637X/790/1/57}

\bibitem[{B.~P. Coy {et~al.}(2025)Coy, Ih, Kite, Koll, Tenthoff, Bean, Mansfield, Zhang, Xue, Kempton, {et~al.}}]{coy2025}
Coy, B.~P., Ih, J., Kite, E.~S., {et~al.} 2025, \bibinfo{title}{Population-level Hypothesis Testing with Rocky Planet Emission Data: A Tentative Trend in the Brightness Temperatures of M-Earths,} The Astrophysical Journal, 987, 22

\bibitem[{I.~J.~M. Crossfield {et~al.}(2022)Crossfield, Malik, Hill, Kane, Foley, Polanski, Coria, Brande, Zhang, Wienke, Kreidberg, Cowan, Dragomir, Gorjian, Mikal-Evans, Benneke, Christiansen, Deming, \& Morales}]{crossfield_gj_2022}
Crossfield, I. J.~M., Malik, M., Hill, M.~L., {et~al.} 2022, \bibinfo{title}{{GJ} 1252b: {A} {Hot} {Terrestrial} {Super}-{Earth} with {No} {Atmosphere},} The Astrophysical Journal Letters, 937, L17, \dodoi{10.3847/2041-8213/ac886b}

\bibitem[{B.-O. Demory {et~al.}(2016)Demory, Gillon, De~Wit, Madhusudhan, Bolmont, Heng, Kataria, Lewis, Hu, Krick, Stamenković, Benneke, Kane, \& Queloz}]{demory_map_2016}
Demory, B.-O., Gillon, M., De~Wit, J., {et~al.} 2016, \bibinfo{title}{A map of the large day–night temperature gradient of a super-{Earth} exoplanet,} Nature, 532, 207, \dodoi{10.1038/nature17169}

\bibitem[{T.~R. Denman {et~al.}(2020)Denman, Leinhardt, Carter, \& Mordasini}]{denman_atmosphere_2020}
Denman, T.~R., Leinhardt, Z.~M., Carter, P.~J., \& Mordasini, C. 2020, \bibinfo{title}{Atmosphere loss in planet–planet collisions,} Monthly Notices of the Royal Astronomical Society, 496, 1166, \dodoi{10.1093/mnras/staa1623}

\bibitem[{L.~N.~R. do~Amaral {et~al.}(2022)do~Amaral, Barnes, Segura, \& Luger}]{do_amaral_contribution_2022}
do~Amaral, L. N.~R., Barnes, R., Segura, A., \& Luger, R. 2022, \bibinfo{title}{The {Contribution} of {M}-dwarf {Flares} to the {Thermal} {Escape} of {Potentially} {Habitable} {Planet} {Atmospheres},} The Astrophysical Journal, 928, 12, \dodoi{10.3847/1538-4357/ac53af}

\bibitem[{C. Dong {et~al.}(2018)Dong, Jin, Lingam, Airapetian, Ma, \& Van Der~Holst}]{dong_atmospheric_2018}
Dong, C., Jin, M., Lingam, M., {et~al.} 2018, \bibinfo{title}{Atmospheric escape from the {TRAPPIST}-1 planets and implications for habitability,} Proceedings of the National Academy of Sciences, 115, 260, \dodoi{10.1073/pnas.1708010115}

\bibitem[{C. Dong {et~al.}(2017)Dong, Lingam, Ma, \& Cohen}]{dong_is_2017}
Dong, C., Lingam, M., Ma, Y., \& Cohen, O. 2017, \bibinfo{title}{Is {Proxima} {Centauri} b {Habitable}: {A} {Study} of {Atmospheric} {Loss},} The Astrophysical Journal Letters, 837, L26, \dodoi{10.3847/2041-8213/aa6438}

\bibitem[{D.~S. Ebel \& S.~T. Stewart(2018)Ebel \& Stewart}]{ebel_elusive_2018}
Ebel, D.~S., \& Stewart, S.~T. 2018, in The {Elusive} {Origin} of {Mercury}, 497--515, \dodoi{10.1017/9781316650684.019}

\bibitem[{L. Elkins-Tanton(2008)Elkins-Tanton}]{elkins-tanton_linked_2008}
Elkins-Tanton, L. 2008, \bibinfo{title}{Linked magma ocean solidification and atmospheric growth for {Earth} and {Mars},} Earth and Planetary Science Letters, 271, 181, \dodoi{10.1016/j.epsl.2008.03.062}

\bibitem[{N.~V. Erkaev {et~al.}(2016)Erkaev, Lammer, Odert, Kislyakova, Johnstone, Güdel, \& Khodachenko}]{erkaev_euv-driven_2016}
Erkaev, N.~V., Lammer, H., Odert, P., {et~al.} 2016, \bibinfo{title}{{EUV}-driven mass-loss of protoplanetary cores with hydrogen-dominated atmospheres: the influences of ionization and orbital distance,} Monthly Notices of the Royal Astronomical Society, 460, 1300, \dodoi{10.1093/mnras/stw935}

\bibitem[{N.~V. Erkaev {et~al.}(2013)Erkaev, Lammer, Odert, Kulikov, Kislyakova, Khodachenko, Güdel, Hanslmeier, \& Biernat}]{erkaev_xuv-exposed_2013}
Erkaev, N.~V., Lammer, H., Odert, P., {et~al.} 2013, \bibinfo{title}{{XUV}-{Exposed}, {Non}-{Hydrostatic} {Hydrogen}-{Rich} {Upper} {Atmospheres} of {Terrestrial} {Planets}. {Part} {I}: {Atmospheric} {Expansion} and {Thermal} {Escape},} Astrobiology, 13, 1011, \dodoi{10.1089/ast.2012.0957}

\bibitem[{M. Fischer-Gödde \& T. Kleine(2017)Fischer-Gödde \& Kleine}]{fischer-godde_ruthenium_2017}
Fischer-Gödde, M., \& Kleine, T. 2017, \bibinfo{title}{Ruthenium isotopic evidence for an inner {Solar} {System} origin of the late veneer,} Nature, 541, 525, \dodoi{10.1038/nature21045}

\bibitem[{E. Gaidos {et~al.}(2023)Gaidos, Claytor, Dungee, Ali, \& Feiden}]{Gaidos2023}
Gaidos, E., Claytor, Z., Dungee, R., Ali, A., \& Feiden, G.~A. 2023, \bibinfo{title}{The TIME Table: rotation and ages of cool exoplanet host stars,} Monthly Notices of the Royal Astronomical Society, 520, 5283, \dodoi{10.1093/mnras/stad343}

\bibitem[{P. Gao \& D. Powell(2021)Gao \& Powell}]{gao_universal_2021}
Gao, P., \& Powell, D. 2021, \bibinfo{title}{A {Universal} {Cloud} {Composition} on the {Nightsides} of {Hot} {Jupiters},} The Astrophysical Journal Letters, 918, L7, \dodoi{10.3847/2041-8213/ac139f}

\bibitem[{P. Gao {et~al.}(2020)Gao, Thorngren, Lee, Fortney, Morley, Wakeford, Powell, Stevenson, \& Zhang}]{gao_aerosol_2020}
Gao, P., Thorngren, D.~P., Lee, E. K.~H., {et~al.} 2020, \bibinfo{title}{Aerosol composition of hot giant exoplanets dominated by silicates and hydrocarbon hazes,} Nature Astronomy, 4, 951, \dodoi{10.1038/s41550-020-1114-3}

\bibitem[{A. García~Muñoz(2023)García~Muñoz}]{garcia_munoz_heating_2023}
García~Muñoz, A. 2023, \bibinfo{title}{Heating and ionization by non-thermal electrons in the upper atmospheres of water-rich exoplanets,} Astronomy \& Astrophysics, 672, A77, \dodoi{10.1051/0004-6361/202245766}

\bibitem[{A. García~Muñoz {et~al.}(2024)García~Muñoz, Asensio~Ramos, \& Faure}]{garcia_munoz_nlte_2024}
García~Muñoz, A., Asensio~Ramos, A., \& Faure, A. 2024, \bibinfo{title}{{NLTE} modelling of water-rich exoplanet atmospheres. {Cooling} and heating rates,} Icarus, 415, 116080, \dodoi{10.1016/j.icarus.2024.116080}

\bibitem[{A. García~Muñoz \& E. Bataille(2024)García~Muñoz \& Bataille}]{garcia_munoz_heating_2024}
García~Muñoz, A., \& Bataille, E. 2024, \bibinfo{title}{Heating, {Excitation}, {Dissociation}, and {Ionization} of {Molecules} by {High}-{Energy} {Photons} in {Planetary} {Atmospheres},} ACS Earth and Space Chemistry, 8, 2652, \dodoi{10.1021/acsearthspacechem.4c00273}

\bibitem[{A. García~Muñoz {et~al.}(2021)García~Muñoz, Fossati, Youngblood, Nettelmann, Gandolfi, Cabrera, \& Rauer}]{munoz_heavy_2021}
García~Muñoz, A., Fossati, L., Youngblood, A., {et~al.} 2021, \bibinfo{title}{A {Heavy} {Molecular} {Weight} {Atmosphere} for the {Super}-{Earth} π {Men} c,} The Astrophysical Journal Letters, 907, L36, \dodoi{10.3847/2041-8213/abd9b8}

\bibitem[{J.~D. Garrett(2021)Garrett}]{SciencePlots}
Garrett, J.~D. 2021, \bibinfo{title}{{garrettj403/SciencePlots},} \dodoi{10.5281/zenodo.4106649}

\bibitem[{T.~P. Greene {et~al.}(2023)Greene, Bell, Ducrot, Dyrek, Lagage, \& Fortney}]{greene_thermal_2023}
Greene, T.~P., Bell, T.~J., Ducrot, E., {et~al.} 2023, \bibinfo{title}{Thermal emission from the {Earth}-sized exoplanet {TRAPPIST}-1 b using {JWST},} Nature, 618, 39, \dodoi{10.1038/s41586-023-05951-7}

\bibitem[{G. Gronoff {et~al.}(2020{\natexlab{a}})Gronoff, Arras, Baraka, Bell, Cessateur, Cohen, Curry, Drake, Elrod, Erwin, Garcia‐Sage, Garraffo, Glocer, Heavens, Lovato, Maggiolo, Parkinson, Simon~Wedlund, Weimer, \& Moore}]{gronoff_atmospheric_2020}
Gronoff, G., Arras, P., Baraka, S., {et~al.} 2020{\natexlab{a}}, \bibinfo{title}{Atmospheric {Escape} {Processes} and {Planetary} {Atmospheric} {Evolution},} Journal of Geophysical Research: Space Physics, 125, e2019JA027639, \dodoi{10.1029/2019JA027639}

\bibitem[{G. Gronoff {et~al.}(2020{\natexlab{b}})Gronoff, Arras, Baraka, Bell, Cessateur, Cohen, Curry, Drake, Elrod, Erwin, Garcia-Sage, Garraffo, Glocer, Heavens, Lovato, Maggiolo, Parkinson, Simon~Wedlund, Weimer, \& Moore}]{Gronoff2020}
Gronoff, G., Arras, P., Baraka, S., {et~al.} 2020{\natexlab{b}}, \bibinfo{title}{Atmospheric {Escape} {Processes} and {Planetary} {Atmospheric} {Evolution},} Journal of Geophysical Research: Space Physics, 125, e2019JA027639, \dodoi{10.1029/2019JA027639}

\bibitem[{J.~T. Gu {et~al.}(2024)Gu, Peng, Ji, Zhang, Yang, Hoyos, Hirschmann, Kite, \& Fischer}]{gu_composition_2024}
Gu, J.~T., Peng, B., Ji, X., {et~al.} 2024, \bibinfo{title}{Composition of {Earth}'s initial atmosphere and fate of accreted volatiles set by core formation and magma ocean redox evolution,} Earth and Planetary Science Letters, 629, 118618, \dodoi{10.1016/j.epsl.2024.118618}

\bibitem[{E.~F. Guinan {et~al.}(2016)Guinan, Engle, \& Durbin}]{guinan_living_2016}
Guinan, E.~F., Engle, S.~G., \& Durbin, A. 2016, \bibinfo{title}{Living with a Red Dwarf: Rotation and X-Ray and Ultraviolet Properties of the Halo Population Kapteyn's Star,} The Astrophysical Journal, 821, 81, \dodoi{10.3847/0004-637X/821/2/81}

\bibitem[{ {Gunell, Herbert} {et~al.}(2018){Gunell, Herbert}, {Maggiolo, Romain}, {Nilsson, Hans}, {Stenberg Wieser, Gabriella}, {Slapak, Rikard}, {Lindkvist, Jesper}, {Hamrin, Maria}, \& {De Keyser, Johan}}]{Gunell2018}
{Gunell, Herbert}, {Maggiolo, Romain}, {Nilsson, Hans}, {et~al.} 2018, \bibinfo{title}{Why an intrinsic magnetic field does not protect a planet against atmospheric escape,} A\&A, 614, L3, \dodoi{10.1051/0004-6361/201832934}

\bibitem[{A.~N. Halliday(2013)Halliday}]{halliday_origins_2013}
Halliday, A.~N. 2013, \bibinfo{title}{The origins of volatiles in the terrestrial planets,} Geochimica et Cosmochimica Acta, 105, 146, \dodoi{10.1016/j.gca.2012.11.015}

\bibitem[{C.~R. Harris {et~al.}(2020)Harris, Millman, van~der Walt, Gommers, Virtanen, Cournapeau, Wieser, Taylor, Berg, Smith, Kern, Picus, Hoyer, van Kerkwijk, Brett, Haldane, del R{\'{i}}o, Wiebe, Peterson, G{\'{e}}rard-Marchant, Sheppard, Reddy, Weckesser, Abbasi, Gohlke, \& Oliphant}]{harris2020array}
Harris, C.~R., Millman, K.~J., van~der Walt, S.~J., {et~al.} 2020, \bibinfo{title}{Array programming with {NumPy},} Nature, 585, 357, \dodoi{10.1038/s41586-020-2649-2}

\bibitem[{Y. Hasegawa \& M.~R. Swain(2024)Hasegawa \& Swain}]{hasegawa_bulk_2024}
Hasegawa, Y., \& Swain, M.~R. 2024, \bibinfo{title}{Bulk and {Atmospheric} {Metallicities} as {Direct} {Probes} of {Sequentially} {Varying} {Accretion} {Mechanisms} of {Gas} and {Solids} {Onto} {Planets},} The Astrophysical Journal Letters, 973, L46, \dodoi{10.3847/2041-8213/ad7957}

\bibitem[{S. Hier‐Majumder \& M.~M. Hirschmann(2017)Hier‐Majumder \& Hirschmann}]{hiermajumder_origin_2017}
Hier‐Majumder, S., \& Hirschmann, M.~M. 2017, \bibinfo{title}{The origin of volatiles in the Earth's mantle,} Geochemistry, Geophysics, Geosystems, 18, 3078, \dodoi{10.1002/2017GC006937}

\bibitem[{M.~M. Hirschmann {et~al.}(2021)Hirschmann, Bergin, Blake, Ciesla, \& Li}]{hirschmann_early_2021}
Hirschmann, M.~M., Bergin, E.~A., Blake, G.~A., Ciesla, F.~J., \& Li, J. 2021, \bibinfo{title}{Early volatile depletion on planetesimals inferred from {C}–{S} systematics of iron meteorite parent bodies,} Proceedings of the National Academy of Sciences, 118, e2026779118, \dodoi{10.1073/pnas.2026779118}

\bibitem[{C.~S.~K. Ho \& V. Van Eylen(2023)Ho \& Van Eylen}]{ho_deep_2023}
Ho, C. S.~K., \& Van Eylen, V. 2023, \bibinfo{title}{A deep radius valley revealed by \textit{{Kepler}} short cadence observations,} Monthly Notices of the Royal Astronomical Society, 519, 4056, \dodoi{10.1093/mnras/stac3802}

\bibitem[{B.~J. Hord {et~al.}(2024)Hord, Kempton, Evans-Soma, Latham, Ciardi, Dragomir, Colón, Ross, Vanderburg, De~Beurs, Collins, Watkins, Bean, Cowan, Daylan, Morley, Ih, Baker, Barkaoui, Batalha, Behmard, Belinski, Benkhaldoun, Benni, Bernacki, Bieryla, Binnenfeld, Bosch-Cabot, Bouchy, Bozza, Brahm, Buchhave, Calkins, Chontos, Clark, Cloutier, Cointepas, Collins, Conti, Crossfield, Dai, De~Leon, Dransfield, Dressing, Dustor, Esquerdo, Evans, Fajardo-Acosta, Fiołka, Forés-Toribio, Frasca, Fukui, Fulton, Furlan, Gan, Gandolfi, Ghachoui, Giacalone, Gilbert, Gillon, Girardin, Gonzales, Grau~Horta, Gregorio, Greklek-McKeon, Guerra, Hartman, Hellier, Helm, Hełminiak, Henning, Hill, Horne, Howard, Howell, Huber, Isopi, Jehin, Jenkins, Jensen, Johnson, Jordán, Kane, Kielkopf, Krushinsky, Lasota, Lee, Lewin, Livingston, Lubin, Lund, Mallia, Mann, Marino, Maslennikova, Massey, Matson, Matthews, Mayo, Mazeh, McLeod, Michaels, Močnik, Mori, Mraz, Muñoz, Narita, Natarajan, Dyregaard~Nielsen, Osborn, Palle,
  Panahi, Papini, Plavchan, Polanski, Popowicz, Pozuelos, Quinn, Radford, Reed, Relles, Rice, Robertson, Rodriguez, Rosenthal, Rubenzahl, Schanche, Schlieder, Schwarz, Sefako, Shporer, Sozzetti, Srdoc, Stockdale, Tarasenkov, Tan, Timmermans, Ting, Van~Zandt, Vignes, Waite, Watanabe, Weiss, Wittrock, Zhou, Ziegler, \& Zucker}]{hord_identification_2024}
Hord, B.~J., Kempton, E. M.-R., Evans-Soma, T.~M., {et~al.} 2024, \bibinfo{title}{Identification of the {Top} {TESS} {Objects} of {Interest} for {Atmospheric} {Characterization} of {Transiting} {Exoplanets} with {JWST},} The Astronomical Journal, 167, 233, \dodoi{10.3847/1538-3881/ad3068}

\bibitem[{R. Hu {et~al.}(2024)Hu, Bello-Arufe, Zhang, Paragas, Zilinskas, Van~Buchem, Bess, Patel, Ito, Damiano, Scheucher, Oza, Knutson, Miguel, Dragomir, Brandeker, \& Demory}]{hu_secondary_2024}
Hu, R., Bello-Arufe, A., Zhang, M., {et~al.} 2024, \bibinfo{title}{A secondary atmosphere on the rocky exoplanet 55 {Cancri} e,} Nature, 630, 609, \dodoi{10.1038/s41586-024-07432-x}

\bibitem[{J.~D. Hunter(2007)Hunter}]{Hunter:2007}
Hunter, J.~D. 2007, \bibinfo{title}{Matplotlib: A 2D graphics environment,} Computing in Science \& Engineering, 9, 90, \dodoi{10.1109/MCSE.2007.55}

\bibitem[{Y. Ito \& M. Ikoma(2021)Ito \& Ikoma}]{ito_hydrodynamic_2021}
Ito, Y., \& Ikoma, M. 2021, \bibinfo{title}{Hydrodynamic escape of mineral atmosphere from hot rocky exoplanet. {I}. {Model} description,} Monthly Notices of the Royal Astronomical Society, 502, 750, \dodoi{10.1093/mnras/staa3962}

\bibitem[{A.~P. Jackson {et~al.}(2012)Jackson, Davis, \& Wheatley}]{jackson_coronal_2012}
Jackson, A.~P., Davis, T.~A., \& Wheatley, P.~J. 2012, \bibinfo{title}{The coronal {X}-ray-age relation and its implications for the evaporation of exoplanets: {X}-ray-age relation and exoplanet evaporation,} Monthly Notices of the Royal Astronomical Society, 422, 2024, \dodoi{10.1111/j.1365-2966.2012.20657.x}

\bibitem[{R.~D. {Jeffries} {et~al.}(2021){Jeffries}, {Jackson}, {Sun}, \& {Deliyannis}}]{2021MNRAS.500.1158J}
{Jeffries}, R.~D., {Jackson}, R.~J., {Sun}, Q., \& {Deliyannis}, C.~P. 2021, \bibinfo{title}{{The effects of rotation on the lithium depletion of G- and K-dwarfs in Messier 35},} \mnras, 500, 1158, \dodoi{10.1093/mnras/staa3141}

\bibitem[{A. Johansen {et~al.}(2023)Johansen, Ronnet, Schiller, Deng, \& Bizzarro}]{johansen_anatomy_2023}
Johansen, A., Ronnet, T., Schiller, M., Deng, Z., \& Bizzarro, M. 2023, \bibinfo{title}{Anatomy of rocky planets formed by rapid pebble accretion: {III}. {Partitioning} of volatiles between planetary core, mantle, and atmosphere,} Astronomy \& Astrophysics, 671, A76, \dodoi{10.1051/0004-6361/202142143}

\bibitem[{B.~W. Johnson \& C. Goldblatt(2018)Johnson \& Goldblatt}]{johnson_earthn_2018}
Johnson, B.~W., \& Goldblatt, C. 2018, \bibinfo{title}{{EarthN}: {A} {New} {Earth} {System} {Nitrogen} {Model},} Geochemistry, Geophysics, Geosystems, 19, 2516, \dodoi{10.1029/2017gc007392}

\bibitem[{R.~E. {Johnson} {et~al.}(2013){Johnson}, {Volkov}, \& {Erwin}}]{Johnson2013}
{Johnson}, R.~E., {Volkov}, A.~N., \& {Erwin}, J.~T. 2013, \bibinfo{title}{{Molecular-kinetic Simulations of Escape from the Ex-planet and Exoplanets: Criterion for Transonic Flow},} \apjl, 768, L4, \dodoi{10.1088/2041-8205/768/1/L4}

\bibitem[{C.~P. Johnstone(2020)Johnstone}]{johnstone_hydrodynamic_2020}
Johnstone, C.~P. 2020, \bibinfo{title}{Hydrodynamic {Escape} of {Water} {Vapor} {Atmospheres} near {Very} {Active} {Stars},} The Astrophysical Journal, 890, 79, \dodoi{10.3847/1538-4357/ab6224}

\bibitem[{C.~P. Johnstone {et~al.}(2021)Johnstone, Bartel, \& Güdel}]{johnstone_active_2021}
Johnstone, C.~P., Bartel, M., \& Güdel, M. 2021, \bibinfo{title}{The active lives of stars: a complete description of rotation and {XUV} evolution of {F}, {G}, {K}, and {M} dwarfs,} Astronomy \& Astrophysics, 649, A96, \dodoi{10.1051/0004-6361/202038407}

\bibitem[{C.~P. Johnstone {et~al.}(2018)Johnstone, Güdel, Lammer, \& Kislyakova}]{johnstone_upper_2018}
Johnstone, C.~P., Güdel, M., Lammer, H., \& Kislyakova, K.~G. 2018, \bibinfo{title}{Upper atmospheres of terrestrial planets: {Carbon} dioxide cooling and the {Earth}’s thermospheric evolution,} Astronomy \& Astrophysics, 617, A107, \dodoi{10.1051/0004-6361/201832776}

\bibitem[{C.~P. Johnstone {et~al.}(2019)Johnstone, Khodachenko, Lüftinger, Kislyakova, Lammer, \& Güdel}]{johnstone_extreme_2019}
Johnstone, C.~P., Khodachenko, M.~L., Lüftinger, T., {et~al.} 2019, \bibinfo{title}{Extreme hydrodynamic losses of {Earth}-like atmospheres in the habitable zones of very active stars,} Astronomy \& Astrophysics, 624, L10, \dodoi{10.1051/0004-6361/201935279}

\bibitem[{D. Jontof-Hutter(2019)Jontof-Hutter}]{jontof-hutter_compositional_2019}
Jontof-Hutter, D. 2019, \bibinfo{title}{The {Compositional} {Diversity} of {Low}-{Mass} {Exoplanets},} Annual Review of Earth and Planetary Sciences, 47, 141, \dodoi{10.1146/annurev-earth-053018-060352}

\bibitem[{J.~A. Kegerreis {et~al.}(2020)Kegerreis, Eke, Catling, Massey, Teodoro, \& Zahnle}]{kegerreis_atmospheric_2020}
Kegerreis, J.~A., Eke, V.~R., Catling, D.~C., {et~al.} 2020, \bibinfo{title}{Atmospheric {Erosion} by {Giant} {Impacts} onto {Terrestrial} {Planets}: {A} {Scaling} {Law} for any {Speed}, {Angle}, {Mass}, and {Density},} The Astrophysical Journal, 901, L31, \dodoi{10.3847/2041-8213/abb5fb}

\bibitem[{E.~M.~R. Kempton {et~al.}(2018)Kempton, Bean, Louie, Deming, Koll, Mansfield, Christiansen, López-Morales, Swain, Zellem, Ballard, Barclay, Barstow, Batalha, Beatty, Berta-Thompson, Birkby, Buchhave, Charbonneau, Cowan, Crossfield, de~Val-Borro, Doyon, Dragomir, Gaidos, Heng, Hu, Kane, Kreidberg, Mallonn, Morley, Narita, Nascimbeni, Pallé, Quintana, Rauscher, Seager, Shkolnik, Sing, Sozzetti, Stassun, Valenti, \& von Essen}]{kempton_framework_2018}
Kempton, E. M.~R., Bean, J.~L., Louie, D.~R., {et~al.} 2018, \bibinfo{title}{A {Framework} for {Prioritizing} the {TESS} {Planetary} {Candidates} {Most} {Amenable} to {Atmospheric} {Characterization},} Publications of the Astronomical Society of the Pacific, 130, 114401, \dodoi{10.1088/1538-3873/aadf6f}

\bibitem[{G.~W. King \& P.~J. Wheatley(2020)King \& Wheatley}]{king_euv_2020}
King, G.~W., \& Wheatley, P.~J. 2020, \bibinfo{title}{{EUV} irradiation of exoplanet atmospheres occurs on {Gyr} time-scales,} Monthly Notices of the Royal Astronomical Society: Letters, 501, L28, \dodoi{10.1093/mnrasl/slaa186}

\bibitem[{G.~W. King {et~al.}(2018)King, Wheatley, Salz, Bourrier, Czesla, Ehrenreich, Kirk, Lecavelier Des~Etangs, Louden, Schmitt, \& Schneider}]{king_xuv_2018}
King, G.~W., Wheatley, P.~J., Salz, M., {et~al.} 2018, \bibinfo{title}{The {XUV} environments of exoplanets from {Jupiter}-size to super-{Earth},} Monthly Notices of the Royal Astronomical Society, \dodoi{10.1093/mnras/sty1110}

\bibitem[{E.~S. Kite \& M.~N. Barnett(2020)Kite \& Barnett}]{kite_exoplanet_2020}
Kite, E.~S., \& Barnett, M.~N. 2020, \bibinfo{title}{Exoplanet secondary atmosphere loss and revival,} Proceedings of the National Academy of Sciences, 117, 18264, \dodoi{10.1073/pnas.2006177117}

\bibitem[{E.~S. Kite {et~al.}(2016)Kite, Fegley~Jr, Schaefer, \& Gaidos}]{kite2016atmosphere}
Kite, E.~S., Fegley~Jr, B., Schaefer, L., \& Gaidos, E. 2016, \bibinfo{title}{Atmosphere-interior exchange on hot, rocky exoplanets,} The Astrophysical Journal, 828, 80

\bibitem[{E.~S. Kite {et~al.}(2009)Kite, Manga, \& Gaidos}]{kite_geodynamics_2009}
Kite, E.~S., Manga, M., \& Gaidos, E. 2009, \bibinfo{title}{Geodynamics and Rate of Volcanism on Massive Earth-like Planets,} The Astrophysical Journal, 700, 1732, \dodoi{10.1088/0004-637X/700/2/1732}

\bibitem[{E.~S. Kite \& L. Schaefer(2021)Kite \& Schaefer}]{kite_water_2021}
Kite, E.~S., \& Schaefer, L. 2021, \bibinfo{title}{Water on {Hot} {Rocky} {Exoplanets},} The Astrophysical Journal Letters, 909, L22, \dodoi{10.3847/2041-8213/abe7dc}

\bibitem[{R.~k. Kopparapu {et~al.}(2013)Kopparapu, Ramirez, Kasting, Eymet, Robinson, Mahadevan, Terrien, Domagal-Goldman, Meadows, \& Deshpande}]{kopparapu_habitable_2013}
Kopparapu, R.~k., Ramirez, R., Kasting, J.~F., {et~al.} 2013, \bibinfo{title}{Habitable {Zones} {Around} {Main}-{Sequence} {Stars}: {New} {Estimates},} The Astrophysical Journal, 765, 131, \dodoi{10.1088/0004-637X/765/2/131}

\bibitem[{Q. Kral {et~al.}(2018)Kral, Wyatt, Triaud, Marino, Thébault, \& Shorttle}]{kral_cometary_2018}
Kral, Q., Wyatt, M.~C., Triaud, A. H. M.~J., {et~al.} 2018, \bibinfo{title}{Cometary impactors on the {TRAPPIST}-1 planets can destroy all planetary atmospheres and rebuild secondary atmospheres on planets f, g, and h,} Monthly Notices of the Royal Astronomical Society, 479, 2649, \dodoi{10.1093/mnras/sty1677}

\bibitem[{V.~A. Krasnopolsky {et~al.}(2004)Krasnopolsky, Maillard, \& Owen}]{krasnopolsky_detection_2004}
Krasnopolsky, V.~A., Maillard, J.~P., \& Owen, T.~C. 2004, \bibinfo{title}{Detection of methane in the martian atmosphere: evidence for life?} Icarus, 172, 537, \dodoi{10.1016/j.icarus.2004.07.004}

\bibitem[{L. Kreidberg {et~al.}(2019)Kreidberg, Koll, Morley, Hu, Schaefer, Deming, Stevenson, Dittmann, Vanderburg, Berardo, Guo, Stassun, Crossfield, Charbonneau, Latham, Loeb, Ricker, Seager, \& Vanderspek}]{kreidberg_absence_2019}
Kreidberg, L., Koll, D. D.~B., Morley, C., {et~al.} 2019, \bibinfo{title}{Absence of a thick atmosphere on the terrestrial exoplanet {LHS} 3844b,} Nature, 573, 87, \dodoi{10.1038/s41586-019-1497-4}

\bibitem[{J. Krissansen-Totton(2023)Krissansen-Totton}]{krissansen-totton_implications_2023}
Krissansen-Totton, J. 2023, \bibinfo{title}{Implications of {Atmospheric} {Nondetections} for {Trappist}-1 {Inner} {Planets} on {Atmospheric} {Retention} {Prospects} for {Outer} {Planets},} The Astrophysical Journal Letters, 951, L39, \dodoi{10.3847/2041-8213/acdc26}

\bibitem[{J. Krissansen-Totton \& J.~J. Fortney(2022)Krissansen-Totton \& Fortney}]{krissansen-totton_predictions_2022}
Krissansen-Totton, J., \& Fortney, J.~J. 2022, \bibinfo{title}{Predictions for {Observable} {Atmospheres} of {Trappist}-1 {Planets} from a {Fully} {Coupled} {Atmosphere}-{Interior} {Evolution} {Model},} The Astrophysical Journal, 933, 115, \dodoi{10.3847/1538-4357/ac69cb}

\bibitem[{J. Krissansen-Totton {et~al.}(2024)Krissansen-Totton, Wogan, Thompson, \& Fortney}]{krissansen-totton_erosion_2024}
Krissansen-Totton, J., Wogan, N., Thompson, M., \& Fortney, J.~J. 2024, \bibinfo{title}{The erosion of large primary atmospheres typically leaves behind substantial secondary atmospheres on temperate rocky planets,} Nature Communications, 15, 8374, \dodoi{10.1038/s41467-024-52642-6}

\bibitem[{S. {Labrosse} {et~al.}(2007){Labrosse}, {Hernlund}, \& {Coltice}}]{Labrosse2007}
{Labrosse}, S., {Hernlund}, J.~W., \& {Coltice}, N. 2007, \bibinfo{title}{{A crystallizing dense magma ocean at the base of the Earth's mantle},} \nat, 450, 866, \dodoi{10.1038/nature06355}

\bibitem[{W.~G. Levine {et~al.}(2024)Levine, Vissapragada, Feinstein, King, Hernandez, Corrales, Greklek-McKeon, \& Knutson}]{levine_exoplanet_2024}
Levine, W.~G., Vissapragada, S., Feinstein, A.~D., {et~al.} 2024, \bibinfo{title}{Exoplanet {Aeronomy}: {A} {Case} {Study} of {WASP}-69 b’s {Variable} {Thermosphere},} The Astronomical Journal, 168, 65, \dodoi{10.3847/1538-3881/ad5354}

\bibitem[{T. Lichtenberg {et~al.}(2021{\natexlab{a}})Lichtenberg, Bower, Hammond, Boukrouche, Sanan, Tsai, \& Pierrehumbert}]{lichtenberg_vertically_2021}
Lichtenberg, T., Bower, D.~J., Hammond, M., {et~al.} 2021{\natexlab{a}}, \bibinfo{title}{Vertically Resolved Magma Ocean-Protoatmosphere Evolution: H$_{2}$, H$_{2}$O, CO$_{2}$, CH$_{4}$, CO, O$_{2}$, and N$_{2}$ as Primary Absorbers,} Journal of Geophysical Research: Planets, 126, e2020JE006711, \dodoi{10.1029/2020JE006711}

\bibitem[{T. Lichtenberg {et~al.}(2021{\natexlab{b}})Lichtenberg, Dra̧żkowska, Schönbächler, Golabek, \& Hands}]{lichtenberg_bifurcation_2021}
Lichtenberg, T., Dra̧żkowska, J., Schönbächler, M., Golabek, G.~J., \& Hands, T.~O. 2021{\natexlab{b}}, \bibinfo{title}{Bifurcation of planetary building blocks during {Solar} {System} formation,} Science, 371, 365, \dodoi{10.1126/science.abb3091}

\bibitem[{T. Lichtenberg {et~al.}(2019)Lichtenberg, Golabek, Burn, Meyer, Alibert, Gerya, \& Mordasini}]{lichtenberg_water_2019}
Lichtenberg, T., Golabek, G.~J., Burn, R., {et~al.} 2019, \bibinfo{title}{A water budget dichotomy of rocky protoplanets from {26Al}-heating,} Nature Astronomy, 3, 307, \dodoi{10.1038/s41550-018-0688-5}

\bibitem[{A.~P. Lincowski {et~al.}(2023)Lincowski, Meadows, Zieba, Kreidberg, Morley, Gillon, Selsis, Agol, Bolmont, Ducrot, Hu, Koll, Lyu, Mandell, Suissa, \& Tamburo}]{lincowski_potential_2023}
Lincowski, A.~P., Meadows, V.~S., Zieba, S., {et~al.} 2023, \bibinfo{title}{Potential {Atmospheric} {Compositions} of {TRAPPIST}-1 c {Constrained} by {JWST}/{MIRI} {Observations} at 15 μm,} The Astrophysical Journal Letters, 955, L7, \dodoi{10.3847/2041-8213/acee02}

\bibitem[{B. Liu {et~al.}(2022)Liu, Johansen, Lambrechts, Bizzarro, \& Haugbølle}]{liu_natural_2022}
Liu, B., Johansen, A., Lambrechts, M., Bizzarro, M., \& Haugbølle, T. 2022, \bibinfo{title}{Natural separation of two primordial planetary reservoirs in an expanding solar protoplanetary disk,} Science Advances, 8, eabm3045, \dodoi{10.1126/sciadv.abm3045}

\bibitem[{L. Liu \& F. Tian(2018)Liu \& Tian}]{Liu_Tian_2018}
Liu, L., \& Tian, F. 2018, \bibinfo{title}{Efficient metal emissions in the upper atmospheres of close-in exoplanets,} Earth and Planetary Physics, 2, 22, \dodoi{10.26464/epp2018003}

\bibitem[{S.~J. Lock \& S.~T. Stewart(2017)Lock \& Stewart}]{lock_structure_2017}
Lock, S.~J., \& Stewart, S.~T. 2017, \bibinfo{title}{The structure of terrestrial bodies: {Impact} heating, corotation limits, and synestias,} Journal of Geophysical Research: Planets, 122, 950, \dodoi{10.1002/2016JE005239}

\bibitem[{G.~V. Looveren {et~al.}(2025)Looveren, Saikia, Herbort, Schleich, Güdel, Johnstone, \& Kislyakova}]{looveren_habitable_2025}
Looveren, G.~V., Saikia, S.~B., Herbort, O., {et~al.} 2025, \bibinfo{title}{Habitable {Zone} and {Atmosphere} {Retention} {Distance} ({HaZARD}) - {Stellar}-evolution-dependent loss models of secondary atmospheres,} Astronomy \& Astrophysics, 694, A310, \dodoi{10.1051/0004-6361/202452998}

\bibitem[{E.~D. Lopez \& J.~J. Fortney(2014)Lopez \& Fortney}]{lopez_understanding_2014}
Lopez, E.~D., \& Fortney, J.~J. 2014, \bibinfo{title}{Understanding the Mass-Radius Relation for Sub-Neptunes: Radius as a Proxy for Composition,} The Astrophysical Journal, 792, 1, \dodoi{10.1088/0004-637X/792/1/1}

\bibitem[{E.~D. Lopez {et~al.}(2012)Lopez, Fortney, \& Miller}]{lopez_how_2012}
Lopez, E.~D., Fortney, J.~J., \& Miller, N. 2012, \bibinfo{title}{How Thermal Evolution and Mass-loss Sculpt Populations of Super-Earths and Sub-Neptunes: Application to the Kepler-11 System and Beyond,} The Astrophysical Journal, 761, 59, \dodoi{10.1088/0004-637X/761/1/59}

\bibitem[{R. Luger \& R. Barnes(2015)Luger \& Barnes}]{luger_extreme_2015}
Luger, R., \& Barnes, R. 2015, \bibinfo{title}{Extreme {Water} {Loss} and {Abiotic} {O}$_{\textrm{2}}$ {Buildup} on {Planets} {Throughout} the {Habitable} {Zones} of {M} {Dwarfs},} Astrobiology, 15, 119, \dodoi{10.1089/ast.2014.1231}

\bibitem[{H. Luo {et~al.}(2024)Luo, O’Rourke, \& Deng}]{Luo2024}
Luo, H., O’Rourke, J.~G., \& Deng, J. 2024, \bibinfo{title}{Radiogenic heating sustains long-lived volcanism and magnetic dynamos in super-Earths,} Science Advances, 10, eado7603, \dodoi{10.1126/sciadv.ado7603}

\bibitem[{R. Luque \& E. Pallé(2022)Luque \& Pallé}]{luque_density_2022}
Luque, R., \& Pallé, E. 2022, \bibinfo{title}{Density, not radius, separates rocky and water-rich small planets orbiting {M} dwarf stars,} Science, 377, 1211, \dodoi{10.1126/science.abl7164}

\bibitem[{R. Luque {et~al.}(2024)Luque, Coy, Xue, Feinstein, Ahrer, Changeat, Zhang, Moran, Bean, Kite, {et~al.}}]{luque2024dark}
Luque, R., Coy, B.~P., Xue, Q., {et~al.} 2024, \bibinfo{title}{A dark, bare rock for TOI-1685 b from a JWST NIRSpec G395H phase curve,} arXiv preprint arXiv:2412.03411

\bibitem[{U. Malamud {et~al.}(2024)Malamud, Podolak, Podolak, \& Bodenheimer}]{malamud_uranus_2024}
Malamud, U., Podolak, M., Podolak, J.~I., \& Bodenheimer, P.~H. 2024, \bibinfo{title}{Uranus and {Neptune} as methane planets: {Producing} icy giants from refractory planetesimals,} Icarus, 421, 116217, \dodoi{10.1016/j.icarus.2024.116217}

\bibitem[{M.~W. Mansfield {et~al.}(2024)Mansfield, Xue, Zhang, Mahajan, Ih, Koll, Bean, Coy, Eastman, Kempton, \& Kite}]{mansfield_no_2024}
Mansfield, M.~W., Xue, Q., Zhang, M., {et~al.} 2024, \bibinfo{title}{No {Thick} {Atmosphere} on the {Terrestrial} {Exoplanet} {Gl} 486b,} arXiv, \dodoi{10.48550/ARXIV.2408.15123}

\bibitem[{B. Marty \& L. Zimmermann(1999)Marty \& Zimmermann}]{marty_volatiles_1999}
Marty, B., \& Zimmermann, L. 1999, \bibinfo{title}{Volatiles ({He}, {C}, {N}, {Ar}) in mid-ocean ridge basalts: assesment of shallow-level fractionation and characterization of source composition,} Geochimica et Cosmochimica Acta, 63, 3619, \dodoi{10.1016/S0016-7037(99)00169-6}

\bibitem[{C. Monaghan {et~al.}(2025)Monaghan, Roy, Benneke, Crossfield, Coulombe, Piaulet-Ghorayeb, Kreidberg, Dressing, Kane, Dragomir, Werner, Parmentier, Christiansen, Morales, Berardo, \& Gorjian}]{monaghan2025}
Monaghan, C., Roy, P.-A., Benneke, B., {et~al.} 2025, \bibinfo{title}{Low 4.5 μm {Dayside} {Emission} {Disfavors} a {Dark} {Bare}-{Rock} scenario for the {Hot} {Super}-{Earth} {TOI}-431 b,} arXiv, \dodoi{10.48550/arXiv.2503.09698}

\bibitem[{K. Moore {et~al.}(2024)Moore, David, Zhang, \& Cowan}]{moore_water_2024}
Moore, K., David, B., Zhang, A.~Y., \& Cowan, N.~B. 2024, \bibinfo{title}{Water {Evolution} and {Inventories} of {Super}-{Earths} {Orbiting} {Late} {M} {Dwarfs},} The Astrophysical Journal, 972, 131, \dodoi{10.3847/1538-4357/ad6444}

\bibitem[{R.~A. Murray-Clay {et~al.}(2009)Murray-Clay, Chiang, \& Murray}]{murray-clay_atmospheric_2009}
Murray-Clay, R.~A., Chiang, E.~I., \& Murray, N. 2009, \bibinfo{title}{Atmospheric {Escape} {From} {Hot} {Jupiters},} The Astrophysical Journal, 693, 23, \dodoi{10.1088/0004-637X/693/1/23}

\bibitem[{A. Nakayama {et~al.}(2022)Nakayama, Ikoma, \& Terada}]{nakayama_survival_2022}
Nakayama, A., Ikoma, M., \& Terada, N. 2022, \bibinfo{title}{Survival of {Terrestrial} {N2}–{O2} {Atmospheres} in {Violent} {XUV} {Environments} through {Efficient} {Atomic} {Line} {Radiative} {Cooling},} The Astrophysical Journal, 937, 72, \dodoi{10.3847/1538-4357/ac86ca}

\bibitem[{ {NASA Exoplanet Archive}(2019){NASA Exoplanet Archive}}]{nasa_exoplanet_archive_confirmed_2019}
{NASA Exoplanet Archive}. 2019, \bibinfo{title}{Confirmed {Planets} {Table},} IPAC, \dodoi{10.26133/NEA1}

\bibitem[{J.~E. Owen(2019)Owen}]{owen_atmospheric_2019}
Owen, J.~E. 2019, \bibinfo{title}{Atmospheric {Escape} and the {Evolution} of {Close}-{In} {Exoplanets},} Annual Review of Earth and Planetary Sciences, 47, 67, \dodoi{10.1146/annurev-earth-053018-060246}

\bibitem[{J.~E. Owen \& F.~C. Adams(2014)Owen \& Adams}]{Owen2014}
Owen, J.~E., \& Adams, F.~C. 2014, \bibinfo{title}{Magnetically controlled mass-loss from extrasolar planets in close orbits,} Monthly Notices of the Royal Astronomical Society, 444, 3761, \dodoi{10.1093/mnras/stu1684}

\bibitem[{J.~E. Owen \& H.~E. Schlichting(2024)Owen \& Schlichting}]{owen_mapping_2024}
Owen, J.~E., \& Schlichting, H.~E. 2024, \bibinfo{title}{Mapping out the parameter space for photoevaporation and core-powered mass-loss,} Monthly Notices of the Royal Astronomical Society, 528, 1615, \dodoi{10.1093/mnras/stad3972}

\bibitem[{J.~E. Owen \& Y. Wu(2016)Owen \& Wu}]{owen_atmospheres_2016}
Owen, J.~E., \& Wu, Y. 2016, \bibinfo{title}{Atmospheres of low-mass planets: the "boil-off",} The Astrophysical Journal, 817, 107, \dodoi{10.3847/0004-637X/817/2/107}

\bibitem[{T. pandas~development team(2020)pandas~development team}]{reback2020pandas}
pandas~development team, T. 2020, \bibinfo{title}{pandas-dev/pandas: Pandas,}, latest Zenodo, \dodoi{10.5281/zenodo.3509134}

\bibitem[{K. Paragas {et~al.}(2025)Paragas, Knutson, Hu, Ehlmann, Alemanno, Helbert, Maturilli, Zhang, Iyer, \& Rossman}]{paragas25}
Paragas, K., Knutson, H.~A., Hu, R., {et~al.} 2025, \bibinfo{title}{A New Spectral Library for Modeling the Surfaces of Hot, Rocky Exoplanets,} The Astrophysical Journal, 981, 130

\bibitem[{H. Parviainen {et~al.}(2023)Parviainen, Luque, \& Palle}]{parviainen_span_2023}
Parviainen, H., Luque, R., \& Palle, E. 2023, \bibinfo{title}{Spright: a probabilistic mass-density-radius relation for small planets,} Monthly Notices of the Royal Astronomical Society, 527, 5693, \dodoi{10.1093/mnras/stad3504}

\bibitem[{E.~K. Pass {et~al.}(2025)Pass, Charbonneau, \& Vanderburg}]{pass_receding_2025}
Pass, E.~K., Charbonneau, D., \& Vanderburg, A. 2025, \bibinfo{title}{The {Receding} {Cosmic} {Shoreline} of {Mid}-to-{Late} {M} {Dwarfs}: {Measurements} of {Active} {Lifetimes} {Worsen} {Challenges} for {Atmosphere} {Retention} by {Rocky} {Exoplanets},} arXiv, \dodoi{10.48550/arXiv.2504.01182}

\bibitem[{B. Peng \& D. Valencia(2024)Peng \& Valencia}]{peng_puffy_2024}
Peng, B., \& Valencia, D. 2024, \bibinfo{title}{Puffy {Venuses}: {The} {Mass}–{Radius} {Impact} of {Carbon}-rich {Atmospheres} on {Lava} {Worlds},} The Astrophysical Journal, 976, 202, \dodoi{10.3847/1538-4357/ad6f03}

\bibitem[{T. Penz {et~al.}(2008)Penz, Micela, \& Lammer}]{penz_influence_2008}
Penz, T., Micela, G., \& Lammer, H. 2008, \bibinfo{title}{Influence of the evolving stellar {X}-ray luminosity distribution on exoplanetary mass loss,} Astronomy and Astrophysics, 477, 309, \dodoi{10.1051/0004-6361:20078364}

\bibitem[{A.~A.~A. Piette {et~al.}(2023)Piette, Gao, Brugman, Shahar, Lichtenberg, Miozzi, \& Driscoll}]{piette_rocky_2023}
Piette, A. A.~A., Gao, P., Brugman, K., {et~al.} 2023, \bibinfo{title}{Rocky {Planet} or {Water} {World}? {Observability} of {Low}-density {Lava} {World} {Atmospheres},} The Astrophysical Journal, 954, 29, \dodoi{10.3847/1538-4357/acdef2}

\bibitem[{M. Plotnykov \& D. Valencia(2024)Plotnykov \& Valencia}]{plotnykov_observation_2024}
Plotnykov, M., \& Valencia, D. 2024, \bibinfo{title}{Observation uncertainty effects on the precision of interior planetary parameters,} Monthly Notices of the Royal Astronomical Society, 530, 3488, \dodoi{10.1093/mnras/stae993}

\bibitem[{T. Preibisch \& E.~D. Feigelson(2005)Preibisch \& Feigelson}]{preibisch_evolution_2005}
Preibisch, T., \& Feigelson, E.~D. 2005, \bibinfo{title}{The {Evolution} of {X}‐{Ray} {Emission} in {Young} {Stars},} The Astrophysical Journal Supplement Series, 160, 390, \dodoi{10.1086/432094}

\bibitem[{R. Ramstad \& S. Barabash(2021)Ramstad \& Barabash}]{ramstad_intrinsic_2021}
Ramstad, R., \& Barabash, S. 2021, \bibinfo{title}{Do {Intrinsic} {Magnetic} {Fields} {Protect} {Planetary} {Atmospheres} from {Stellar} {Winds}?} Space Science Reviews, 217, 36, \dodoi{10.1007/s11214-021-00791-1}

\bibitem[{S.~N. Raymond {et~al.}(2004)Raymond, Quinn, \& Lunine}]{raymond_making_2004}
Raymond, S.~N., Quinn, T., \& Lunine, J.~I. 2004, \bibinfo{title}{Making other earths: dynamical simulations of terrestrial planet formation and water delivery,} Icarus, 168, 1, \dodoi{10.1016/j.icarus.2003.11.019}

\bibitem[{S. Redfield {et~al.}(2024)Redfield, Batalha, Benneke, Biller, Espinoza, France, Konopacky, Kreidberg, Rauscher, \& Sing}]{redfield_report_2024}
Redfield, S., Batalha, N., Benneke, B., {et~al.} 2024, \bibinfo{title}{Report of the {Working} {Group} on {Strategic} {Exoplanet} {Initiatives} with {HST} and {JWST},} arXiv, \dodoi{10.48550/arXiv.2404.02932}

\bibitem[{T. {Reinhold} {et~al.}(2020){Reinhold}, {Shapiro}, {Solanki}, {Montet}, {Krivova}, {Cameron}, \& {Amazo-G{\'o}mez}}]{2020Sci...368..518R}
{Reinhold}, T., {Shapiro}, A.~I., {Solanki}, S.~K., {et~al.} 2020, \bibinfo{title}{{The Sun is less active than other solar-like stars},} Science, 368, 518, \dodoi{10.1126/science.aay3821}

\bibitem[{B. Reynard \& C. Sotin(2023)Reynard \& Sotin}]{reynard_carbon-rich_2023}
Reynard, B., \& Sotin, C. 2023, \bibinfo{title}{Carbon-rich icy moons and dwarf planets,} Earth and Planetary Science Letters, 612, 118172, \dodoi{10.1016/j.epsl.2023.118172}

\bibitem[{I. Ribas {et~al.}(2005)Ribas, Guinan, Gudel, \& Audard}]{ribas_evolution_2005}
Ribas, I., Guinan, E.~F., Gudel, M., \& Audard, M. 2005, \bibinfo{title}{Evolution of the {Solar} {Activity} over {Time} and {Effects} on {Planetary} {Atmospheres}. {I}. {High}‐{Energy} {Irradiances} (1–1700 {A}),} The Astrophysical Journal, 622, 680, \dodoi{10.1086/427977}

\bibitem[{I. Ribas {et~al.}(2016)Ribas, Bolmont, Selsis, Reiners, Leconte, Raymond, Engle, Guinan, Morin, Turbet, Forget, \& Anglada-Escudé}]{ribas_habitability_2016}
Ribas, I., Bolmont, E., Selsis, F., {et~al.} 2016, \bibinfo{title}{The habitability of {Proxima} {Centauri} b: {I}. {Irradiation}, rotation and volatile inventory from formation to the present,} Astronomy \& Astrophysics, 596, A111, \dodoi{10.1051/0004-6361/201629576}

\bibitem[{D.~R. Rice {et~al.}(2025)Rice, Huang, Steffen, \& Vazan}]{rice_uncertainties_2025}
Rice, D.~R., Huang, C., Steffen, J.~H., \& Vazan, A. 2025, \bibinfo{title}{Uncertainties in the {Inference} of {Internal} {Structure}: {The} {Case} of {TRAPPIST}-1 f,} The Astrophysical Journal, 986, 2, \dodoi{10.3847/1538-4357/add34b}

\bibitem[{K. Righter \& D.~P. O’Brien(2011)Righter \& O’Brien}]{righter_terrestrial_2011}
Righter, K., \& O’Brien, D.~P. 2011, \bibinfo{title}{Terrestrial planet formation,} Proceedings of the National Academy of Sciences, 108, 19165, \dodoi{10.1073/pnas.1013480108}

\bibitem[{L.~A. Rogers(2015)Rogers}]{rogers_most_2015}
Rogers, L.~A. 2015, \bibinfo{title}{Most 1.6 Earth-Radius Planets are not Rocky,} The Astrophysical Journal, 801, 41, \dodoi{10.1088/0004-637X/801/1/41}

\bibitem[{L.~A. Rogers \& S. Seager(2010)Rogers \& Seager}]{rogers_framework_2010}
Rogers, L.~A., \& Seager, S. 2010, \bibinfo{title}{A Framework for Quantifying the Degeneracies of Exoplanet Interior Compositions,} The Astrophysical Journal, 712, 974, \dodoi{10.1088/0004-637X/712/2/974}

\bibitem[{D.~C. Rubie {et~al.}(2015)Rubie, Jacobson, Morbidelli, O’Brien, Young, de~Vries, Nimmo, Palme, \& Frost}]{rubie_accretion_2015}
Rubie, D.~C., Jacobson, S.~A., Morbidelli, A., {et~al.} 2015, \bibinfo{title}{Accretion and differentiation of the terrestrial planets with implications for the compositions of early-formed {Solar} {System} bodies and accretion of water,} Icarus, 248, 89, \dodoi{10.1016/j.icarus.2014.10.015}

\bibitem[{H. Sakuraba {et~al.}(2021)Sakuraba, Kurokawa, Genda, \& Ohta}]{sakuraba_numerous_2021}
Sakuraba, H., Kurokawa, H., Genda, H., \& Ohta, K. 2021, \bibinfo{title}{Numerous chondritic impactors and oxidized magma ocean set {Earth}’s volatile depletion,} Scientific Reports, 11, 20894, \dodoi{10.1038/s41598-021-99240-w}

\bibitem[{P. {Saxena} {et~al.}(2019){Saxena}, {Killen}, {Airapetian}, {Petro}, {Curran}, \& {Mandell}}]{2019ApJ...876L..16S}
{Saxena}, P., {Killen}, R.~M., {Airapetian}, V., {et~al.} 2019, \bibinfo{title}{{Was the Sun a Slow Rotator? Sodium and Potassium Constraints from the Lunar Regolith},} \apjl, 876, L16, \dodoi{10.3847/2041-8213/ab18fb}

\bibitem[{L. Schaefer {et~al.}(2016)Schaefer, Wordsworth, Berta-Thompson, \& Sasselov}]{schaefer_predictions_2016}
Schaefer, L., Wordsworth, R.~D., Berta-Thompson, Z., \& Sasselov, D. 2016, \bibinfo{title}{Predictions of the atmospheric composition of GJ 1132b,} The Astrophysical Journal, 829, 63, \dodoi{10.3847/0004-637X/829/2/63}

\bibitem[{H.~E. Schlichting \& S. Mukhopadhyay(2018)Schlichting \& Mukhopadhyay}]{schlichting_atmosphere_2018}
Schlichting, H.~E., \& Mukhopadhyay, S. 2018, \bibinfo{title}{Atmosphere {Impact} {Losses},} Space Science Reviews, 214, 34, \dodoi{10.1007/s11214-018-0471-z}

\bibitem[{S. Seabold \& J. Perktold(2010)Seabold \& Perktold}]{seabold2010statsmodels}
Seabold, S., \& Perktold, J. 2010, in 9th Python in Science Conference

\bibitem[{F. Selsis {et~al.}(2007)Selsis, Kasting, Levrard, Paillet, Ribas, \& Delfosse}]{selsis_habitable_2007}
Selsis, F., Kasting, J.~F., Levrard, B., {et~al.} 2007, \bibinfo{title}{Habitable planets around the star {Gl} 581?} Astronomy \& Astrophysics, 476, 1373, \dodoi{10.1051/0004-6361:20078091}

\bibitem[{F. Selsis {et~al.}(2023)Selsis, Leconte, Turbet, Chaverot, \& Bolmont}]{selsis_cool_2023}
Selsis, F., Leconte, J., Turbet, M., Chaverot, G., \& Bolmont, E. 2023, \bibinfo{title}{A cool runaway greenhouse without surface magma ocean,} Nature, 620, 287, \dodoi{10.1038/s41586-023-06258-3}

\bibitem[{C.~A. Sinclair {et~al.}(2020)Sinclair, Wyatt, Morbidelli, \& Nesvorný}]{sinclair_evolution_2020}
Sinclair, C.~A., Wyatt, M.~C., Morbidelli, A., \& Nesvorný, D. 2020, \bibinfo{title}{Evolution of the {Earth}’s atmosphere during {Late} {Veneer} accretion,} Monthly Notices of the Royal Astronomical Society, 499, 5334, \dodoi{10.1093/mnras/staa3210}

\bibitem[{P.~A. Sossi(2021)Sossi}]{sossi_atmospheres_2021}
Sossi, P.~A. 2021, \bibinfo{title}{Atmospheres in the baking,} Nature Astronomy, 5, 535, \dodoi{10.1038/s41550-021-01353-9}

\bibitem[{E. Stafne \& J. Becker(2024)Stafne \& Becker}]{stafne_predicting_2024}
Stafne, E., \& Becker, J. 2024, \bibinfo{title}{Predicting {CO2} {Gas} {Layer} {Thickness} due to {Mantle} {Outgassing} in the {Exoplanet} {Census},} Research Notes of the AAS, 8, 176, \dodoi{10.3847/2515-5172/ad5f30}

\bibitem[{R.~J. Strangeway {et~al.}(2010)Strangeway, Russell, Luhmann, Moore, Foster, Barabash, \& Nilsson}]{Strangeway2010}
Strangeway, R.~J., Russell, C.~T., Luhmann, J.~G., {et~al.} 2010, in AGU Fall Meeting Abstracts, Vol. 2010, SM33B--1893.
\newblock \url{https://ui.adsabs.harvard.edu/abs/2010AGUFMSM33B1893S}

\bibitem[{C. Swastik {et~al.}(2023)Swastik, Banyal, Narang, Unni, Banerjee, Manoj, \& Sivarani}]{swastik_age_2023}
Swastik, C., Banyal, R.~K., Narang, M., {et~al.} 2023, \bibinfo{title}{Age {Distribution} of {Exoplanet} {Host} {Stars}: {Chemical} and {Kinematic} {Age} {Proxies} from {GAIA} {DR3},} The Astronomical Journal, 166, 91, \dodoi{10.3847/1538-3881/ace782}

\bibitem[{F. Tian(2009)Tian}]{tian_thermal_2009}
Tian, F. 2009, \bibinfo{title}{Thermal Escape from Super Earth Atmospheres in the Habitable Zones of M Stars,} The Astrophysical Journal, 703, 905, \dodoi{10.1088/0004-637X/703/1/905}

\bibitem[{F. Tian(2015{\natexlab{a}})Tian}]{tian_history_2015}
Tian, F. 2015{\natexlab{a}}, \bibinfo{title}{History of water loss and atmospheric {O2} buildup on rocky exoplanets near {M} dwarfs,} Earth and Planetary Science Letters, 432, 126, \dodoi{10.1016/j.epsl.2015.09.051}

\bibitem[{F. Tian(2015{\natexlab{b}})Tian}]{tian_atmospheric_2015}
Tian, F. 2015{\natexlab{b}}, \bibinfo{title}{Atmospheric {Escape} from {Solar} {System} {Terrestrial} {Planets} and {Exoplanets},} Annual Review of Earth and Planetary Sciences, 43, 459, \dodoi{10.1146/annurev-earth-060313-054834}

\bibitem[{F. Tian \& S. Ida(2015)Tian \& Ida}]{tian_water_2015}
Tian, F., \& Ida, S. 2015, \bibinfo{title}{Water contents of {Earth}-mass planets around {M} dwarfs,} Nature Geoscience, 8, 177, \dodoi{10.1038/ngeo2372}

\bibitem[{F. Tian {et~al.}(2008{\natexlab{a}})Tian, Kasting, Liu, \& Roble}]{tian_hydrodynamic_2008}
Tian, F., Kasting, J.~F., Liu, H.-L., \& Roble, R.~G. 2008{\natexlab{a}}, \bibinfo{title}{Hydrodynamic planetary thermosphere model: 1. {Response} of the {Earth}'s thermosphere to extreme solar {EUV} conditions and the significance of adiabatic cooling,} Journal of Geophysical Research: Planets, 113, \dodoi{10.1029/2007JE002946}

\bibitem[{F. Tian {et~al.}(2009)Tian, Kasting, \& Solomon}]{tian_thermal_2009-1}
Tian, F., Kasting, J.~F., \& Solomon, S.~C. 2009, \bibinfo{title}{Thermal escape of carbon from the early {Martian} atmosphere,} Geophysical Research Letters, 36, 2008GL036513, \dodoi{10.1029/2008GL036513}

\bibitem[{F. Tian {et~al.}(2008{\natexlab{b}})Tian, Solomon, Qian, Lei, \& Roble}]{tian_hydrodynamic_2008-1}
Tian, F., Solomon, S.~C., Qian, L., Lei, J., \& Roble, R.~G. 2008{\natexlab{b}}, \bibinfo{title}{Hydrodynamic planetary thermosphere model: 2. {Coupling} of an electron transport/energy deposition model,} Journal of Geophysical Research (Planets), 113, E07005, \dodoi{10.1029/2007JE003043}

\bibitem[{L. Tu {et~al.}(2015)Tu, Johnstone, Güdel, \& Lammer}]{tu_extreme_2015}
Tu, L., Johnstone, C.~P., Güdel, M., \& Lammer, H. 2015, \bibinfo{title}{The extreme ultraviolet and {X}-ray {Sun} in {Time}: {High}-energy evolutionary tracks of a solar-like star,} Astronomy \& Astrophysics, 577, L3, \dodoi{10.1051/0004-6361/201526146}

\bibitem[{E. Vald{\'e}s {et~al.}(2025)Vald{\'e}s, Demory, Diamond-Lowe, Mendon{\c{c}}a, August, Fortune, Allen, Kitzmann, Gressier, Hooton, {et~al.}}]{valdes2025hot}
Vald{\'e}s, E., Demory, B.-O., Diamond-Lowe, H., {et~al.} 2025, \bibinfo{title}{Hot Rocks Survey II: The thermal emission of TOI-1468 b reveals a hot bare rock,} arXiv preprint arXiv:2503.19772

\bibitem[{D. Valencia {et~al.}(2025)Valencia, Moro-Martin, \& Teske}]{Valencia_2025}
Valencia, D., Moro-Martin, A., \& Teske, J. 2025, Diversity of exoplanets (Elsevier), 19–49, \dodoi{10.1016/b978-0-323-99762-1.00139-x}

\bibitem[{D. Valencia {et~al.}(2007)Valencia, Sasselov, \& O’Connell}]{valencia_detailed_2007}
Valencia, D., Sasselov, D.~D., \& O’Connell, R.~J. 2007, \bibinfo{title}{Detailed {Models} of {Super}‐{Earths}: {How} {Well} {Can} {We} {Infer} {Bulk} {Properties}?} The Astrophysical Journal, 665, 1413, \dodoi{10.1086/519554}

\bibitem[{O. Vilhu(1984)Vilhu}]{vilhu_nature_1984}
Vilhu, O. 1984, \bibinfo{title}{The nature of magnetic activity in lower main sequence stars.,} Astronomy and Astrophysics, 133, 117.
\newblock \url{https://ui.adsabs.harvard.edu/abs/1984A&A...133..117V}

\bibitem[{P. Virtanen {et~al.}(2020)Virtanen, Gommers, Oliphant, Haberland, Reddy, Cournapeau, Burovski, Peterson, Weckesser, Bright, {van der Walt}, Brett, Wilson, Millman, Mayorov, Nelson, Jones, Kern, Larson, Carey, Polat, Feng, Moore, {VanderPlas}, Laxalde, Perktold, Cimrman, Henriksen, Quintero, Harris, Archibald, Ribeiro, Pedregosa, {van Mulbregt}, \& {SciPy 1.0 Contributors}}]{2020SciPy-NMeth}
Virtanen, P., Gommers, R., Oliphant, T.~E., {et~al.} 2020, \bibinfo{title}{{{SciPy} 1.0: Fundamental Algorithms for Scientific Computing in Python},} Nature Methods, 17, 261, \dodoi{10.1038/s41592-019-0686-2}

\bibitem[{P. Wachiraphan {et~al.}(2024)Wachiraphan, Berta-Thompson, Diamond-Lowe, Winters, Murray, Zhang, Xue, Morley, Rosario-Franco, \& Duvvuri}]{wachiraphan_thermal_2024}
Wachiraphan, P., Berta-Thompson, Z.~K., Diamond-Lowe, H., {et~al.} 2024, \bibinfo{title}{The {Thermal} {Emission} {Spectrum} of the {Nearby} {Rocky} {Exoplanet} {LTT} {1445A} b from {JWST} {MIRI}/{LRS},} arXiv, \dodoi{10.48550/ARXIV.2410.10987}

\bibitem[{W. Wagner(1988)Wagner}]{wagner_observations_1988}
Wagner, W. 1988, \bibinfo{title}{Observations of 1–8 Å solar {X}-ray variability during solar cycle 21,} Advances in Space Research, 8, 67, \dodoi{10.1016/0273-1177(88)90173-1}

\bibitem[{A.~J. Watson {et~al.}(1981)Watson, Donahue, \& Walker}]{watson_dynamics_1981}
Watson, A.~J., Donahue, T.~M., \& Walker, J.~C. 1981, \bibinfo{title}{The dynamics of a rapidly escaping atmosphere: {Applications} to the evolution of {Earth} and {Venus},} Icarus, 48, 150, \dodoi{10.1016/0019-1035(81)90101-9}

\bibitem[{B.~E. Wood {et~al.}(2021)Wood, Müller, Redfield, Konow, Vannier, Linsky, Youngblood, Vidotto, Jardine, Alvarado-Gómez, \& Drake}]{wood_new_2021}
Wood, B.~E., Müller, H.-R., Redfield, S., {et~al.} 2021, \bibinfo{title}{New {Observational} {Constraints} on the {Winds} of {M} dwarf {Stars},} The Astrophysical Journal, 915, 37, \dodoi{10.3847/1538-4357/abfda5}

\bibitem[{T.~N. Woods {et~al.}(1996)Woods, Prinz, Rottman, London, Crane, Cebula, Hilsenrath, Brueckner, Andrews, White, VanHoosier, Floyd, Herring, Knapp, Pankratz, \& Reiser}]{woods_validation_1996}
Woods, T.~N., Prinz, D.~K., Rottman, G.~J., {et~al.} 1996, \bibinfo{title}{Validation of the {UARS} solar ultraviolet irradiances: {Comparison} with the {ATLAS} 1 and 2 measurements,} Journal of Geophysical Research: Atmospheres, 101, 9541, \dodoi{10.1029/96JD00225}

\bibitem[{R. Wordsworth(2015)Wordsworth}]{wordsworth2015}
Wordsworth, R. 2015, \bibinfo{title}{Atmospheric heat redistribution and collapse on tidally locked rocky planets,} The Astrophysical Journal, 806, 180

\bibitem[{R. Wordsworth {et~al.}(2017)Wordsworth, Kalugina, Lokshtanov, Vigasin, Ehlmann, Head, Sanders, \& Wang}]{wordsworth_transient_2017}
Wordsworth, R., Kalugina, Y., Lokshtanov, S., {et~al.} 2017, \bibinfo{title}{Transient reducing greenhouse warming on early {Mars},} Geophysical Research Letters, 44, 665, \dodoi{10.1002/2016GL071766}

\bibitem[{R. Wordsworth \& L. Kreidberg(2022)Wordsworth \& Kreidberg}]{wordsworth_atmospheres_2022}
Wordsworth, R., \& Kreidberg, L. 2022, \bibinfo{title}{Atmospheres of {Rocky} {Exoplanets},} Annual Review of Astronomy and Astrophysics, 60, 159, \dodoi{10.1146/annurev-astro-052920-125632}

\bibitem[{R.~D. Wordsworth(2016)Wordsworth}]{wordsworth_atmospheric_2016}
Wordsworth, R.~D. 2016, \bibinfo{title}{Atmospheric nitrogen evolution on {Earth} and {Venus},} Earth and Planetary Science Letters, 447, 103, \dodoi{10.1016/j.epsl.2016.04.002}

\bibitem[{R.~D. Wordsworth {et~al.}(2018)Wordsworth, Schaefer, \& Fischer}]{wordsworth_redox_2018}
Wordsworth, R.~D., Schaefer, L.~K., \& Fischer, R.~A. 2018, \bibinfo{title}{Redox {Evolution} via {Gravitational} {Differentiation} on {Low}-mass {Planets}: {Implications} for {Abiotic} {Oxygen}, {Water} {Loss}, and {Habitability},} The Astronomical Journal, 155, 195, \dodoi{10.3847/1538-3881/aab608}

\bibitem[{M.~C. Wyatt {et~al.}(2020)Wyatt, Kral, \& Sinclair}]{wyatt_susceptibility_2020}
Wyatt, M.~C., Kral, Q., \& Sinclair, C.~A. 2020, \bibinfo{title}{Susceptibility of planetary atmospheres to mass-loss and growth by planetesimal impacts: the impact shoreline,} Monthly Notices of the Royal Astronomical Society, 491, 782, \dodoi{10.1093/mnras/stz3052}

\bibitem[{S. Xu \& A. Bonsor(2021)Xu \& Bonsor}]{xu_exogeology_2021}
Xu, S., \& Bonsor, A. 2021, \bibinfo{title}{Exogeology from {Polluted} {White} {Dwarfs},} arXiv, \dodoi{10.48550/arXiv.2108.08384}

\bibitem[{Q. Xue {et~al.}(2024)Xue, Bean, Zhang, Mahajan, Ih, Eastman, Lunine, Mansfield, Coy, Kempton, Koll, \& Kite}]{xue_jwst_2024}
Xue, Q., Bean, J.~L., Zhang, M., {et~al.} 2024, \bibinfo{title}{{JWST} {Thermal} {Emission} of the {Terrestrial} {Exoplanet} {GJ} 1132b,} The Astrophysical Journal Letters, 973, L8, \dodoi{10.3847/2041-8213/ad72e9}

\bibitem[{T. Yoshida {et~al.}(2022)Yoshida, Terada, Ikoma, \& Kuramoto}]{yoshida_less_2022}
Yoshida, T., Terada, N., Ikoma, M., \& Kuramoto, K. 2022, \bibinfo{title}{Less {Effective} {Hydrodynamic} {Escape} of {H2}–{H2O} {Atmospheres} on {Terrestrial} {Planets} {Orbiting} {Pre}-main-sequence {M} {Dwarfs},} The Astrophysical Journal, 934, 137, \dodoi{10.3847/1538-4357/ac7be7}

\bibitem[{T. Yoshida {et~al.}(2024)Yoshida, Terada, \& Kuramoto}]{yoshida_suppression_2024}
Yoshida, T., Terada, N., \& Kuramoto, K. 2024, \bibinfo{title}{Suppression of hydrodynamic escape of an {H2}-rich early {Earth} atmosphere by radiative cooling of carbon oxides,} Progress in Earth and Planetary Science, 11, 59, \dodoi{10.1186/s40645-024-00666-3}

\bibitem[{K.~J. Zahnle \& D.~C. Catling(2017)Zahnle \& Catling}]{zahnle_cosmic_2017}
Zahnle, K.~J., \& Catling, D.~C. 2017, \bibinfo{title}{The {Cosmic} {Shoreline}: {The} {Evidence} that {Escape} {Determines} which {Planets} {Have} {Atmospheres}, and what this {May} {Mean} for {Proxima} {Centauri} {B},} The Astrophysical Journal, 843, 122, \dodoi{10.3847/1538-4357/aa7846}

\bibitem[{K.~J. Zahnle \& J.~F. Kasting(2023)Zahnle \& Kasting}]{zahnle_elemental_2023}
Zahnle, K.~J., \& Kasting, J.~F. 2023, \bibinfo{title}{Elemental and isotopic fractionation as fossils of water escape from {Venus},} Geochimica et Cosmochimica Acta, 361, 228, \dodoi{10.1016/j.gca.2023.09.023}

\bibitem[{L. Zeng {et~al.}(2019)Zeng, Jacobsen, Sasselov, Petaev, Vanderburg, Lopez-Morales, Perez-Mercader, Mattsson, Li, Heising, Bonomo, Damasso, Berger, Cao, Levi, \& Wordsworth}]{zeng_growth_2019}
Zeng, L., Jacobsen, S.~B., Sasselov, D.~D., {et~al.} 2019, \bibinfo{title}{Growth model interpretation of planet size distribution,} Proceedings of the National Academy of Sciences, 116, 9723, \dodoi{10.1073/pnas.1812905116}

\bibitem[{J. {Zhang} \& L.~A. {Rogers}(2022){Zhang} \& {Rogers}}]{Zhang2022}
{Zhang}, J., \& {Rogers}, L.~A. 2022, \bibinfo{title}{{Thermal Evolution and Magnetic History of Rocky Planets},} \apj, 938, 131, \dodoi{10.3847/1538-4357/ac8e65}

\bibitem[{M. Zhang {et~al.}(2022{\natexlab{a}})Zhang, Cauley, Knutson, France, Kreidberg, Oklopčić, Redfield, \& Shkolnik}]{zhang_more_2022}
Zhang, M., Cauley, P.~W., Knutson, H.~A., {et~al.} 2022{\natexlab{a}}, \bibinfo{title}{More {Evidence} for {Variable} {Helium} {Absorption} from {HD} 189733b,} The Astronomical Journal, 164, 237, \dodoi{10.3847/1538-3881/ac9675}

\bibitem[{M. Zhang {et~al.}(2023)Zhang, Dai, Bean, Knutson, \& Rescigno}]{zhang_outflowing_2023}
Zhang, M., Dai, F., Bean, J.~L., Knutson, H.~A., \& Rescigno, F. 2023, \bibinfo{title}{Outflowing {Helium} from a {Mature} {Mini}-{Neptune},} The Astrophysical Journal, 953, L25, \dodoi{10.3847/2041-8213/aced51}

\bibitem[{M. Zhang {et~al.}(2022{\natexlab{b}})Zhang, Knutson, Wang, Dai, \& Barragán}]{zhang_escaping_2022}
Zhang, M., Knutson, H.~A., Wang, L., Dai, F., \& Barragán, O. 2022{\natexlab{b}}, \bibinfo{title}{Escaping {Helium} from {TOI} 560.01, a {Young} {Mini}-{Neptune},} The Astronomical Journal, 163, 67, \dodoi{10.3847/1538-3881/ac3fa7}

\bibitem[{M. Zhang {et~al.}(2024)Zhang, Hu, Inglis, Dai, Bean, Knutson, Lam, Goffo, \& Gandolfi}]{zhang_gj_2024}
Zhang, M., Hu, R., Inglis, J., {et~al.} 2024, \bibinfo{title}{{GJ} 367b {Is} a {Dark}, {Hot}, {Airless} {Sub}-{Earth},} The Astrophysical Journal Letters, 961, L44, \dodoi{10.3847/2041-8213/ad1a07}

\bibitem[{S. Zieba {et~al.}(2022)Zieba, Zilinskas, Kreidberg, Nguyen, Miguel, Cowan, Pierrehumbert, Carone, Dang, Hammond, {et~al.}}]{zieba2022}
Zieba, S., Zilinskas, M., Kreidberg, L., {et~al.} 2022, \bibinfo{title}{K2 and Spitzer phase curves of the rocky ultra-short-period planet K2-141 b hint at a tenuous rock vapor atmosphere,} Astronomy \& Astrophysics, 664, A79

\bibitem[{S. Zieba {et~al.}(2023)Zieba, Kreidberg, Ducrot, Gillon, Morley, Schaefer, Tamburo, Koll, Lyu, Acuña, Agol, Iyer, Hu, Lincowski, Meadows, Selsis, Bolmont, Mandell, \& Suissa}]{zieba_no_2023}
Zieba, S., Kreidberg, L., Ducrot, E., {et~al.} 2023, \bibinfo{title}{No thick carbon dioxide atmosphere on the rocky exoplanet {TRAPPIST}-1 c,} Nature, 620, 746, \dodoi{10.1038/s41586-023-06232-z}

\end{thebibliography}

\appendix

\section{Carbon Distribution During Magma Ocean Solidification}\label{sec:app}
\renewcommand{\thefigure}{A\arabic{figure}}
\setcounter{figure}{0}

We distribute volatile between the magma, single-species atmosphere, and mantle reservoirs during magma ocean solidification as follows. 

For nitrogen-rich cases, owing to the extremely low solubility of \ch{N2} in solids, the mass of nitrogen in the residual mantle can be neglected \citep{hiermajumder_origin_2017}.

For carbon-rich cases, we assume that as the magma cools and crystallizes, the atmosphere remains in equilibrium with the magma ocean surface according to the solubility law \citep{lichtenberg_vertically_2021}. Additionally, we assume that the initial mass of the magma ocean is 2/3 of the planetary mass, and we only consider the carbon distributed between magma and atmosphere but neglect the partition into fe-alloy core \citep{blanchard_metalsilicate_2022}. The carbon mass distribution is calculated as a function of the changing magma mass during crystallization. As the magma mass decreases, carbon is redistributed between the atmosphere, magma ocean, and solid mantle. 

\subsection*{Variables and Constants}

The key variables used in our model include \( C_{\text{atm}} \), \( C_{\text{magma}} \), and \( C_{\text{mantle}} \), which represent the carbon mass in the atmosphere, magma, and mantle, respectively, with the total carbon mass \( C_0 \) remaining constant. The magma ocean mass is denoted as \( M_{\text{magma}} \), and the carbon concentration in the magma is given by \( f = {C_{\text{magma}}}/{M_{\text{magma}}} \). The partition coefficient for mantle-magma exchange is represented by \( D_{\text{mm}} \) (assumed constant). The solubility law linking surface pressure to carbon concentration in magma is parameterized by the constants \( \alpha \) and \( \beta \), following \cite{lichtenberg_vertically_2021}, such that \( P_{\text{surf}} = (f/\alpha)^\beta \). To simplify calculations, we define a factor \( k_{Ps} = \frac{12}{\text{MMW}} \cdot \frac{4\pi R_p^2}{g} \), where \( R_p \) is the planetary radius, \( g \) is the gravitational acceleration, and the \( 12/\text{MMW} \) factor accounts for the mean molecular ratio of carbon to atmospheric species. MMW=44 for \ch{CO2} and MMW=16 for \ch{CH4}. This allows the atmospheric carbon mass to be expressed as \( C_{\text{atm}} = k_{Ps} \cdot P_{\text{surf}} \).

\subsection*{Governing Equations}

1. \textbf{Mass Conservation}:
   \begin{equation}
       C_{\text{atm}} + C_{\text{magma}} + C_{\text{mantle}} = C_0
       \label{eq:mass_conservation}
   \end{equation}

2. \textbf{Carbon in atmosphere} \cite{lichtenberg_vertically_2021}:
   \begin{equation}
       C_{\text{atm}} = k_{Ps} \left(\frac{f}{\alpha} \right)^\beta
       \label{eq:atm_equilibrium}
   \end{equation}
   This equation reflects the equilibrium between the atmospheric carbon and the carbon dissolved in the magma ocean surface. For \ch{CO2}, \(\alpha = 1.937 \times 10^{-15}  \text{Pa}^{-1}; \beta = 0.714\), and for \ch{CH4}, \(\alpha = 9.937 \times 10^{-14} \text{Pa}^{-1}; \beta = 1.000\)\citep{lichtenberg_vertically_2021}

3. \textbf{Carbon in Magma}:
   \begin{equation}
       {C_{\text{magma}}} = f{M_{\text{magma}}}
       \label{eq:carbon_fraction}
   \end{equation}

4. \textbf{Mantle Carbon Change}:
   \begin{equation}
       dC_{\text{mantle}} =- D_{\text{mm}} f \, dM_{\text{magma}}
       \label{eq:mantle_change}
   \end{equation}
   This equation accounts for the partition of carbon from the magma into the solid mantle during solidification, with $D_{\text{mm}}$ representing the partitioning behavior.

\subsection*{Derivation of the Differential Equation}

Starting from the mass conservation equation \ref{eq:mass_conservation}, differentiating both sides and rearanging:

\begin{equation}
    {dC_{\text{mantle}}} = -{dC_{\text{atm}}} - {dC_{\text{magma}}}
    \label{eq:mantle_derivative}
\end{equation}


Differentiating equation \ref{eq:carbon_fraction}:

\begin{equation}
    {dC_{\text{magma}}}= f\cdot dM_{\text{magma}} + {dM_{\text{magma}}}\cdot {df}
    \label{eq:magma_carbon_derivative}
\end{equation}

From equation \ref{eq:atm_equilibrium}, express $C_{\text{atm}}$ and differentiate:

\begin{equation}
    {dC_{\text{atm}}} = k_{Ps} \alpha^{-\beta} \beta f^{\beta - 1} df
    \label{eq:atm_carbon_derivative}
\end{equation}

Define a constant for simplification $A = k_{Ps} \alpha^{-\beta} \beta$

Substitute equations \ref{eq:magma_carbon_derivative} and \ref{eq:atm_carbon_derivative} into equation \ref{eq:mantle_derivative}:

\begin{equation}
f \left( D_{\text{mm}} + 1 \right) dM_{\text{magma}} + \left( M_{\text{magma}} + A f^{\beta - 1} \right) df = 0
\end{equation}

\subsection*{Solving ODE}

Let us consider $M_{\text{magma}}$ as a function of $f$:

\begin{equation}
\frac{dM_{\text{magma}}}{df} = -\frac{M_{\text{magma}} }{f \left( D_{\text{mm}} + 1 \right)} - \frac{A f^{\beta - 2}}{\left( D_{\text{mm}} + 1 \right)}
\end{equation}

The ODE has the standard linear form $\frac{dM_{\text{magma}}}{df} + P(f) M_{\text{magma}} = Q(f)$, where $P(f) = \frac{1}{f \left( D_{\text{mm}} + 1 \right)}$ and $Q(f) = -\frac{A f^{\beta - 2}}{D_{\text{mm}} + 1}$. The integrating factor $\mu(f)$ is:

\begin{equation}
\mu(f) = \exp\left( \int P(f) df \right) = \exp\left( \int \frac{1}{f \left( D_{\text{mm}} + 1 \right)} df \right) = f^{\frac{1}{D_{\text{mm}} + 1}}
\end{equation}

Multiply both sides by the integrating factor and simplify:

\begin{equation}
\frac{d}{df} \left( f^{\frac{1}{D_{\text{mm}} + 1}} M_{\text{magma}} \right) = -\frac{A f^{\beta - 2 + \frac{1}{D_{\text{mm}} + 1}}}{D_{\text{mm}} + 1}
\end{equation}

Integrate with respect to $f$ and let $n = \beta - 2 + \frac{1}{D_{\text{mm}} + 1}$ for simplification:

\begin{equation}
    f^{\frac{1}{D_{\text{mm}} + 1}} M_{\text{magma}} = -\frac{A}{\left( D_{\text{mm}} + 1 \right) \left( n + 1 \right)} f^{n + 1} + \text{const}
    \label{eq: equation_w_constant}
\end{equation}

where the integral constant (const) can be solved by the assumed initial condition.

\subsection*{Integral Constant Determined by Initial Condition}

We assume the initial magma mass is $M_{\text{magma,0}} = 2/3 M_p$, and the carbon is only distributed between magma and the atmosphere. With the solubility law, we can write down the mass conservation equation as:

\begin{equation}
    C_0 = C_{\text{atm,0}} + C_{\text{magma,0}} = k_{Ps}\left( \frac{f_0}{\alpha}\right)^\beta + f_0M_{\text{magma,0}}
\end{equation}

The equation can be solved numerically with the given total carbon mass $C_0$. And plug it back into equation \ref{eq: equation_w_constant}, we can get the value of const.

Rearranging equation \ref{eq: equation_w_constant}, the final expression of $M_{\text{magma}}$ in terms of $f$ is:

\begin{equation}
M_{\text{magma}} = \frac{A f^{\beta - 1}}{\left( D_{\text{mm}} - 1 \right) \left( \beta - 1 + \frac{1}{1 - D_{\text{mm}}} \right)} + \text{const} f^{-\frac{1}{1 - D_{\text{mm}}}}
\end{equation}

where:

\begin{itemize}
    \item \( A = \frac{12}{\text{MMW}} \cdot \frac{4\pi R_p^2 }{g} \cdot \beta \alpha^{-\beta} \)
    \item \( \text{const} = f_0^{\frac{1}{D_{\text{mm}} + 1}} M_{\text{magma,0}} +  \frac{A}{\left( D_{\text{mm}} + 1 \right) \left( n + 1 \right)} f_0^{n + 1} \) 
\end{itemize}

We have derived the differential equation considering how the carbon partitions into the solid mantle with the solidification of magma ocean \( dC_{\text{mantle}} = -f \, D_{\text{mm}} \, dM_{\text{magma}} \) and solved it to find how \( M_{\text{magma}} \) evolves with \( f\). Using this solution, we can determine the evolution of all other variables with a given \( M_{\text{magma}}\). 

\begin{figure}[h]
    \centering
    \includegraphics[width=0.9\linewidth]{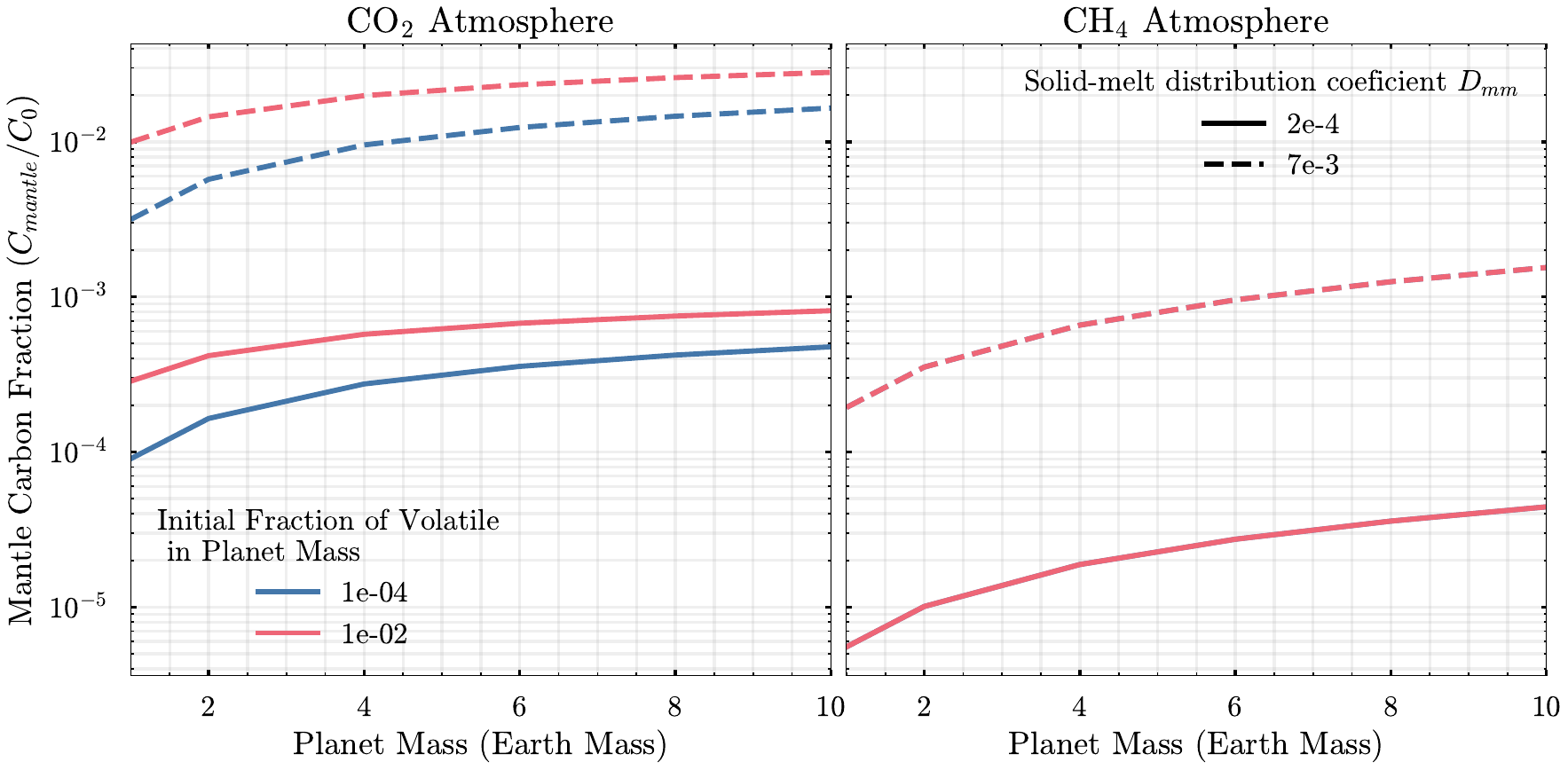}
    \caption{The fraction of carbon sequestered in the mantle relative to the total carbon mass after complete magma ocean solidification, shown as a function of planetary mass, for a pure-\ch{CO2} atmosphere (left) and a pure-\ch{CH4} atmosphere (right). The fraction increases with planetary mass due to the solubility law, which governs surface pressure, and the reduced atmospheric column mass at higher gravity with a given surface pressure. The blue line represents the lowest initial carbon case in our study, where the carbon-to-planetary mass ratio is \(10^{-4} M_p\), while the pink line corresponds to the most carbon-rich case, \(10^{-2} M_p\). A higher initial carbon concentration in the magma ocean results in more carbon being retained in the mantle post-solidification. Solid lines indicate a conservative solid-melt partition coefficient, while dashed lines represent a higher partition coefficient, following \citet{elkins-tanton_linked_2008}.}
    \label{fig:C_distributions}
\end{figure}

When setting \( M_{\text{magma}}=0\), that is the phase after fully solidification of the primordial magma ocean. We show the distribution of carbon in different reservoirs in Fig.~\ref{fig:C_distributions}.

\newpage
\renewcommand{\thetable}{B\arabic{table}}
\setcounter{table}{0}

\section{Appendix Table}

\begin{longtable}{llllllllllllllll}
\toprule
\caption{Atmospheric Retention Targets Ranked by Priority Metric for \ch{CO2}-Dominated Atmospheres}\\
\hline
Planet & \multicolumn{3}{c}{Priority Metric} &  $R_p$ & $M_p$ & $v_{esc}$ & $\rho$ \footnote{the density scaled to that of a planet with Earth-like composition} & $M_*$ & $T_{eq}$ & \textbf{K}-mag & $(R_{p}/R_{s})^{2}$ & ESM\footnote{Emission spectroscopy metric from \cite{kempton_framework_2018}} & TSM\footnote{Transmission spectroscopy metric from \cite{kempton_framework_2018}}\\
& [\ch{CO2}] & [\ch{CH4}] & [\ch{N2}]&  $(R_\oplus)$ & $(M_\oplus)$ & (km/s) & ($\rho^{*}_{\oplus}$) & $(M_\odot)$ & (K)  & & (ppm) &  & \\
\midrule

K2-91 b & -0.02 & -1.07 & -0.16 & 1.22 & 2.05$^{\dagger}$ & 14.5$^{\dagger}$ & 1$^{\dagger}$ & 0.38 & 761 & 10.6 & 1217 & 3.3 & 6$^{\dagger}$ \\
TOI-1416 b & -0.05 & -1.36 & -0.17 & 1.62 & 3.48 & 16.4 & 0.65 & 0.80 & 1515 & 7.7 & 351 & 9.6 & 82 \\
LHS 1678 d & -0.09 & -0.71 & -0.87 & 0.98 & 0.94$^{\dagger}$ & 11.0$^{\dagger}$ & 1$^{\dagger}$ & 0.34 & 483 & 8.3 & 747 & 1.4 & 13$^{\dagger}$ \\
K2-415 b & -0.11 & -0.79 & -0.77 & 1.01 & 1.06$^{\dagger}$ & 11.4$^{\dagger}$ & 1$^{\dagger}$ & 0.16 & 412 & 9.9 & 2243 & 1.2 & 14$^{\dagger}$ \\
TOI-1444 b & -0.11 & -1.69 & -0.44 & 1.42 & 3.58 & 17.8 & 1.00 & 0.95 & 2130 & 9.1 & 204 & 4.5 & 5 \\
LTT 3780 b & -0.15 & -1.30 & -0.38 & 1.32 & 2.46 & 15.2 & 0.90 & 0.38 & 902 & 8.2 & 1022 & 13.1 & 18 \\
GJ 1132 b$^{*}$ & -0.15 & -1.03 & -0.27 & 1.19 & 1.84 & 13.9 & 0.98 & 0.19 & 583 & 8.3 & 2441 & 9.7 & 30 \\
TOI-6086 b & -0.15 & -1.05 & -0.30 & 1.18 & 1.81$^{\dagger}$ & 13.9$^{\dagger}$ & 1$^{\dagger}$ & 0.25 & 633 & 10.0 & 1745 & 4.2 & 11$^{\dagger}$ \\
Kepler-446 b & -0.17 & -1.01 & -0.41 & 1.15 & 1.64$^{\dagger}$ & 13.4$^{\dagger}$ & 1$^{\dagger}$ & 0.19 & 558 & 12.8 & 2316 & 0.9 & 4$^{\dagger}$ \\
TOI-1685 b$^{*}$ & -0.19 & -1.42 & -0.47 & 1.47 & 3.03 & 16.1 & 0.79 & 0.45 & 1090 & 8.8 & 873 & 11.9 & 12 \\
TOI-431 b & -0.19 & -1.67 & -0.40 & 1.28 & 3.07 & 17.3 & 1.20 & 0.78 & 1862 & 6.7 & 258 & 15.6 & 16 \\
TOI-1860 b & -0.19 & -1.39 & -0.24 & 1.31 & 2.66$^{\dagger}$ & 16.0$^{\dagger}$ & 1$^{\dagger}$ & 0.99 & 1885 & 6.8 & 163 & 8.0 & 12$^{\dagger}$ \\
TRAPPIST-1 b$^{*}$ & -0.23 & -1.04 & -0.72 & 1.12 & 1.37 & 12.4 & 0.94 & 0.09 & 397 & 10.3 & 7367 & 3.8 & 29 \\
Wolf 327 b & -0.27 & -1.52 & -0.53 & 1.24 & 2.53 & 16.0 & 1.13 & 0.41 & 1087 & 8.4 & 784 & 12.5 & 13 \\
LP 791-18 b & -0.28 & -1.28 & -0.33 & 1.21 & 2.00$^{\dagger}$ & 14.4$^{\dagger}$ & 1$^{\dagger}$ & 0.14 & 616 & 10.6 & 3727 & 6.8 & 15$^{\dagger}$ \\
TOI-5720 b & -0.33 & -1.17 & -0.86 & 1.09 & 1.36$^{\dagger}$ & 12.5$^{\dagger}$ & 1$^{\dagger}$ & 0.38 & 706 & 9.3 & 681 & 2.9 & 8$^{\dagger}$ \\
LHS 3844 b$^{*}$ & -0.34 & -1.54 & -0.62 & 1.30 & 2.61$^{\dagger}$ & 15.8$^{\dagger}$ & 1$^{\dagger}$ & 0.15 & 804 & 9.2 & 3995 & 28.9 & 35$^{\dagger}$ \\
GJ 3473 b & -0.35 & -1.21 & -0.56 & 1.26 & 1.86 & 13.6 & 0.83 & 0.36 & 772 & 8.8 & 1014 & 6.7 & 14 \\
LHS 1678 c & -0.36 & -0.96 & -1.20 & 0.94 & 0.81$^{\dagger}$ & 10.4$^{\dagger}$ & 1$^{\dagger}$ & 0.34 & 533 & 8.3 & 688 & 1.9 & 15$^{\dagger}$ \\
TOI-1450 A b & -0.36 & -1.11 & -1.07 & 1.13 & 1.26 & 11.8 & 0.84 & 0.48 & 722 & 7.6 & 460 & 4.4 & 14 \\
LP 791-18 d & -0.48 & -1.10 & -1.29 & 1.03 & 0.90 & 10.4 & 0.83 & 0.14 & 432 & 10.6 & 2702 & 1.4 & 15 \\
LHS 475 b & -0.54 & -1.21 & -1.37 & 0.97 & 0.92$^{\dagger}$ & 10.9$^{\dagger}$ & 1$^{\dagger}$ & 0.27 & 594 & 7.7 & 977 & 5.4 & 27$^{\dagger}$ \\
TRAPPIST-1 d & -0.56 & -1.14 & -1.61 & 0.79 & 0.39 & 7.8 & 0.90 & 0.09 & 286 & 10.3 & 3673 & 0.3 & 26 \\
TOI-544 b & -0.61 & -1.46 & -0.78 & 2.02 & 2.89 & 13.4 & 0.29 & 0.63 & 999 & 7.8 & 882 & 12.9 & 176 \\
GJ 806 b & -0.66 & -1.54 & -0.90 & 1.33 & 1.90 & 13.4 & 0.72 & 0.41 & 939 & 6.5 & 867 & 23.9 & 44 \\
TOI-178 b & -0.68 & -1.46 & -0.99 & 1.15 & 1.50 & 12.8 & 0.92 & 0.65 & 1040 & 8.7 & 263 & 2.7 & 6 \\
TOI-6008 b & -0.84 & -1.50 & -1.46 & 1.03 & 1.11$^{\dagger}$ & 11.6$^{\dagger}$ & 1$^{\dagger}$ & 0.23 & 706 & 9.5 & 1523 & 6.4 & 19$^{\dagger}$ \\
TOI-1807 b & -0.86 & -1.81 & -0.69 & 1.50 & 2.44 & 14.3 & 0.62 & 0.80 & 1694 & 7.6 & 338 & 11.8 & 19 \\
TOI-1442 b & -0.99 & -1.88 & -1.19 & 1.17 & 1.76$^{\dagger}$ & 13.7$^{\dagger}$ & 1$^{\dagger}$ & 0.29 & 1072 & 10.1 & 1197 & 9.4 & 13$^{\dagger}$ \\
TOI-540 b & -1.09 & -1.69 & -1.95 & 0.90 & 0.70$^{\dagger}$ & 9.9$^{\dagger}$ & 1$^{\dagger}$ & 0.16 & 611 & 8.9 & 1909 & 6.8 & 38$^{\dagger}$ \\
SPECULOOS-3 b & -1.10 & -1.81 & -1.95 & 0.98 & 0.93$^{\dagger}$ & 10.9$^{\dagger}$ & 1$^{\dagger}$ & 0.10 & 552 & 10.5 & 5303 & 7.7 & 35$^{\dagger}$ \\
L 98-59 b & -1.10 & -1.61 & -2.05 & 0.85 & 0.47 & 8.3 & 0.84 & 0.31 & 620 & 7.1 & 624 & 5.0 & 41 \\
GJ 1252 b$^{*}$ & -1.24 & -2.00 & -1.93 & 1.19 & 1.32 & 11.8 & 0.74 & 0.38 & 1086 & 7.9 & 782 & 16.3 & 32 \\
TOI-500 b & -1.26 & -2.03 & -1.69 & 1.17 & 1.42 & 12.3 & 0.85 & 0.74 & 1615 & 7.7 & 249 & 8.5 & 16 \\
TOI-561 b & -1.45 & -2.36 & -1.55 & 1.40 & 2.02 & 13.5 & 0.66 & 0.81 & 2303 & 8.4 & 231 & 7.9 & 14 \\
K2-183 b & -1.52 & -2.27 & -1.79 & 1.10 & 1.41$^{\dagger}$ & 12.6$^{\dagger}$ & 1$^{\dagger}$ & 0.94 & 2452 & 11.0 & 134 & 1.5 & 3$^{\dagger}$ \\
GJ 367 b$^{*}$ & -1.75 & -2.38 & -2.60 & 0.70 & 0.63 & 10.6 & 1.96 & 0.46 & 1364 & 5.8 & 196 & 16.4 & 31 \\
TOI-4527.01 & -1.82 & -2.43 & -2.79 & 0.91 & 0.72$^{\dagger}$ & 10.0$^{\dagger}$ & 1$^{\dagger}$ & 0.48 & 1359 & 7.0 & 294 & 12.9 & 31$^{\dagger}$ \\
GJ 238 b & -1.85 & -2.35 & -2.75 & 0.57 & 0.14$^{\dagger}$ & 5.6$^{\dagger}$ & 1$^{\dagger}$ & 0.42 & 758 & 7.0 & 145 & 2.0 & 26 \\
LHS 1678 b & -1.87 & -2.39 & -2.82 & 0.69 & 0.27$^{\dagger}$ & 7.1$^{\dagger}$ & 1$^{\dagger}$ & 0.34 & 866 & 8.3 & 364 & 3.9 & 28$^{\dagger}$ \\
K2-223 b & -1.87 & -2.43 & -2.72 & 0.89 & 0.67$^{\dagger}$ & 9.7$^{\dagger}$ & 1$^{\dagger}$ & 1.06 & 2271 & 9.8 & 67 & 1.0 & 4$^{\dagger}$ \\
K2-157 b & -1.91 & -2.53 & -2.59 & 1.00 & 1.01$^{\dagger}$ & 11.2$^{\dagger}$ & 1$^{\dagger}$ & 0.94 & 2601 & 11.0 & 109 & 1.3 & 3$^{\dagger}$ \\
K2-137 b & -3.11 & -3.60 & -4.04 & 0.64 & 0.22$^{\dagger}$ & 6.5$^{\dagger}$ & 1$^{\dagger}$ & 0.46 & 1471 & 10.9 & 409 & 3.5 & 18$^{\dagger}$ \\
\bottomrule
\caption{Planets located below the cosmic shorelines for \( f_{\text{initial}} = 10^{-4} \) are marked with an asterisk (\(^*\)). The scores for each atmospheric composition represent the difference in the base-10 logarithm of instellation between the planet and the cosmic shorelines for a 1 wt\% volatile mass fraction at a given planetary mass. More negative values are more atmosphere-favorable. $^{\dagger}$These planets lack a measured mass value and we assume an Earth-like composition for these calculations. $^{*}$These planets have been suggested to have no thick atmosphere based on thermal emission observations \citep{xue_jwst_2024,luque2024dark,greene_thermal_2023,kreidberg_absence_2019,crossfield_gj_2022,zhang_gj_2024}. We do not take the individual constrained stellar age into account, and the score is obtained by assuming an stellar age distribution from \cite{berger_gaiakepler_2020}.}
\label{tab:targets_appendix}
\addtocounter{table}{-1}
\end{longtable}

\clearpage

\begin{table*}[htpb!]
    \centering
    \resizebox{\textwidth}{!}{
    \begin{tabular}{l|c|l}
        \hline
        & \multicolumn{2}{c}{Assumptions} \\
        \hline
        & \cite{tian_thermal_2009-1}, \cite{tian_thermal_2009} 
        & a. Neutral \& Ion species included \\
        & & b. Jeans escape (hydrodynamic thermosphere) \\
        & & c. Heating: Collisions, chemical reactions, photoelectrons, radiative line heating \\
        & & d. Cooling: Recombination, molecular, Ly-$\alpha$, and atomic oxygen 63 $\mu$m emission \\
        \ch{CO2} & \multicolumn{2}{l}{1. Stellar bolometric luminosity: \cite{baraffe_new_2015} (no uncertainties included)} \\
         (Thermal)& \multicolumn{2}{l}{2. X-ray flux: \cite{jackson_coronal_2012} or \cite{selsis_habitable_2007}} \\
        & \multicolumn{2}{l}{3. EUV-to-X-ray ratio and uncertainties: \cite{king_euv_2020}} \\
        & \multicolumn{2}{l}{4. \textbf{Loss rate obtained using three different interpolation approaches} (see Fig. \ref{fig:CO2-interpolation})} \\
        \hline
        & \cite{chin_role_2024} 
        & a. 3D Block-Adaptive-Tree Solar Wind Roe-type Upwind Scheme (BATS-R-US) \\
        & & b. Multispecies magnetohydrodynamic (MS-MHD) model \\
        & & c. Ion species: \ch{H+}, \ch{O+}, \ch{O2+}, \ch{CO2+} and ionospheric photochemistry \\
        & & d. Unmagnetized planets assumed \\
        \ch{CO2} & \multicolumn{2}{l}{1. Stellar bolometric luminosity: \cite{baraffe_new_2015} (no uncertainties included)} \\
         (Non-Thermal)& \multicolumn{2}{l}{2. X-ray flux: \cite{jackson_coronal_2012} or \cite{selsis_habitable_2007}} \\
        & \multicolumn{2}{l}{3. Stellar mass loss rate scaled to X-ray: \cite{wood_new_2021} (no uncertainties included)} \\
        & \multicolumn{2}{l}{4. Loss rate scaled to orbital distance and stellar mass loss rate: \cite{dong_atmospheric_2018} (no uncertainties included)} \\
        \hline
        & \cite{johnstone_hydrodynamic_2020} 
        & a. Neutral \& Ion species included \\
        & & b. Transonic hydrodynamic \\
        & & c. Heating: Collisions, $\sim$500 chemical reactions, photoelectrons, radiative line heating \\
        & & d. Cooling: Recombination, molecular, Ly-$\alpha$, and atomic oxygen 63 $\mu$m emission \\
        \ch{H2O} (Thermal) & \multicolumn{2}{l}{1. Stellar bolometric luminosity: \cite{baraffe_new_2015} (no uncertainties included)} \\
        & \multicolumn{2}{l}{2. X-ray flux: \cite{jackson_coronal_2012} or \cite{selsis_habitable_2007}} \\
        & \multicolumn{2}{l}{3. EUV-to-X-ray ratio and uncertainties: \cite{king_euv_2020}} \\
        \hline
        & \cite{nakayama_survival_2022} 
        & a. Neutral \& Ion species included \\
        & & b. Jeans escape (hydrodynamic thermosphere)  \\
        & & c. Heating: Collisions, $\sim$500 chemical reactions, photoelectrons, radiative line heating \\
        & & d. Cooling: Recombination, molecular, \textbf{atomic line cooling of N, C, and O} \\
        \ch{N2/O2} (N22) & \multicolumn{2}{l}{1. Stellar bolometric luminosity: \cite{baraffe_new_2015} (no uncertainties included)} \\
        (Thermal)& \multicolumn{2}{l}{2. X-ray flux: \cite{jackson_coronal_2012} or \cite{selsis_habitable_2007}} \\
        & \multicolumn{2}{l}{3. EUV-to-X-ray ratio and uncertainties: \cite{king_euv_2020}} \\
        \hline
        & \cite{chatterjee_novel_2024} 
        & a. Neutral \& Ion species + \textbf{Ambipolar motion} \\
        & & b. Analytic approximations to transonic hydrodynamic and hydrostatic escape \\
        & & b. Heating: Parameterized \\
        & & c. Cooling: Recombination, \textbf{atomic line cooling of N and O} \\
        & & d. Introduces a new regime of global ion outflow controlled by a collisional-radiative thermostat \\
        \ch{N2/O2} (C24) & \multicolumn{2}{l}{1. Stellar bolometric luminosity: \cite{baraffe_new_2015} (no uncertainties included)} \\
        (Thermal)& \multicolumn{2}{l}{2. X-ray flux: \cite{jackson_coronal_2012} or \cite{selsis_habitable_2007}} \\
        & \multicolumn{2}{l}{3. EUV-to-X-ray ratio and uncertainties: \cite{king_euv_2020}} \\
        & \multicolumn{2}{l}{4. Includes uncertainty in the energy-limited loss regime (see Table \ref{tab:CP24})} \\
        \hline
        & Energy-Limited Estimate 
        & a. All uncertainties included in escape efficiency factor $\epsilon$ ($0.05<\epsilon<0.5$) \\
        \ch{CH4} & \multicolumn{2}{l}{1. Stellar bolometric luminosity: \cite{baraffe_new_2015} (no uncertainties included)} \\
        & \multicolumn{2}{l}{2. X-ray flux: \cite{jackson_coronal_2012} or \cite{selsis_habitable_2007}} \\
        & \multicolumn{2}{l}{3. EUV-to-X-ray ratio and uncertainties: \cite{king_euv_2020}} \\
        \hline
    \end{tabular}
    }
    \caption{Summary of model components for different atmospheric compositions. This includes key assumptions regarding species considered, heating and cooling mechanisms, and stellar input parameters.}
    \label{tab:model_assumptions}
\end{table*}
\clearpage

\newpage
\newpage

\section{Appendix Figures}\label{app:figures}
\renewcommand{\thefigure}{C\arabic{figure}}
\setcounter{figure}{0}

\renewcommand{\thetable}{C\arabic{table}}
\setcounter{table}{0}

\begin{figure*}[h!]
    \centering
    \includegraphics[width=1\linewidth]{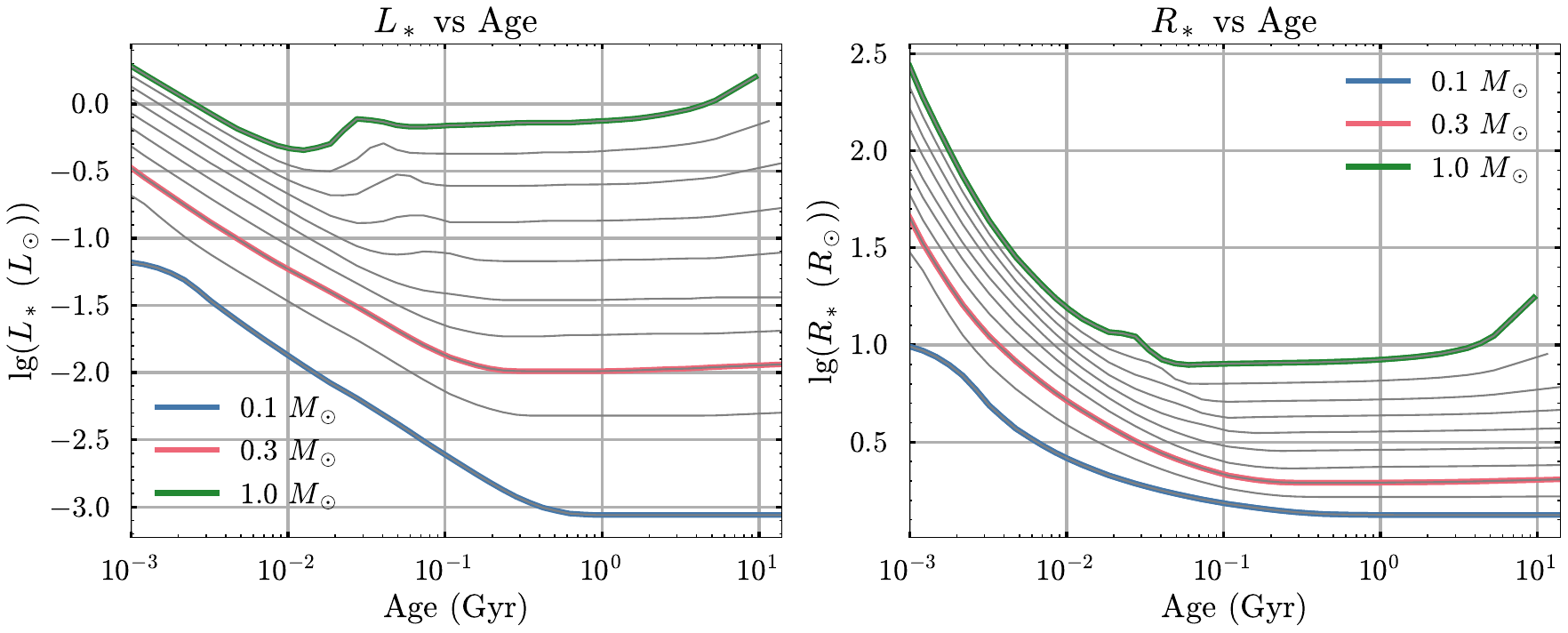}
    \caption{Stellar luminosity ($L_{bol}$) normalized to the Sun’s luminosity at the present-day ($L_\odot$), and stellar radius ($R_*$) normalized to the Sun’s radius ($R_\odot$) versus time from the model of \cite{baraffe_new_2015} for star masses ranging from 0.1 – 1.0 $M_\odot$.}
    \label{fig:baraffel}
\end{figure*}

\begin{figure*}[h!]
    \centering
    \includegraphics[width=1\linewidth]{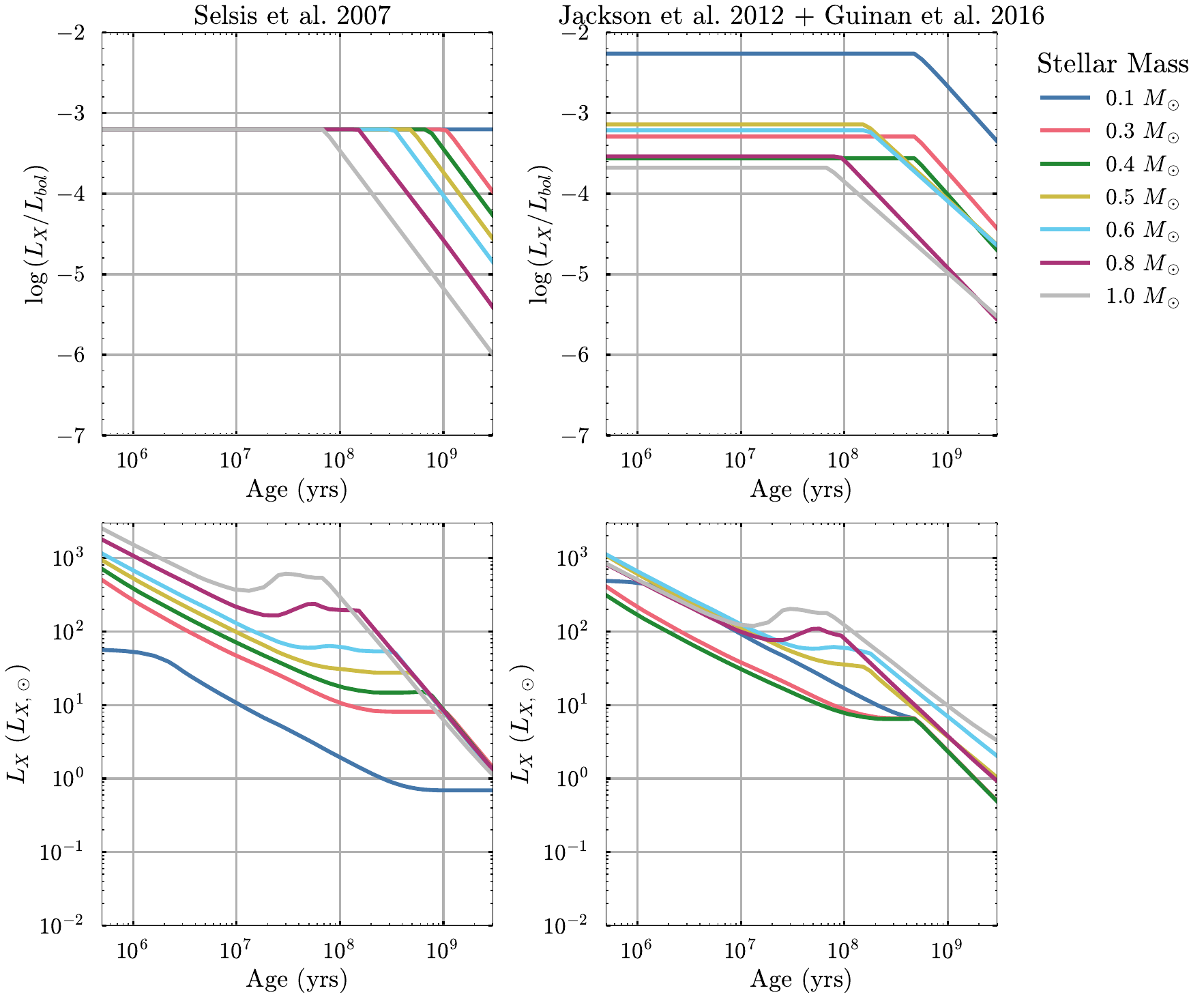}
    \caption{Upper panels: The ratio of X-ray luminosity to bolometric luminosity evolves over time for stellar masses from 0.1 $M_\odot$ to 1 $M_\odot$. The left panel presents the parameterization from \citep{selsis_habitable_2007}, while the right panel shows the results obtained by combining data from \citep{jackson_coronal_2012} and \citep{guinan_living_2016}. Lower panels: The X-ray luminosity scaled to solar X-ray flux evolves over time. In the lower right panel, the lines for \( M_* < 0.5 M_\odot \) overlap because \cite{guinan_living_2016} presents a unified X-ray evolution for M0–M5 stars after saturation phase. Note that $L_{x,\odot}/L_{bol,\odot} = 10^{-6.1}$.}
    \label{fig:X-ray}
\end{figure*}

\begin{figure*}[h!]
    \centering
    \includegraphics[width=1\linewidth]{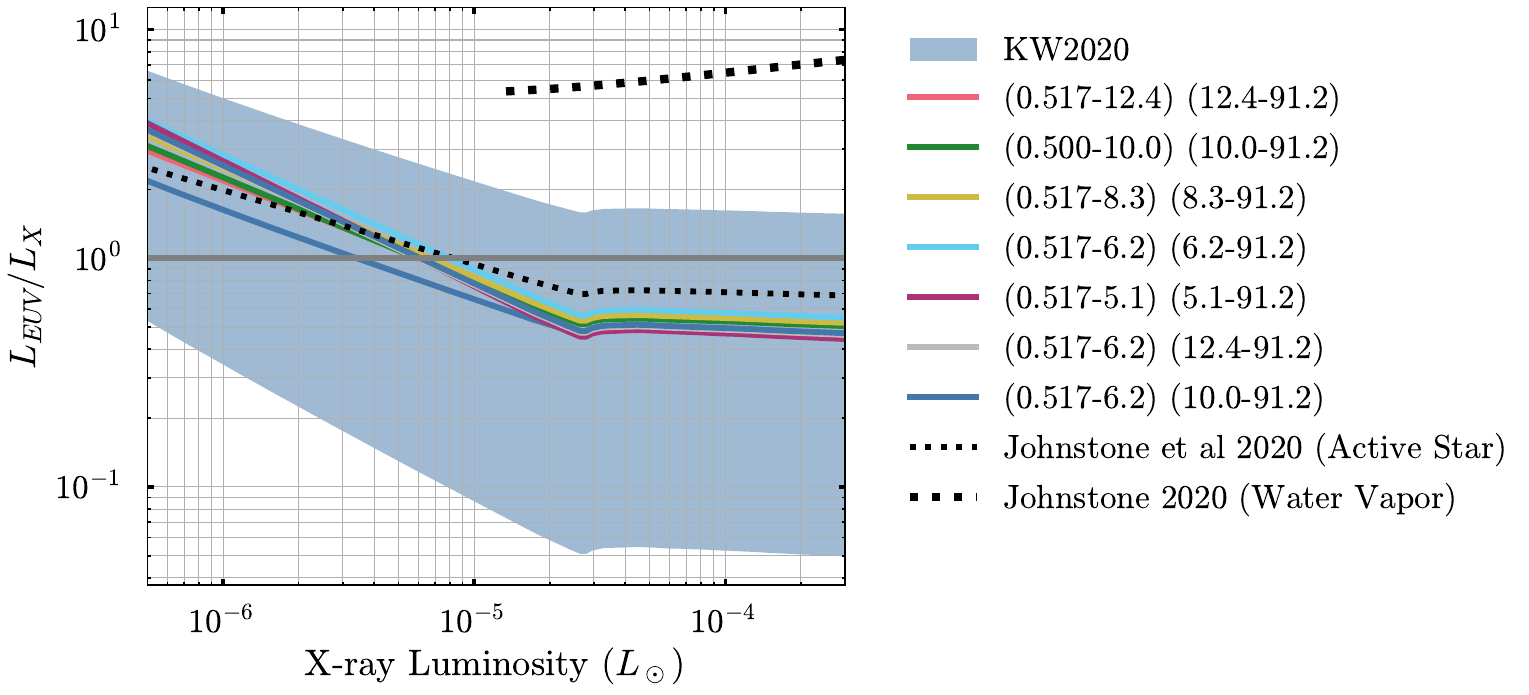}
    \caption{The ratio of EUV luminosity to X-ray luminosity is shown as a function of X-ray flux at Stellar surface. The X-ray flux is calculated for a 0.5 solar-mass star evolving from 1 Myr to 1 Gyr. The blue line and shaded region represent the results from \cite{king_euv_2020} including uncertainties. Additional colorful lines, labeled with wavelength ranges in \r{A}, are obtained from Table 4 of \cite{king_xuv_2018}, where the left bracket indicates the X-ray range and the right bracket indicates the EUV range.The dotted black line is derived from \cite{johnstone_active_2021}.}
    \label{fig:EUV}
\end{figure*}

\begin{figure*}[h!]
    \centering
    \includegraphics[width=1\linewidth]{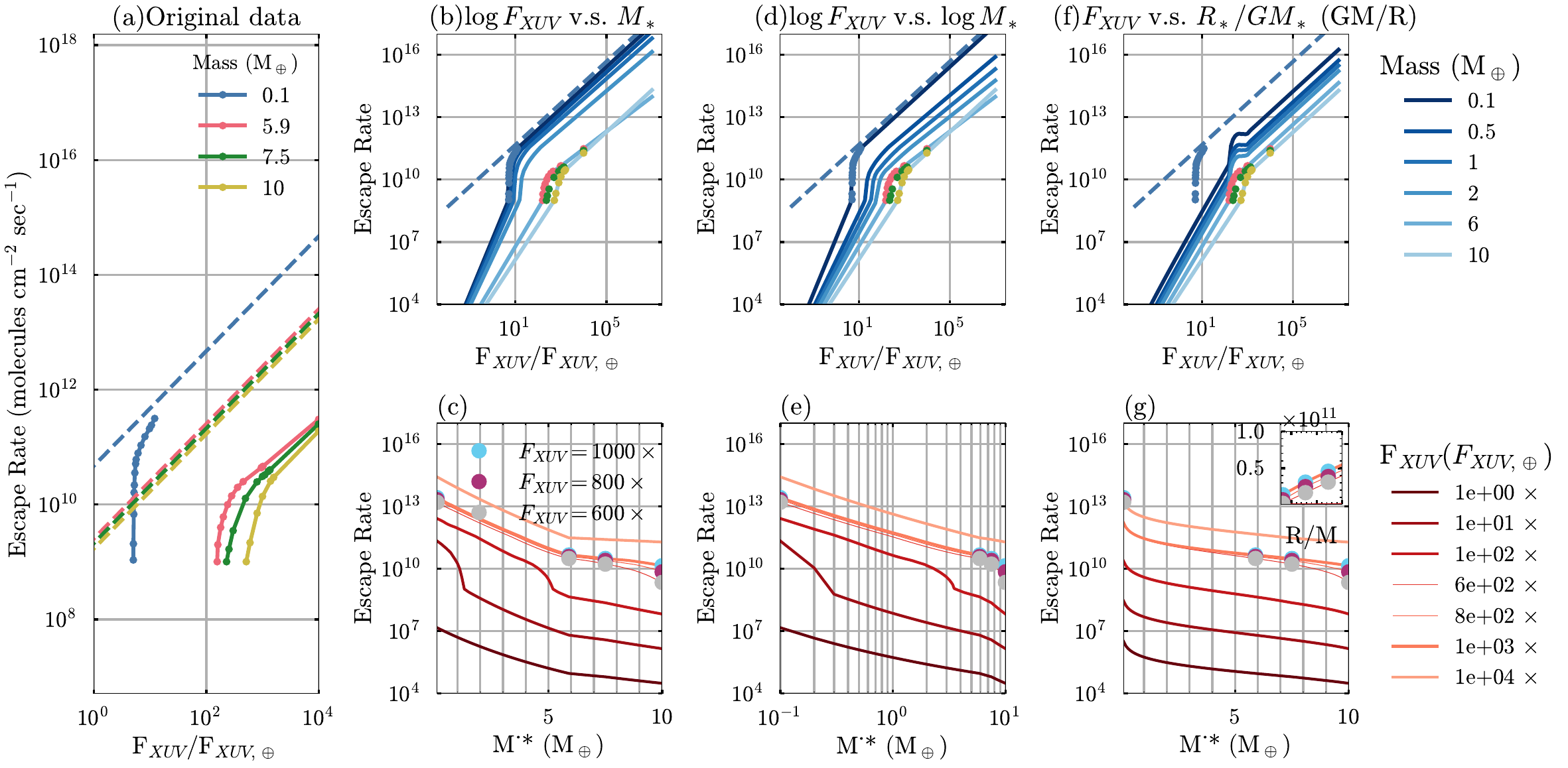}
    \caption{\ch{CO2} escape rates are sensitive to interpolation approaches. Leftmost panel: Atmospheric escape rate as a function of XUV flux, scaled to Earth's present-day value. The dots and connected lines represent output from \cite{tian_thermal_2009} and \cite{tian_thermal_2009-1}. The rightmost points for super-Earths are extrapolated by \cite{tian_thermal_2009}. The dashed line shows the energy-limited escape rate assuming an efficiency of 0.1. We interpolate their model output and our results for near-Earth-sized planets are sensitive to the interpolate approach used. (b-g) show this sensitivity. In (b)\&(c), we interpolate \(\log{F_{\text{xuv}}}\) with planetary mass \(M_p\). (d)\&(e) interpolates \(\log{F_{\text{xuv}}}\) with the logarithm of planetary mass (\(\log{M_p}\)). (f)\&(g) excludes Mars data and assumes the escape rate scales with gravitational potential energy (\(GM_p/R_p\)), interpolating between \(F_{\text{xuv}}\) and \(R_p/(GM_p)\). The upper panels illustrate how the escape rate varies with XUV flux for different planetary masses, while the lower panels show how the escape rate changes with planetary mass for various XUV flux levels. The dots represent benchmark output obtained from the leftmost panel, which are used for interpolation. Overall, the linear-planetary-mass provides higher escape rate estimates for near-Earth-sized planets compared to the other methods.}
    \label{fig:CO2-interpolation}
\end{figure*}

\begin{figure*}[h!]
    \centering
    \includegraphics[width=1\linewidth]{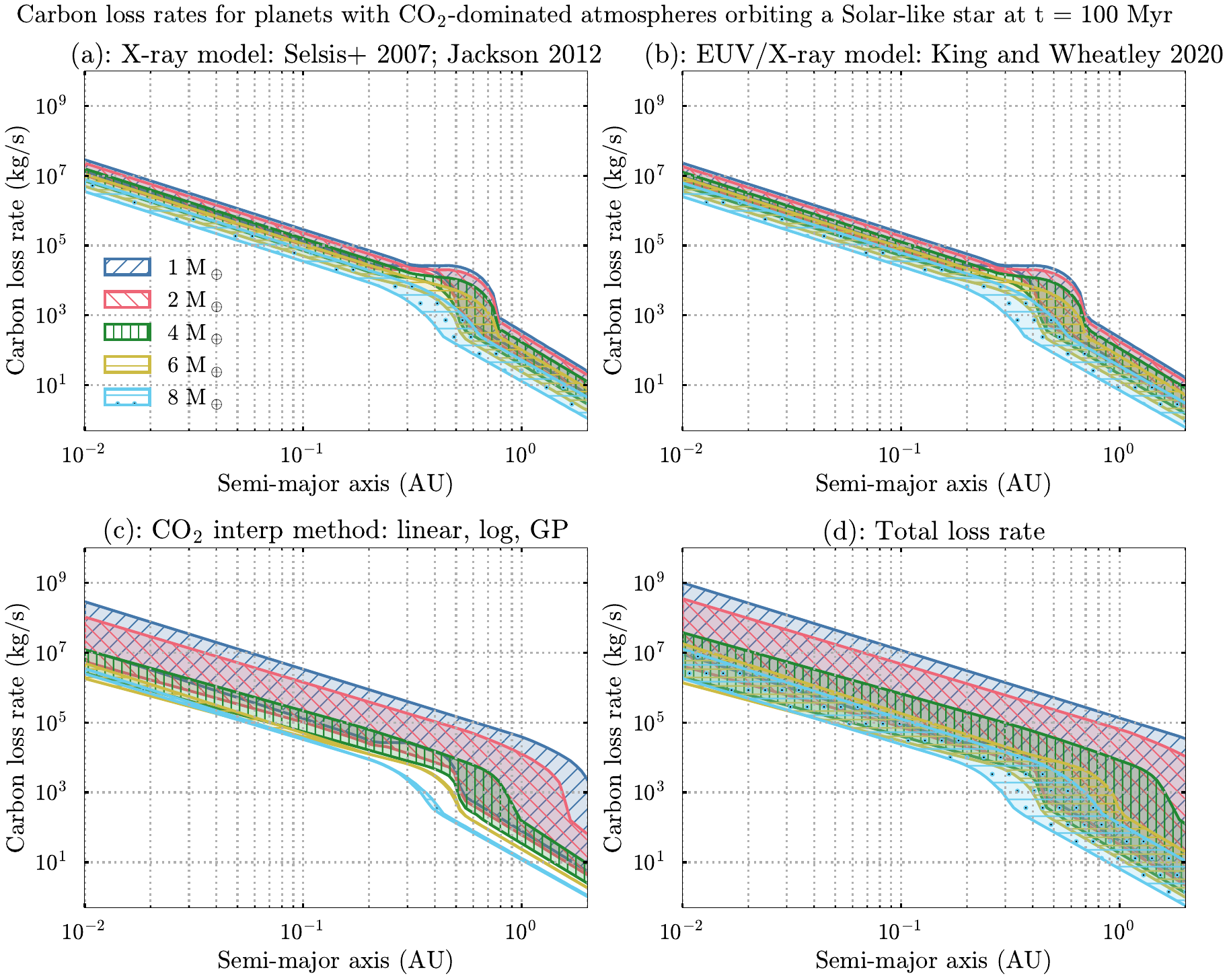}
    \caption{Carbon loss rate as a function of semi-major axis for planets with \ch{CO2}-dominated atmospheres orbiting a Solar-like star at \( t = 1 \) Myr. The shaded region represents the range between the maximum and minimum values under certain assumptions. The uncertainties are broken down as follows: (a): Using the center values of \(\gamma\) for EUV extrapolation and a interpolation of \ch{CO2} escape rates to \( M_p \) (linear), the uncertainty comes solely from the choice of X-ray model.  (b): Adopting the J12+G16 X-ray model and linear escape rate interpolation, the uncertainty arises only from the EUV extrapolation.  (c): Using the J12+G16 X-ray model and center EUV extrapolation values, the uncertainty stems only from escape rate interpolation.  (d): Incorporating all uncertainties. The uncertainties from the X-ray model and EUV extrapolation are of the same order of magnitude. The uncertainty from the interpolation method of \ch{CO2} escape rates is more significant for near-Earth-sized planets but negligible for super-Earths. The total uncertainty results in the loss rates for an 8 Earth-radius planet overlapping with those of a 1 Earth-radius planet.}
    \label{fig:CO2-unc}
\end{figure*}

\begin{figure*}[h!]
    \centering
    \includegraphics[width=0.9\linewidth]{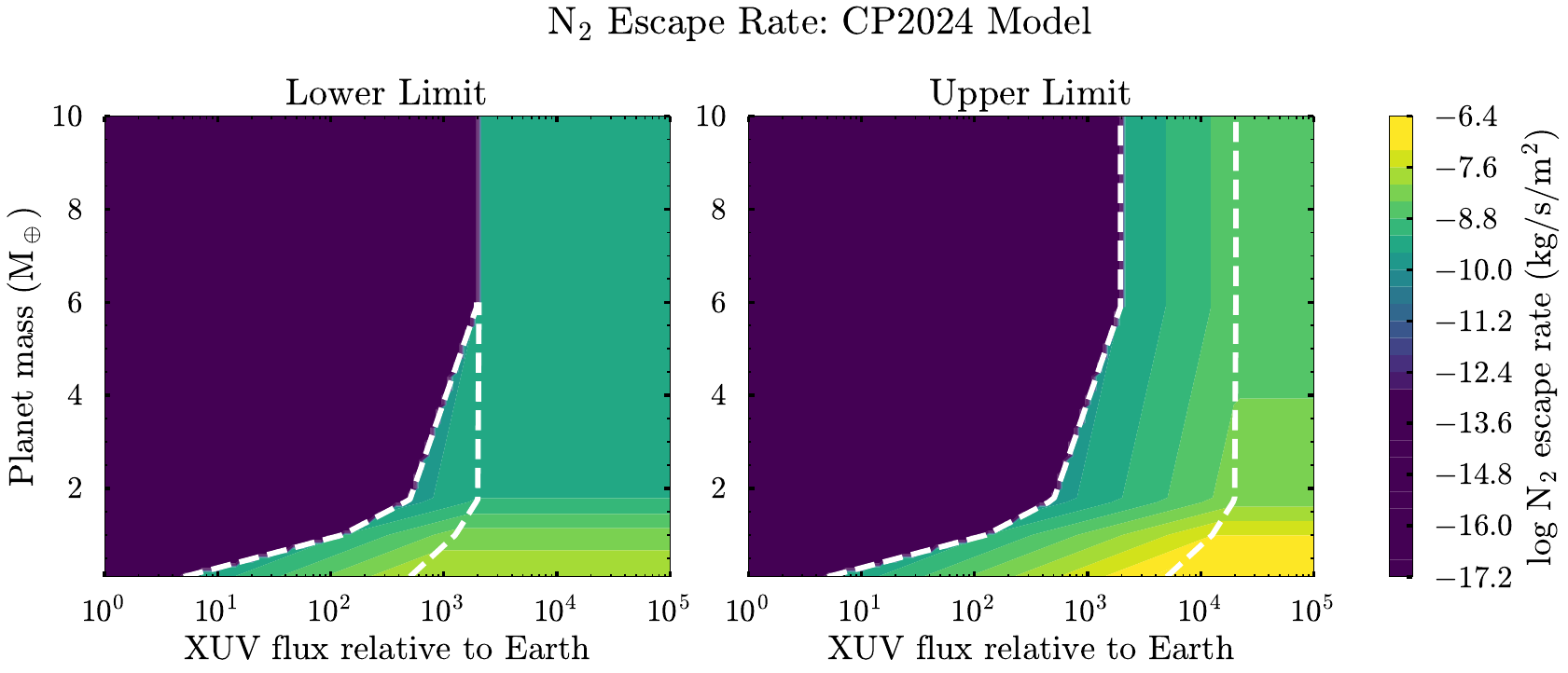}
    \caption{The loss rate per surface area for an \(\text{N}_2/\text{O}_2\) atmosphere is shown as a function of XUV flux and planetary mass, interpolated/extrapolated from \cite{chatterjee_novel_2024}. The dark purple region indicates where hydrostatic escape occurs, with the escape rate fixed at \(10^{-13}\) kg/s/m\(^2\). The white dashed lines show bounds to the energy-limited regime.}
    \label{fig:N2O2-loss-rate}
\end{figure*}

\begin{figure*}[h!]
    \centering
    \includegraphics[width=0.5\linewidth]{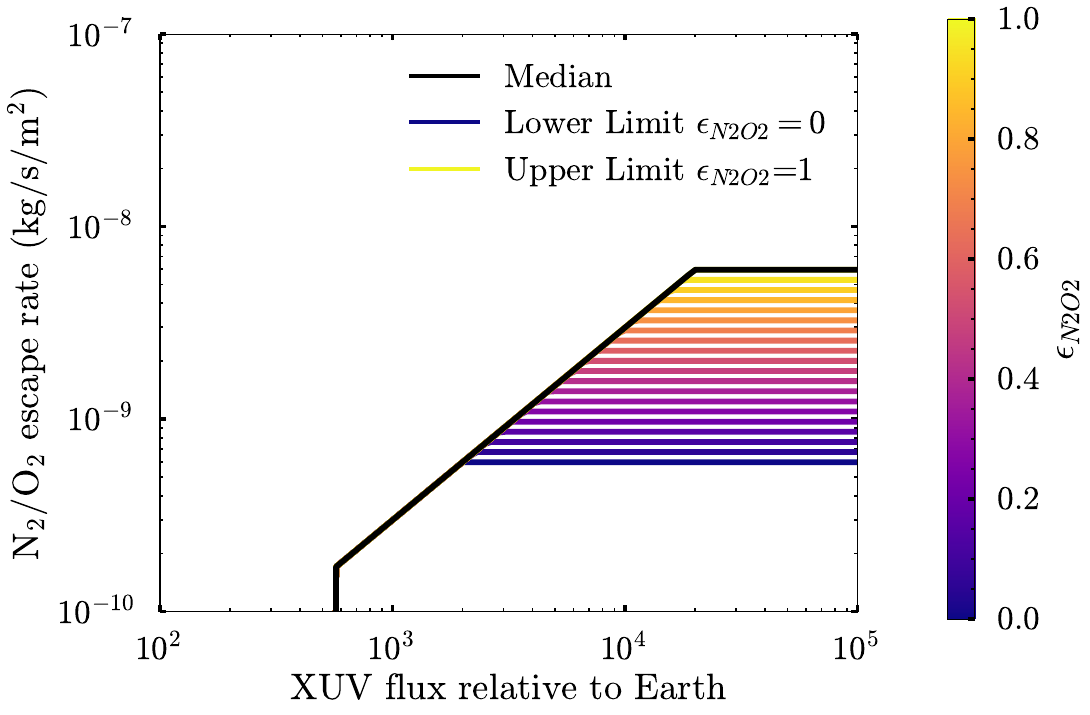}
    \caption{The \ch{N2/O2} loss rate as a function of XUV flux. In our Monte Carlo simulations, we randomly sample the \(\epsilon_{N2O2}\) from a uniform distribution between 0 and 1. The variations in escape rates resulting from this sampling are illustrated by the colorful lines.}
    \label{fig:N2O2-loss-rate-frac}
\end{figure*}

\begin{figure*}
    \centering
    \includegraphics[width=0.9\linewidth]{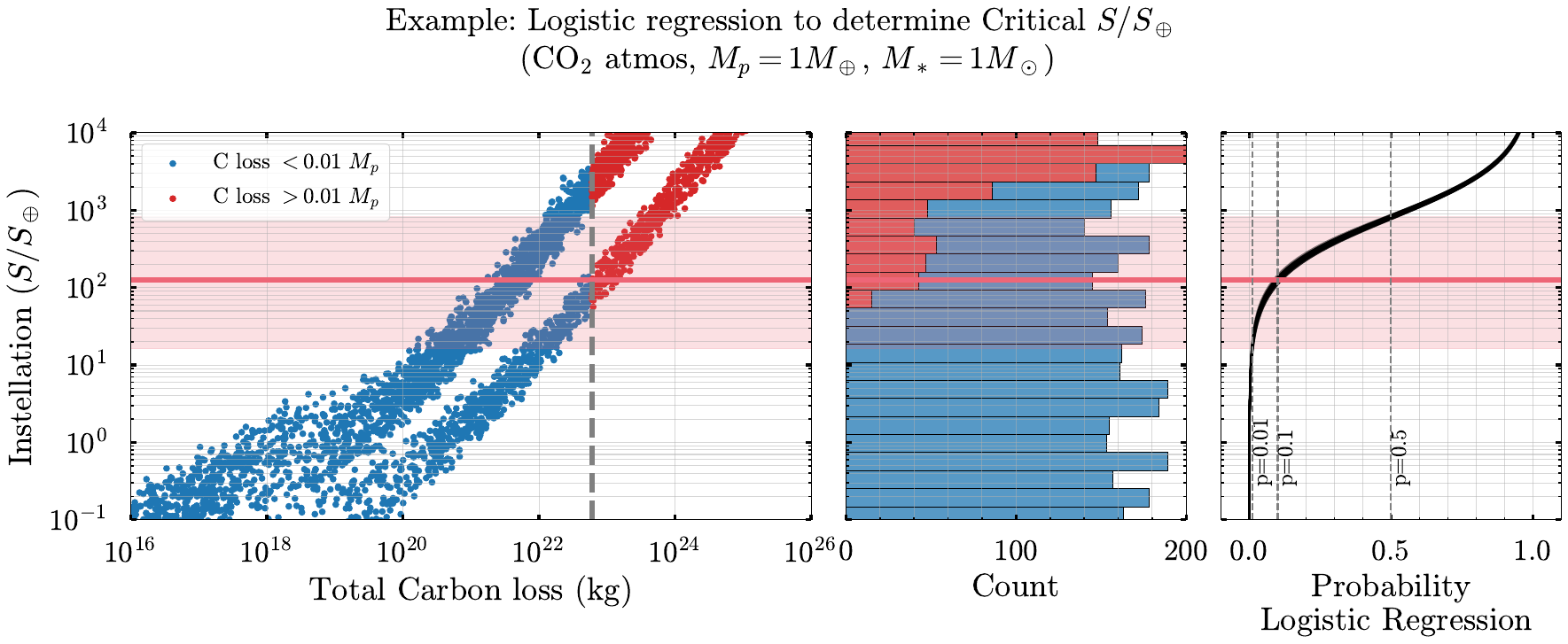}
    \includegraphics[width=0.9\linewidth]{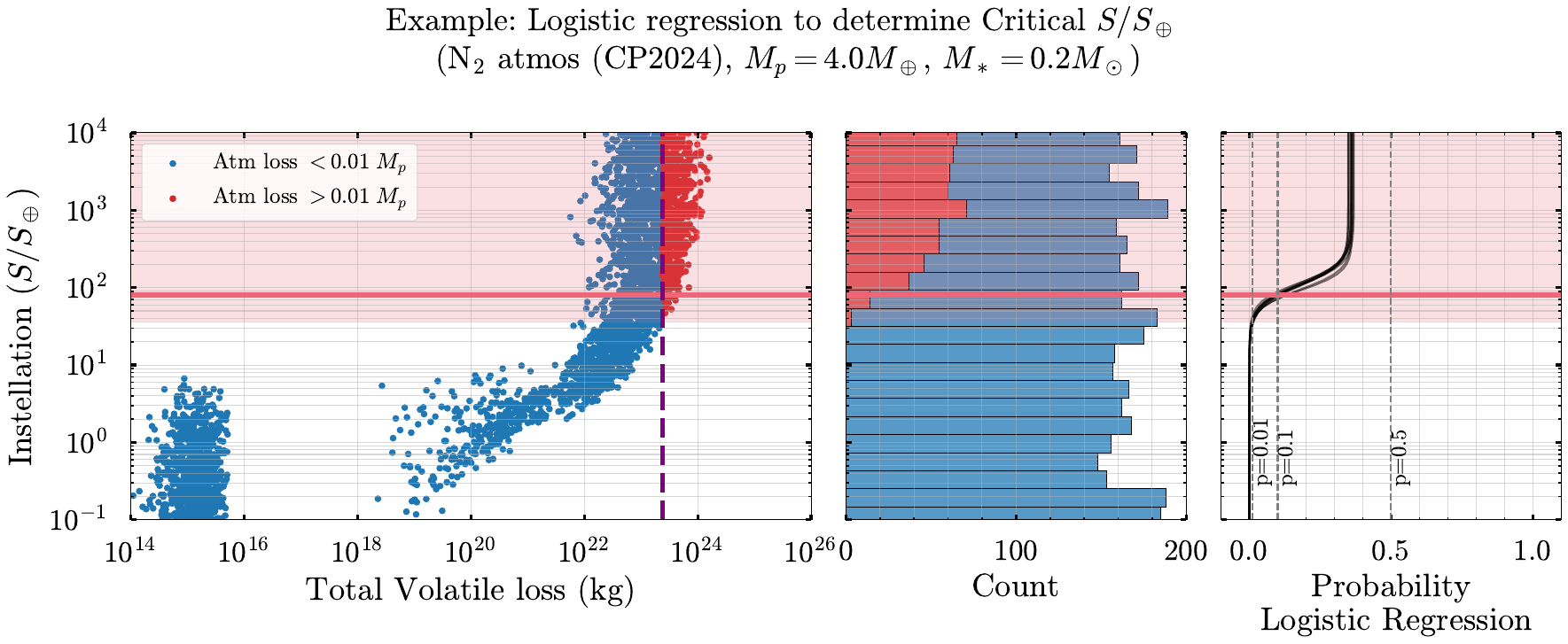}
    \caption{Example Monte Carlo simulation results for an Earth-mass planet with a \ch{CO2} atmosphere orbiting a $1\,M_\odot$ star (Upper) and a 4 Earth-mass planet with a \ch{N2}-dominated atmosphere orbiting a $0.2\,M_\odot$ star (Lower). Left: Bolometric instellation, scaled to earth's value\((S / S_\oplus)\), plotted against the total carbon mass lost. Blue dots indicate Monte Carlo outcomes in which the planet retains its atmosphere (with a initial carbon mass of \(0.01\,M_\oplus\)), while orange dots indicate outcomes in which the atmosphere is stripped. The middle green line marks the critical instellation at \(p_0 = 0.9\), meaning there is a 50\% probability of atmospheric retention at that flux level. The shaded region spans from \(p_0=0.99\) (lower boundary) to \(p_0=0.5\) (upper boundary), illustrating instellation thresholds for 99\% and 50\% retention probability, respectively. To obtain a robust estimate of the critical instellation \(S^*\) at a given \(p_0\), we draw 5,000-point bootstrap samples from our 10,000-point Monte Carlo dataset (in 10 iterations), refit the logistic model each time, and then average the resulting \(S^*\) values. Right: Histograms of the instellation values for planets that retain atmosphere (blue) versus those that lose it (orange). For our implementation of the CP2024 \ch{N2}-escape, in the volatile rich limit the probability of atmospheric loss does not reach one due to the plateau in escape rates for the highest XUV fluxes. To capture this behavior, we apply a three-parameter logistic fitting. In contrast, for \ch{CO2} and other atmospheres where the escape rate continues to increase with XUV flux, the loss probability approaches one at infinite high XUV flux, so it is reduced to a 2 parameters fitting. }
    \label{fig:logistic}
\end{figure*}

\begin{figure*}
    \centering
    \includegraphics[width=1.0\linewidth]{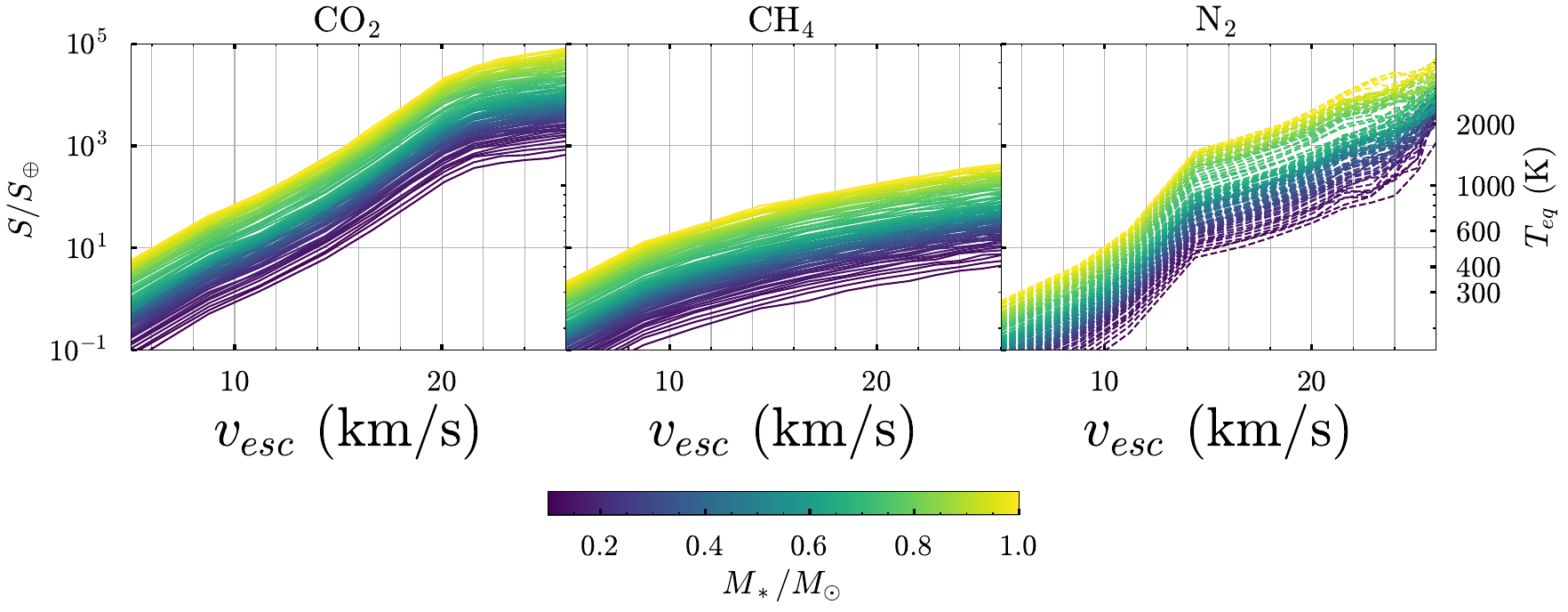}
    \caption{Cosmic shorelines as functions of planetary mass and instellation scaled to Earth's value \((S / S_\oplus)\). Each line corresponds to results computed for stellar masses within a range of 0.01~\(M_\odot\). For example, the lowest (darkest) line represents stellar masses between 0.10 and 0.11~\(M_\odot\). The priority scores listed in Table~\ref{tab:targets} and Table~\ref{tab:targets_appendix} are derived from these shoreline calculations.}
    \label{fig:CS_0d01st_mass}
\end{figure*}

\begin{figure*}
    \centering
    \includegraphics[width=0.5\linewidth]{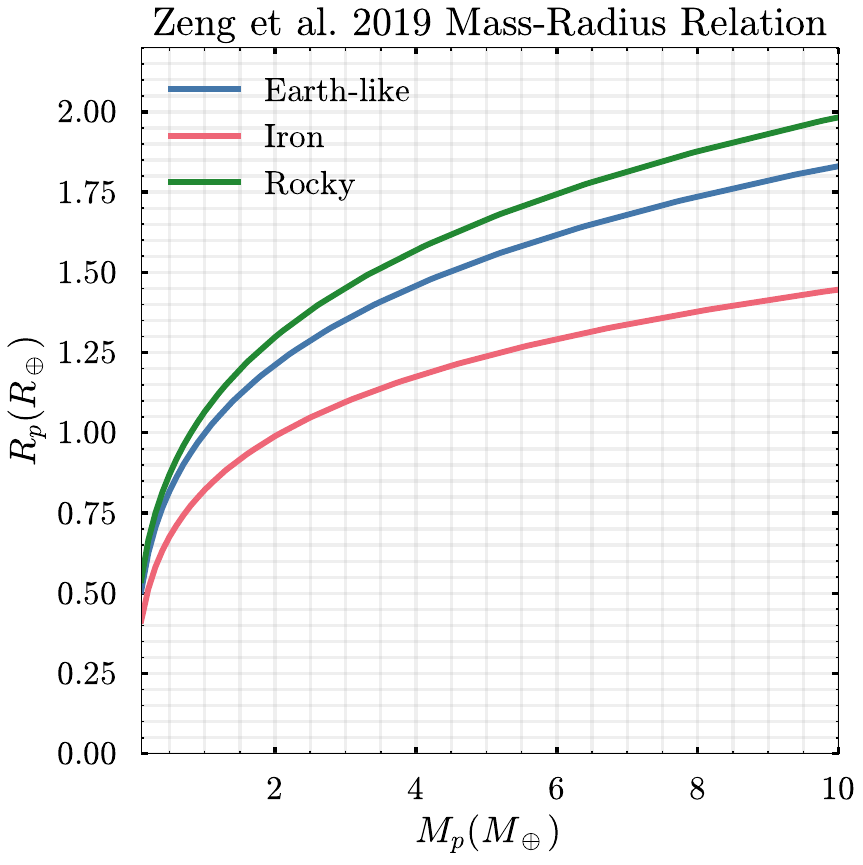}
    \caption{The mass-radius relationship for different planetary compositions: pure iron (100\% Fe), Earth-like rocky (32.5\% Fe + 67.5\% \ch{MgSiO3}), and pure rock (100\% \ch{MgSiO3}). }
    \label{fig:MR}
\end{figure*}

\begin{figure*}
    \centering
    \includegraphics[width=0.5\linewidth]{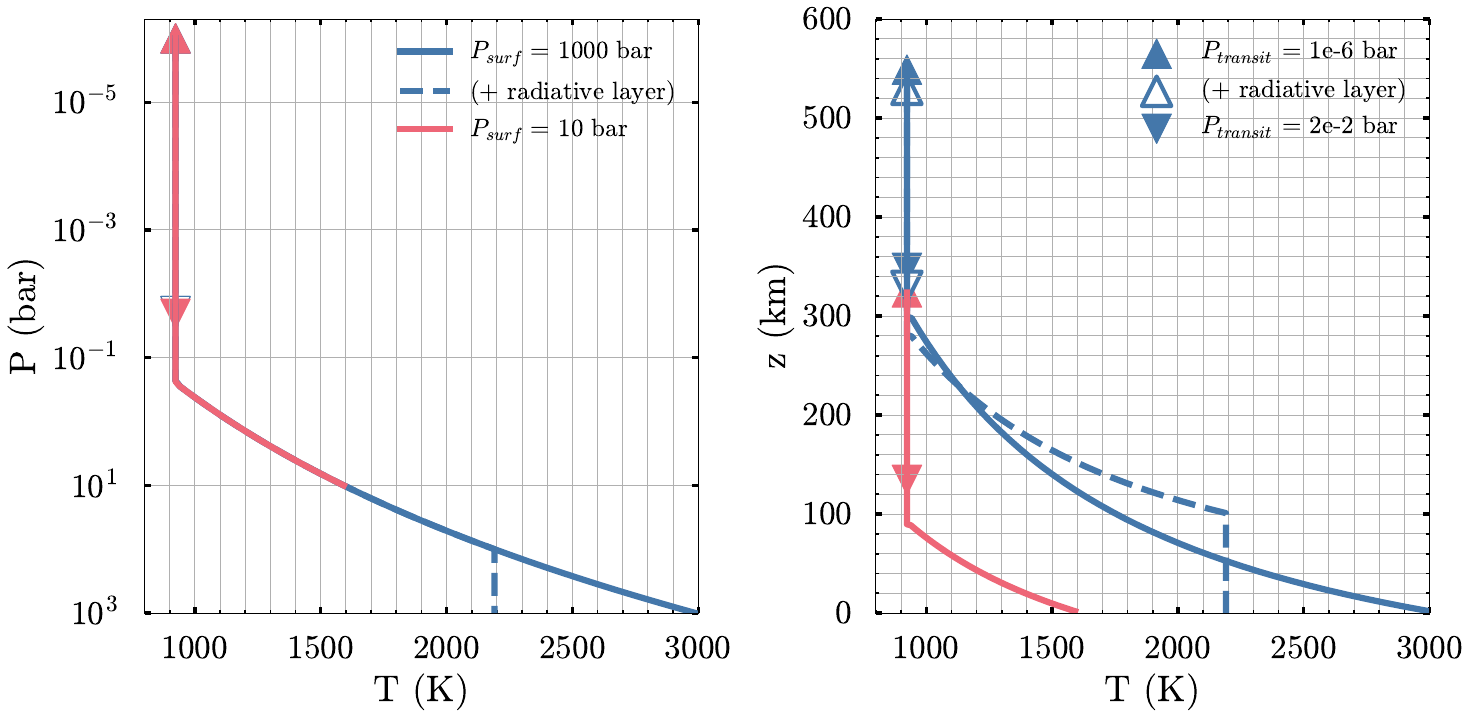}
    \caption{Left Panel: Temperature-Pressure profiles for atmospheres with surface pressures of 10 bar (pink) and 100 bar (blue). The solid lines correspond to the two-layer model (Scenario (a) in Fig. \ref{fig:1D-profile}), where the atmosphere consists of a convective lower layer and an isothermal upper layer beyond the radiative-convective boundary (RCB). The dashed lines represent the three-layer model (Scenario (b) in Fig. \ref{fig:1D-profile}), where an additional radiative layer forms near the surface. Right Panel: Altitude-temperature profiles for the same cases shown in the left panel. Triangular markers indicate the pressure level probed by the transit method, with filled symbols representing the two-layer model (Scenario (a)) and open symbols representing the three-layer model (Scenario (b)).}
    \label{fig:TP-profile}
\end{figure*}

\begin{figure*}
    \centering
    \includegraphics[width=0.9\linewidth]{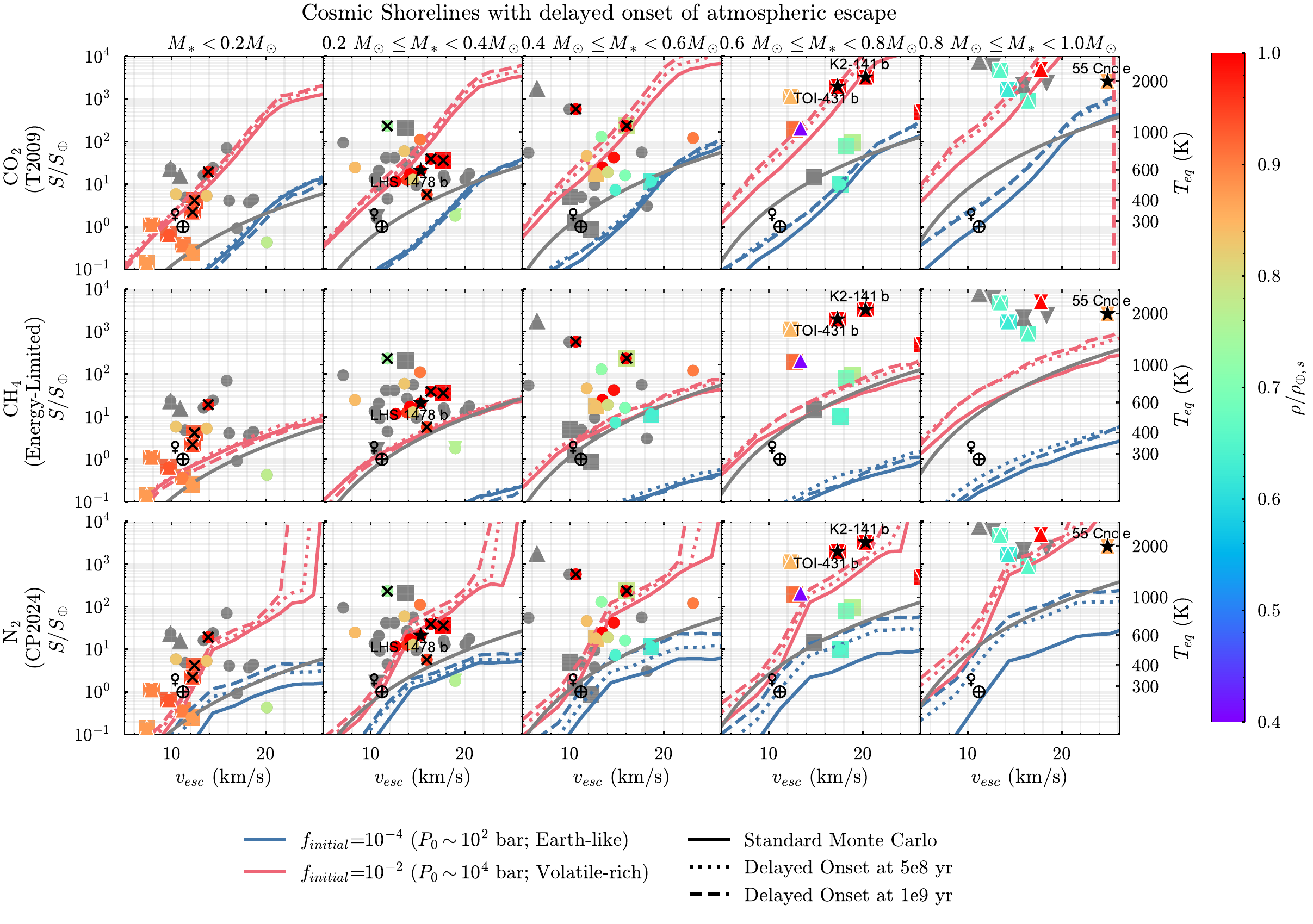}
    \caption{Same configuration and markers as Fig. \ref{fig:CS_pl_vesp}. Here we plot only the 90\% cosmic shoreline for two initial volatile fractions, but instead of sampling the escape onset from Table \ref{tab:monte_carlo}, we delay and fix the start of escape at 500 Myr and 1 Gyr. This illustrates the potential retention of the atmosphere outgassed from the volatiles sequestered in the solid mantle after magma-ocean crystallization, under the assumptions that the initial volatile atmosphere and dissolved magma‑ocean volatiles were completely removed and that escape is not supply‑limited.}
    \label{fig:CS_delay}
\end{figure*}

\begin{figure*}
    \centering
    \includegraphics[width=0.9\linewidth]{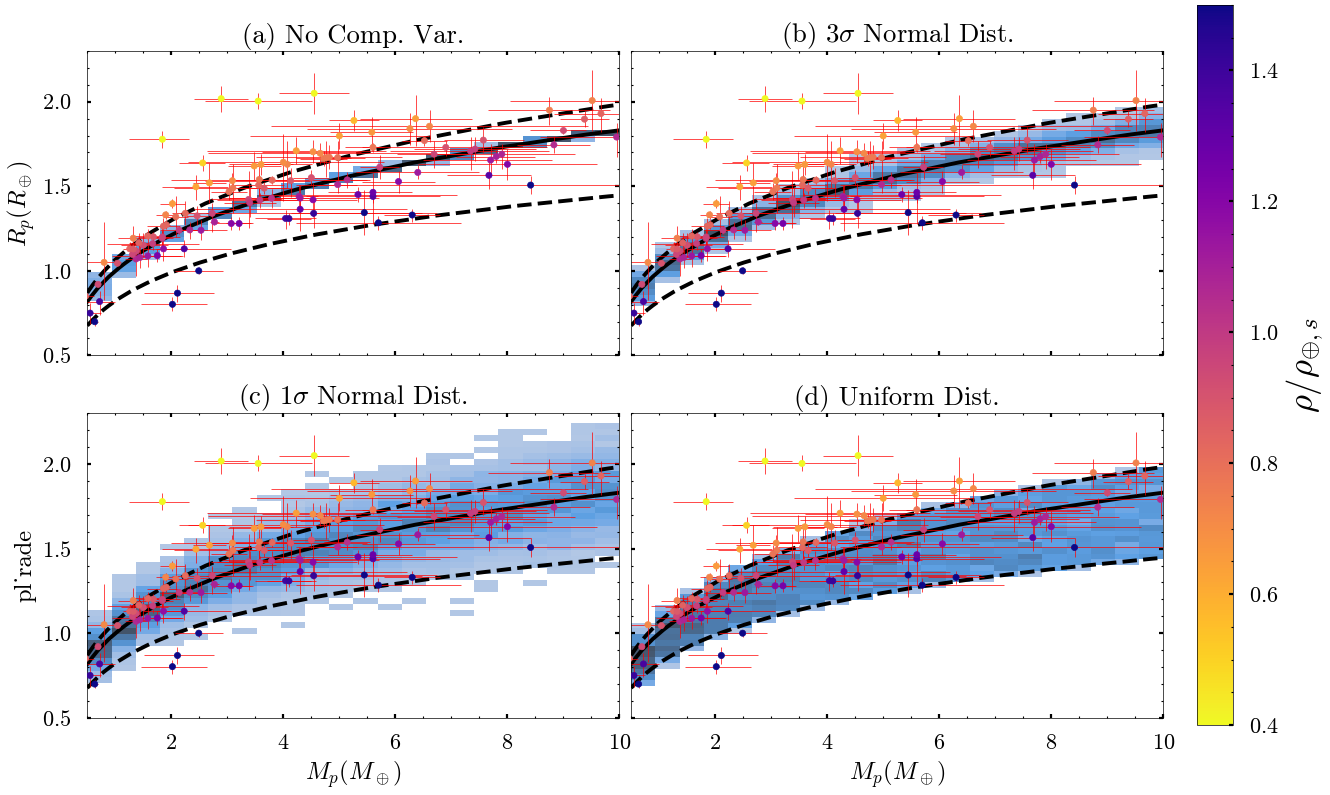}
    \caption{Mass-Radius Relationship for Solid-Body Planets Under Three Composition Assumptions: Given a planetary mass (\(M_p\)), the radius is assigned based on different composition assumptions: (a) a fixed Earth-like composition (\( R_{\text{earth-like}}(M_p) \)), (b) a normal distribution centered at \( R_{\text{earth-like}}(M_p) \) with a standard deviation equal to one-third of the difference between \( R_{\text{earth-like}}(M_p) \) and \( R_{\text{silicate}}(M_p) \), (c) a normal distribution centered at \( R_{\text{earth-like}}(M_p) \) with a standard deviation equal the difference between \( R_{\text{earth-like}}(M_p) \) and \( R_{\text{silicate}}(M_p) \), and (d) a uniform distribution spanning the range between \( R_{\text{silicate}}(M_p) \) (pure \ch{MgSiO3}) and \( R_{\text{iron}}(M_p) \) (pure \ch{Fe}). The shaded blue region represents a 2D histogram of our Monte Carlo samples, with darker colors indicating higher number density. The  dots are observed rocky planet candidates from the NASA Exoplanet Archive, with error bars representing measurement uncertainties. The colors represent the measured density relative to Earth's composition.}
    \label{fig:MR_solid}
\end{figure*}

\begin{figure*}
    \centering
    \includegraphics[width=0.9\linewidth]{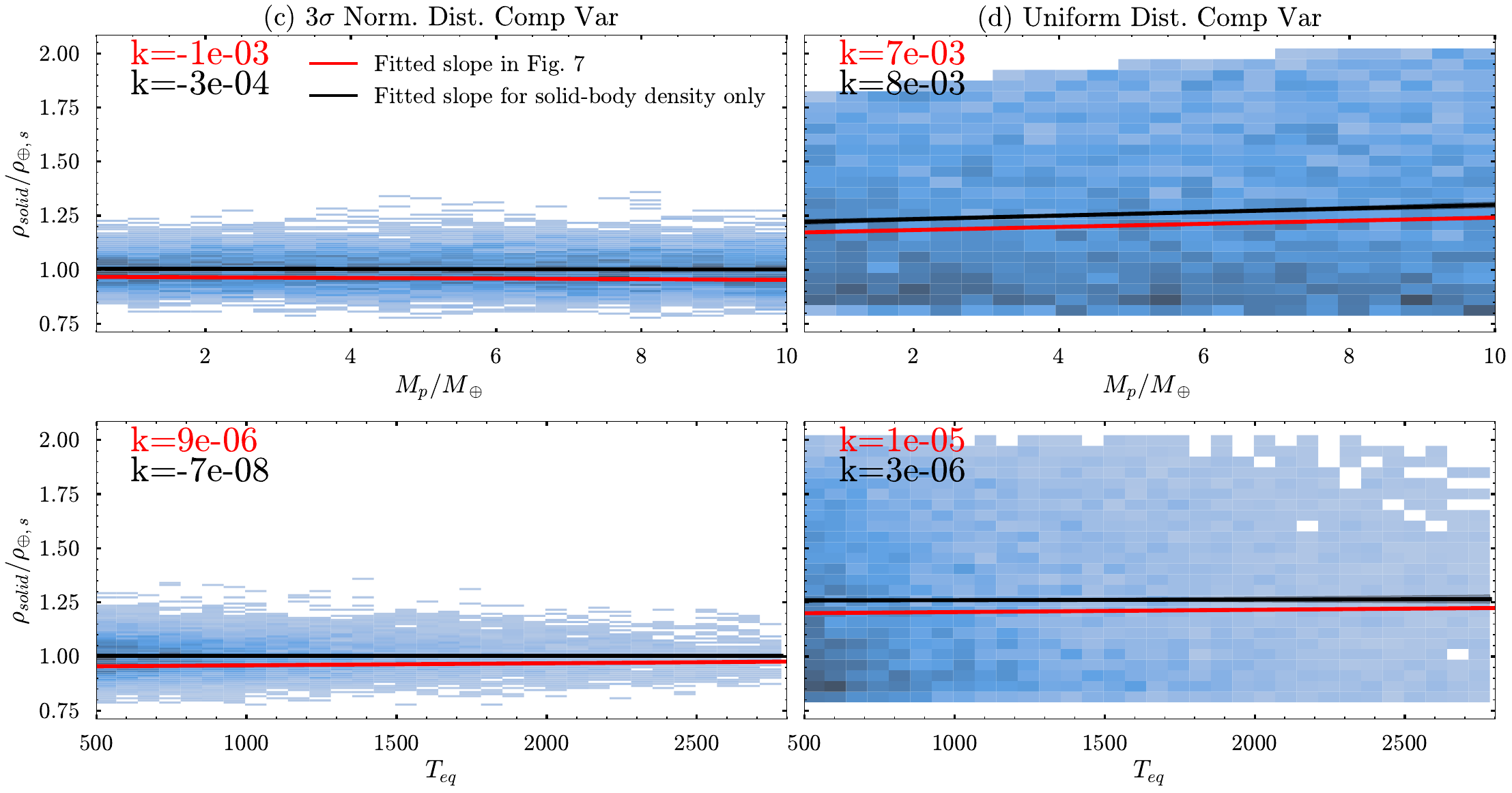}
    \caption{Trend of the Scaled Density of the Solid Body Only:  The upper row shows 2D histograms of the scaled density of the solid body (\(\rho_{solid}/\rho^*_{\oplus}\)) as a function of planetary mass (\(M_p\)), while the lower row shows its relationship with equilibrium temperature (\(T_{eq}\)). The left column (c) assumes that solid density variations follow a normal distribution, with a standard deviation equal to 1/3 of the difference between \( R_{\text{earth-like}}(M_p) \) and \( R_{\text{silicate}}(M_p) \). The right column (d) corresponds to a uniform distribution spanning the range between \( R_{\text{silicate}}(M_p) \) (pure \ch{MgSiO3}) and \( R_{\text{iron}}(M_p) \) (pure \ch{Fe}). The black lines and black text indicate the results of linear regression for the solid-body scaled density, while the red ones represent the linear regression of the total density derived from the transit radius (see Fig. \ref{fig:rho_ratio_pattern}). In (c), the slope of the solid-body density trend is an order of magnitude lower than that of the final total density, suggesting an insignificant contribution from variations in the solid composition. However, in (d), the solid-body density trend significantly influences the overall density slope, even changes the sign for slope of \(\rho_{solid}/\rho^*_{\oplus}-M_p\). We also tested uniformly sampling density instead of radius, but the trend caused by the solid portion alone persists.}
    \label{fig:solid-only}
\end{figure*}

Appendix C compiles supplementary figures (\ref{fig:baraffel}--\ref{fig:solid-only}) illustrating the models inputs, escape parameterizations, and diagnostic behaviors underlying the results presented in the main text.

Figure \ref{fig:baraffel} shows evolution of stellar luminosity and radius based on the evolutionary tracks of \citet{baraffe_new_2015}.

Figure \ref{fig:X-ray} compares two parameterizations of the X-ray to bolometric luminosity ratio as a function of stellar age, highlighting the divergence between the Selsis (2007) and Jackson \& Guinan (2012, 2016) models, as well as the resulting X-ray flux evolution.

Figure \ref{fig:EUV} presents the ratio of EUV to X-ray luminosity and its adopted uncertainty range, compared with empirical relations from several previous studies.

Figure \ref{fig:CO2-interpolation} show how the sensitivity of \ch{CO2} atmospheric escape rates to different interpolation schemes of the \citet{tian_thermal_2009} data. 

Figure \ref{fig:CO2-unc} quantifies the uncertainty in carbon loss rates for \ch{CO2}-atmosphere resulting from three major factors—X-ray model, EUV extrapolation, and interpolation method—showing that the interpolation choice dominates for small planets.

Figure \ref{fig:N2O2-loss-rate} presents escape rates for \ch{N2/O2} atmospheres from \citet{chatterjee_novel_2024}, identifying transitions between hydrostatic and energy-limited regimes as a function of XUV flux and planetary mass.

Figure \ref{fig:N2O2-loss-rate-frac} shows how randomizing the energy-limited efficiency parameter $\epsilon_{N2O2}$ introduces spread in the \ch{N2/O2} loss rates.

Figure \ref{fig:logistic} provides examples of Monte Carlo logistic-regression fits used to determine the critical instellation.

Figure \ref{fig:CS_0d01st_mass} illustrates the dependence of the “cosmic shorelines” on stellar mass, computed in interval of 0.01 $M_\odot$.

Figure \ref{fig:MR} summarizes the mass-radius relations assumed for three different planetary compositions (iron-rich, Earth-like, and silicate-only), which form the solid radius for the  transit radius calculations.

Figure \ref{fig:TP-profile} compares temperature–pressure profiles from two-layer and three-layer radiative–convective models for thick \ch{CO2} atmospheres, illustrating how an additional near-surface radiative zone modifies the thermal structure and observable transit level.

Figure \ref{fig:CS_delay} tests delayed atmospheric escape onset times (500 Myr and 1 Gyr) to illustrate the potential for atmospheric revival from volcanism outgassing.

Figure \ref{fig:MR_solid} shows the statistical mass–radius distributions adopted for Monte Carlo sampling, along with observed rocky planets for reference.

Figure \ref{fig:solid-only} isolates the density trend of the solid portion of planets, demonstrating that composition-induced density variations have a smaller effect than volatile content on the total inferred density trends.

\end{CJK*}
\end{document}